\documentclass[aip,jcp,amsmath,amssymb,floatfix,citeautoscript,reprint]{revtex4-1}

\usepackage{graphicx}
\usepackage{subcaption}
\usepackage{dcolumn}
\usepackage{bm}
\usepackage{braket}
\usepackage[usenames,dvipsnames]{color}
\usepackage{ragged2e}
\usepackage[labelsep=period,format=plain,justification=justified,font=footnotesize]{caption}
\usepackage{times}
\usepackage{txfonts}
\usepackage{multirow}
\usepackage{makecell}
\usepackage{comment}
\usepackage{array}
\usepackage{tabu}
\usepackage[version=3]{mhchem} 
\usepackage{amssymb}

\DeclareCaptionJustification{myjust}{\justifying}
\newcommand{\uR}{{\bf R}}
\newcommand{\uP}{{\bf P}}
\newcommand{\uF}{{\bf F}}

\begin{document}

\title{Elucidating the proton transport pathways in liquid imidazole with first-principles molecular dynamics}

\author{Zhuoran Long} 
\affiliation{Department of Chemistry, New York University, New York, NY 10003, USA}

\author{Austin O. Atsango}
\affiliation{Department of Chemistry, Stanford University, Stanford, California 94305, USA}

\author{Joseph A. Napoli}
\affiliation{Department of Chemistry, Stanford University, Stanford, California 94305, USA}

\author{Thomas E. Markland}
\email{tmarkland@stanford.edu}
\affiliation{Department of Chemistry, Stanford University, Stanford, California 94305, USA}

\author{Mark E. Tuckerman}
\email{mark.tuckerman@nyu.edu}
\affiliation{Department of Chemistry, New York University, New York, NY 10003, USA}
\affiliation{Courant Institute of Mathematical Science, New York University, New York, NY 10012, USA}
\affiliation{NYU-ECNU Center for Computational Chemistry at NYU Shanghai, 3663 Zhongshan Road North, Shanghai 200062, China}

\date{\today}

\begin{abstract}
Imidazole is a promising anhydrous proton conductor with a high conductivity comparable to that of water at a similar temperature relative to its melting point. Previous theoretical studies of the mechanism of proton transport in imidazole have relied either on empirical models or on {\it ab initio} trajectories that have been too short to draw significant conclusions. Here, we present the results of {\it ab initio} molecular dynamics simulations of an excess proton in liquid imidazole reaching 1 nanosecond in total simulation time. We find that the proton transport is dominated by structural diffusion, and the diffusion constant of the proton defect is $\sim$8 times higher than the self-diffusion of the imidazole molecules. By using correlation function analysis, we decompose the mechanism for proton transport into a series of first-order processes and show that the proton transport mechanism occurs over three distinct time and length scales. Although the mechanism at intermediate times is dominated by hopping along pseudo one-dimensional chains, at longer times, the overall rate of diffusion is limited by the reformation of these chains, thus providing a more complete picture of the traditional, idealized Grotthuss structural diffusion mechanism. 
\end{abstract}

{\maketitle}

Proton transport in hydrogen-bonded media is a fundamental process in a myriad of systems relevant to chemistry, physics, and biology. In these media, proton transport can occur via a series of intermolecular proton transfer reactions, a process referred to as structural diffusion or the "Grotthuss mechanism"~\cite{Grot_1806,Marx_Rev_2006}, which allows charge migration to occur much faster than diffusion of the molecules themselves. While structural diffusion phenomena have been extensively studied in aqueous media~\cite{Tuckerman_JPC_1995,Agmon_CPL,Tuckerman_JCP_1995,Tuckerman_Nature_Hp,Marx_Rev_2006,Tuckerman_rev_2010,Agmon_rev_2016,Tuckerman_Nature_Hp}, proton transport in anhydrous hydrogen-bonded environments has received much less attention. However, non-aqueous liquids such as methanol~\cite{Morrone_02,Reed_08,Park_15}, phosphoric acid~\cite{Watanabe_10,Kreuer_12,Kreuer_13,Banerjee_14}, and liquid imidazole~\cite{Voth_09,Shen_12,Martinelli_16}, as well as solid proton conductors such as cesium hydrogen sulfate~\cite{Matvienko_04,Marzari_07}, cesium dihydrogen phosphate~\cite{Haile_01,MET_08,Grey_13,Santamaria_19}, solid imidazole~\cite{Seifert_01,Parrinello_04,Iannuzzi_06}, and solid oxides/perovskites~\cite{Kreuer_96,Kreuer_99,Kreuer_03,Islam_10} have also been shown experimentally and theoretically to support structural diffusion mechanisms. The wide variety of anhydrous systems capable of efficiently transporting protons thus provides the opportunity to use factors such as temperature, functionalization, confinement, phase (solid vs. liquid), and causticity to tailor the mechanism of proton transport and engineer new proton-conducting systems with a range of electrochemical applications. 

A particularly important class of structural proton conductors are hydrogen-bonded organic species such as imidazole and its derivatives, which can be functionalized to create a variety of potentially useful proton-conducting systems for use in membranes~\cite{Kreuer_01,Maier_01,Kreuer_04,Maier_06} or other confined environments~\cite{Kim_13,Kitagawa_13,Xu_14,Ren_19}. These systems contain pseudo one-dimensional hydrogen-bonded chains, similar to those often associated with Grotthuss' original picture~\cite{Grot_1806,Marx_Rev_2006}, that undergo thermal rupturing and reforming, thereby altering the members of the chain and thus providing a continually evolving series of pathways for protons to diffuse through the liquid~\cite{Daycock_68,Kawada_70}. An ideal tool to uncover the molecular level details of these processes is {\it ab initio} molecular dynamics (AIMD) simulations~\cite{CP_85,Tuckerman_AIMD_rev,Marx_book}, which treat the chemical bond-breaking and forming that occurs during individual proton transfer events by solving for the electronic structure of the system ``on the fly". However, studying proton transport phenomena in hydrogen-bonded systems such as imidazole requires reaching timescales (usually several hundreds of picoseconds) necessary to capture the slow hydrogen bond rearrangements that lead to long-range proton transport. This has, up to now, been a significant challenge for AIMD, particularly for system sizes needed to accommodate extended hydrogen-bonded chains.

Here we perform nanosecond AIMD simulations of proton transport in liquid imidazole at the DFT level by employing a multiple time-stepping scheme~\cite{tuck3,tom3,Marsalek2016} that utilizes a density-functional tight-binding reference Hamiltonian to accelerate the calculations (see Supporting Information (SI)). This allows us to probe three distinct timescales and uncover the details of the structural proton transport mechanism in this hydrogen-bonded, organic, anhydrous system. The shortest timescale is shown to be associated with proton hopping times between imidazole molecules, while the intermediate timescale corresponds to the migration of the proton defect along extended hydrogen-bonded chains that retain their composition during this exploration. A third, significantly longer timescale ($\sim$30 ps) is seen to arise from hydrogen bond rearrangement within the system such that new extended chains form, thus changing the hydrogen bond network connectivity and leaving the proton defect without a pathway back to its original site. This third time scale and the associated hydrogen bond network rearrangement process provides a missing piece of the traditional Grotthuss structural diffusion process, as will be discussed in more detail after presentation of the analysis. It is important to note that collecting a statistically significant number of events at such a long timescale would not have been feasible without the multiple time-stepping approach.

\begin{figure}[b]
    \includegraphics[width=0.45\textwidth]{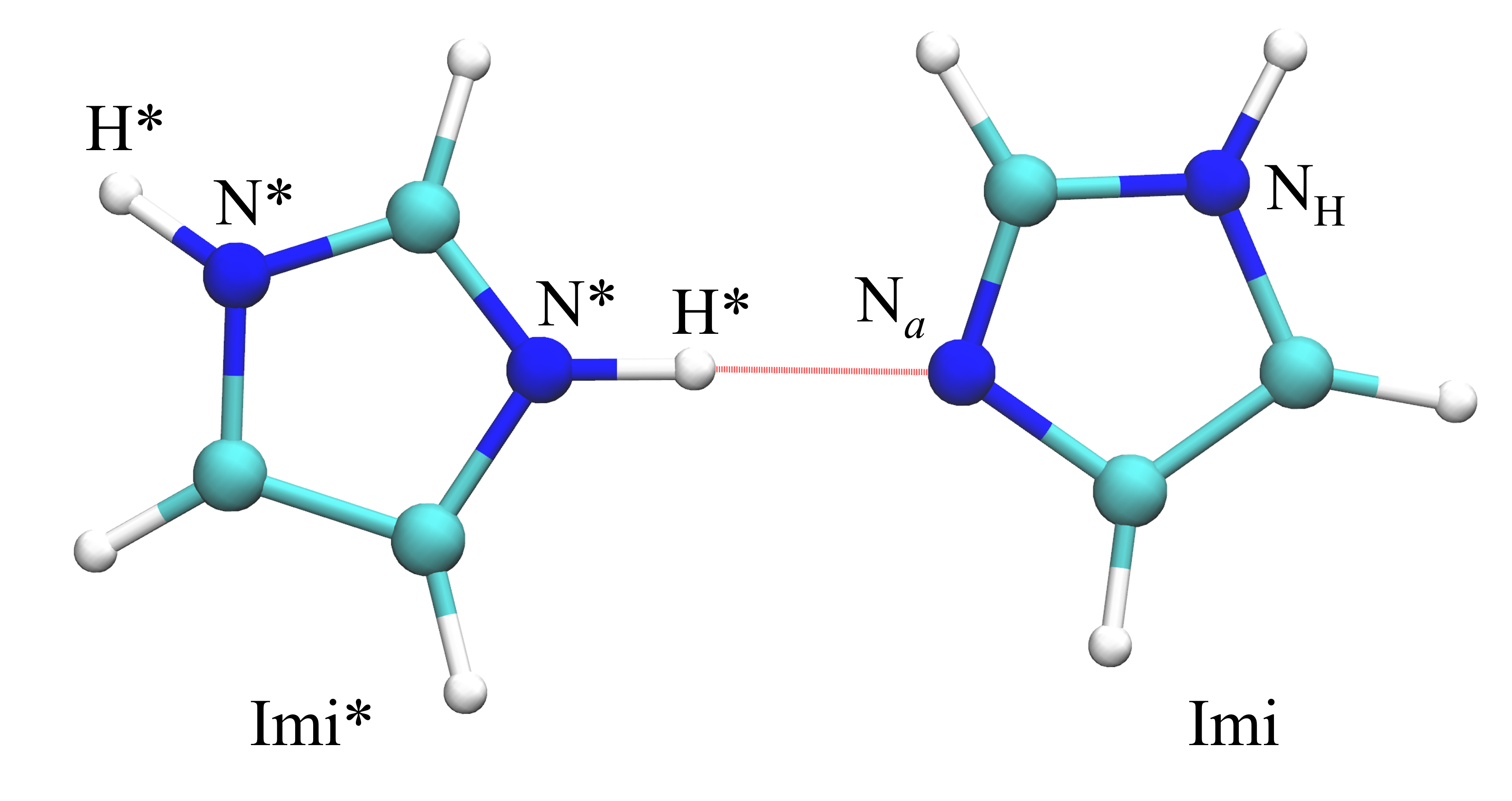}
    \caption{Snapshot showing a hydrogen-bonded pair of Imi$^*$-Imi from our simulations. Also shown are atomic and molecular labels used to describe proton transport.}
    \label{fig:zundel_labeled}
\end{figure}

We begin by analyzing the impact of structural diffusion on proton transport in imidazole. Fig.~\ref{fig:zundel_labeled} shows the nomenclature used throughout the discussion. The protonated imidazole molecule in any MD configuration is referred to as Imi$^{*}$ or the ``charge defect", the latter representing the fact that hydrogen bonds emanating from either side of the molecule have opposite polarity. From our AIMD simulations, we obtain a diffusion coefficient of 0.47~\AA$^2$/ps for the Imi$^*$ or charge defect in liquid imidazole at 384 K, which is 20 K above imidazole's melting point. The AIMD self-diffusion coefficient of ordinary imidazole molecules is found to be 0.06~\AA$^2$/ps. This large (8 fold) difference in diffusion coefficients indicates that proton transport in imidazole is dominated by structural diffusion. The mean-square displacement plots from which these coefficients are extracted are provided in the SI. The 8 fold enhancement is comparable to the corresponding $\sim$4.5 fold enhancement in the diffusion coefficients extracted from conductivity experiments just above the melting point\cite{kreuer1998imidazole}. In contrast, liquid water at 293 K (also 20 K above its melting point) has a self-diffusion coefficient 3-4 times that of imidazole, which is consistent with its $\sim$3-fold lower viscosity\cite{icsc_data}. However, the diffusion constant of a proton defect in water is only $\sim$4 times greater than the self-diffusion constant of ordinary water molecules~\cite{Agmon_CPL}.
Therefore, the importance of structural diffusion for proton defects in imidazole at 384 K is even more pronounced than it is in water at 300 K. 

The extent of proton sharing and delocalization between a protonated imidazole and its neighboring molecules can be quantified and compared to that of a protonated water molecule in an aqueous environment. For this, we employ the proton sharing coordinate, which is defined as $\delta = d_{{\rm N}^*{\rm H}} - d_{{\rm N}_a{\rm H}}$ where $d_{{\rm N}^*{\rm H}}$ and $d_{{\rm N}_a{\rm H}}$  are the covalent and hydrogen bond distances between Imi$^*$ and its closest hydrogen-bonded neighbors (see Fig.~\ref{fig:zundel_labeled}).  Closest neighbors are taken from the H$^*$ atoms on both sides of Imi$^*$.  Comparing the probability distribution $P(\delta)$ obtained from AIMD simulations using the same exchange correlation functional (see SI Section~\ref{sec:comp_details}) along this coordinate for imidazole at 384 K and water at 300 K (27 K above its melting point)\cite{Napoli2018}, qualitatively similar distributions of the proton position are observed. However, details such as the locations of the maxima and the depth of the central minimum differ.  This is due to the fact that imidazole forms weaker hydrogen bonds than water, which makes sharing the proton more difficult, with a free energy cost of 1.6~kcal/mol (compared to only 0.9~kcal/mol for an excess proton in water) to move the proton to $\delta = 0$, i.e. equidistant from the two heavy atoms. Here, the free energy is computed from $P(\delta)$ using the usual formula $F(\delta) = -k_{\rm B}T\ln P(\delta)$. Each proton transfer barrier should be viewed relative to the thermal energy, $k_{\rm B}T$, (0.60 kcal/mol at 300 K versus 0.76 kcal/mol at 384 K), which differs by a factor of 1.3 between imidazole and water. Interestingly, from an ab initio path integral molecular dynamics (AI-PIMD) simulation of imidazole using ring polymer contraction\cite{Markland2008,Markland2008a,Marsalek2016}, which includes nuclear quantum effects, the ZPE along the N-H coordinate leads to a reduction of 1.3 kcal/mol in the barrier along the proton sharing coordinate $\delta$, leaving a barrier of 0.3 kcal/mol, which is below the thermal energy at that temperature. As will be discussed below, this reduction in the barrier affects the ``rattling" of the proton between the charge defect and its neighbors, but has a smaller effect on the processes needed for proton diffusion over longer length scales.

\begin{figure}[b]
    \includegraphics[width=0.45\textwidth]{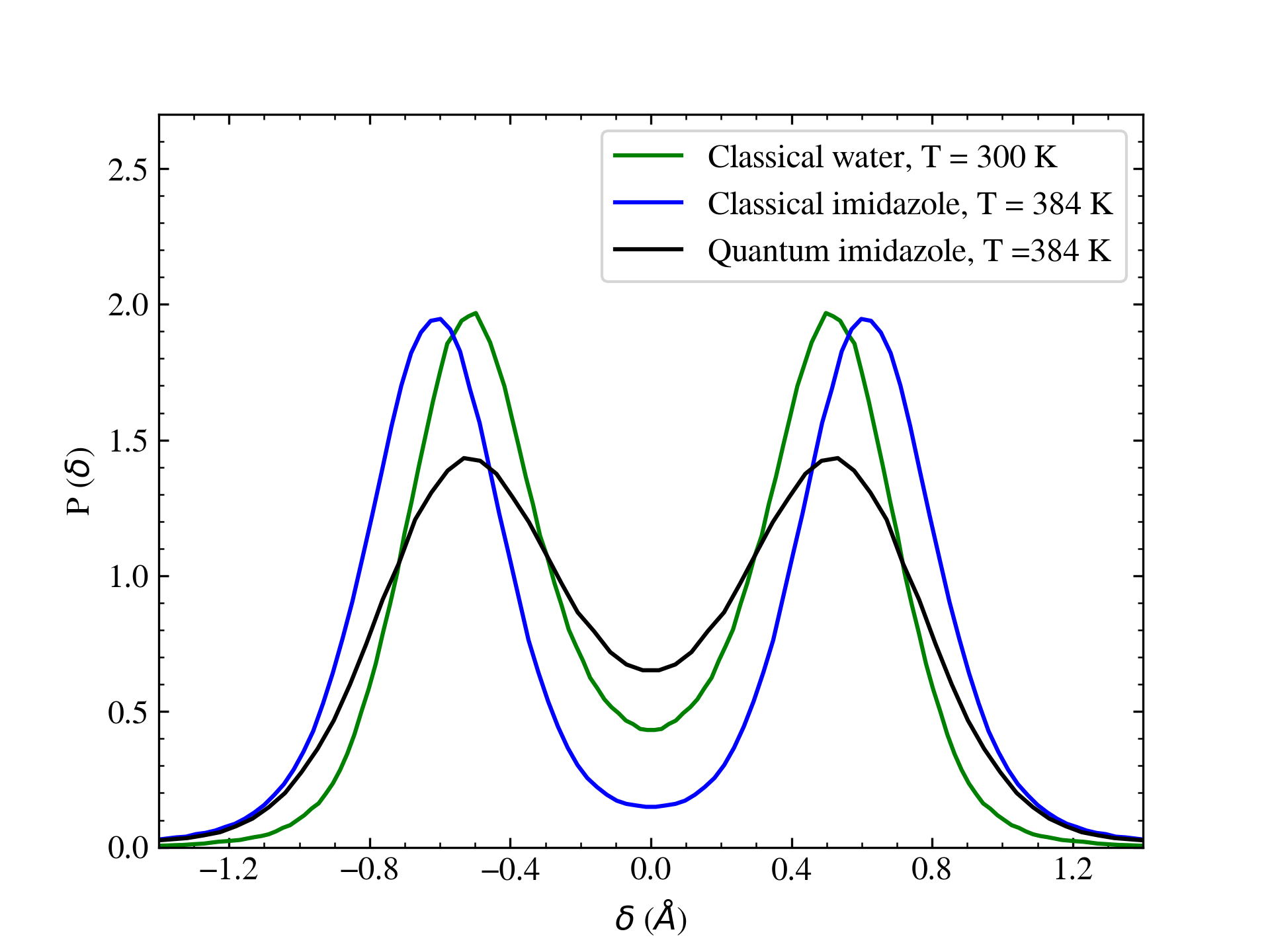}
    \caption{Comparison of the probability distribution along the proton sharing coordinate $\delta$ for classical AIMD simulations of water at 300 K (green line), as well as both classical (blue line) and AI-PIMD (black line) simulations of imidazole at 384 K.}
    \label{fig:delta_values}
\end{figure}

A much larger contribution to the structural diffusion mechanism is observed in our {\it ab initio} MD simulations ($\sim$8 fold) than in previous simulations based on empirical potentials\cite{Voth_09}, where only a 1.4 fold enhancement arising from the structural diffusion mechanism was observed. Therefore, it is worthwhile to investigate the molecular mechanisms that underlie the large enhancement of the structural diffusion mechanism. To accomplish this, we study intermolecular hydrogen bonding patterns. Owing to its two nitrogen groups, imidazole is able to form two hydrogen bonds: a donor and acceptor for ordinary imidazole and two donors for Imi$^{*}$. These two hydrogen bonds allow imidazole molecules in the liquid to form hydrogen-bonded chains, an example of which is illustrated in Fig.~\ref{fig:snapshots}. We will argue that diffusion of the protons along these chains is fast but limited to a finite distance, with long-range diffusion requiring the chains to break and reform.

\begin{figure*}
\centering
    \begin{subfigure}{0.3\textwidth}
        \includegraphics[width=1\textwidth]{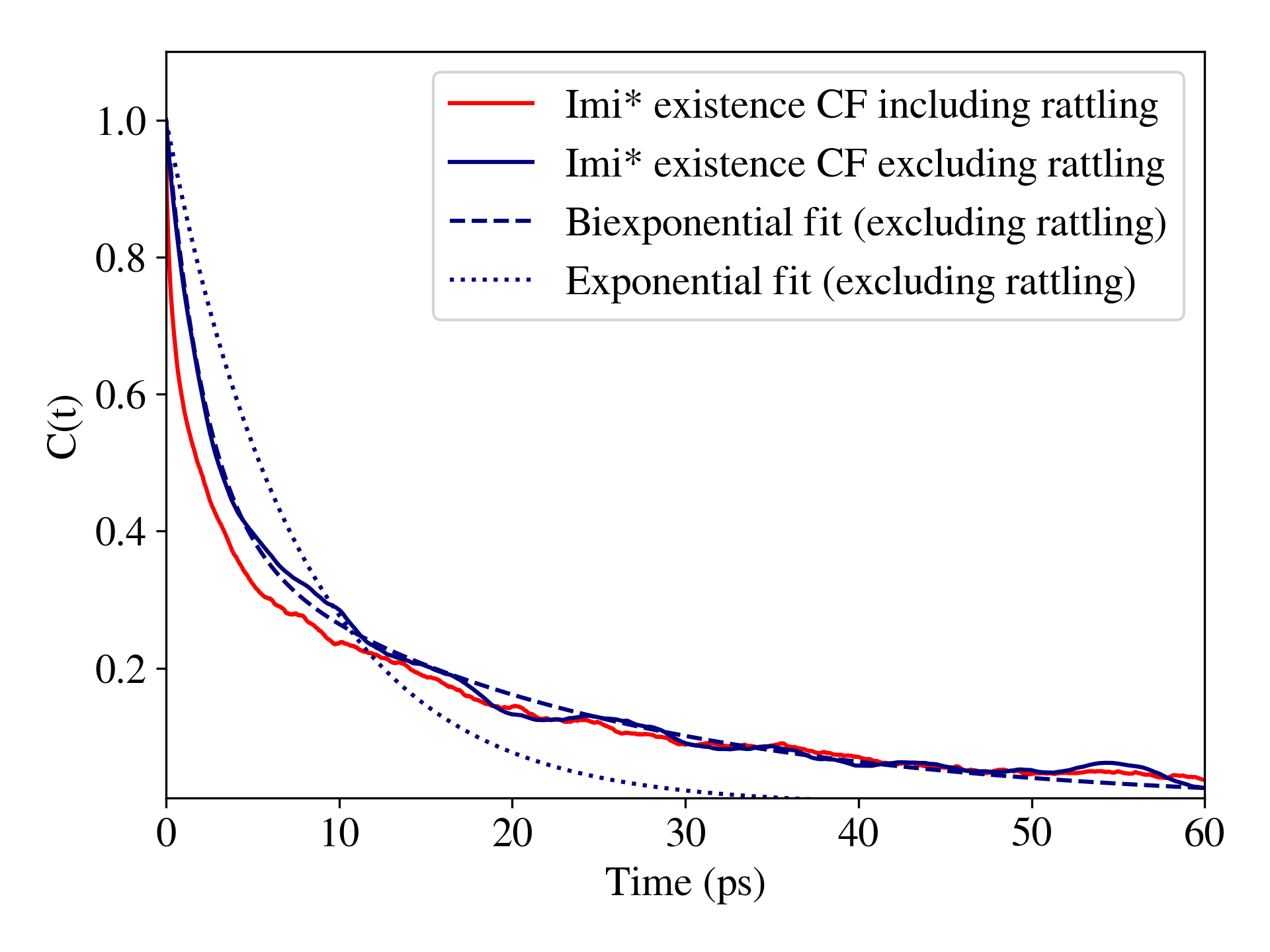}
        \caption{}
        \label{fig:cf_biexp}
    \end{subfigure}
        \begin{subfigure}{0.3\textwidth}
        \includegraphics[width=1\textwidth]{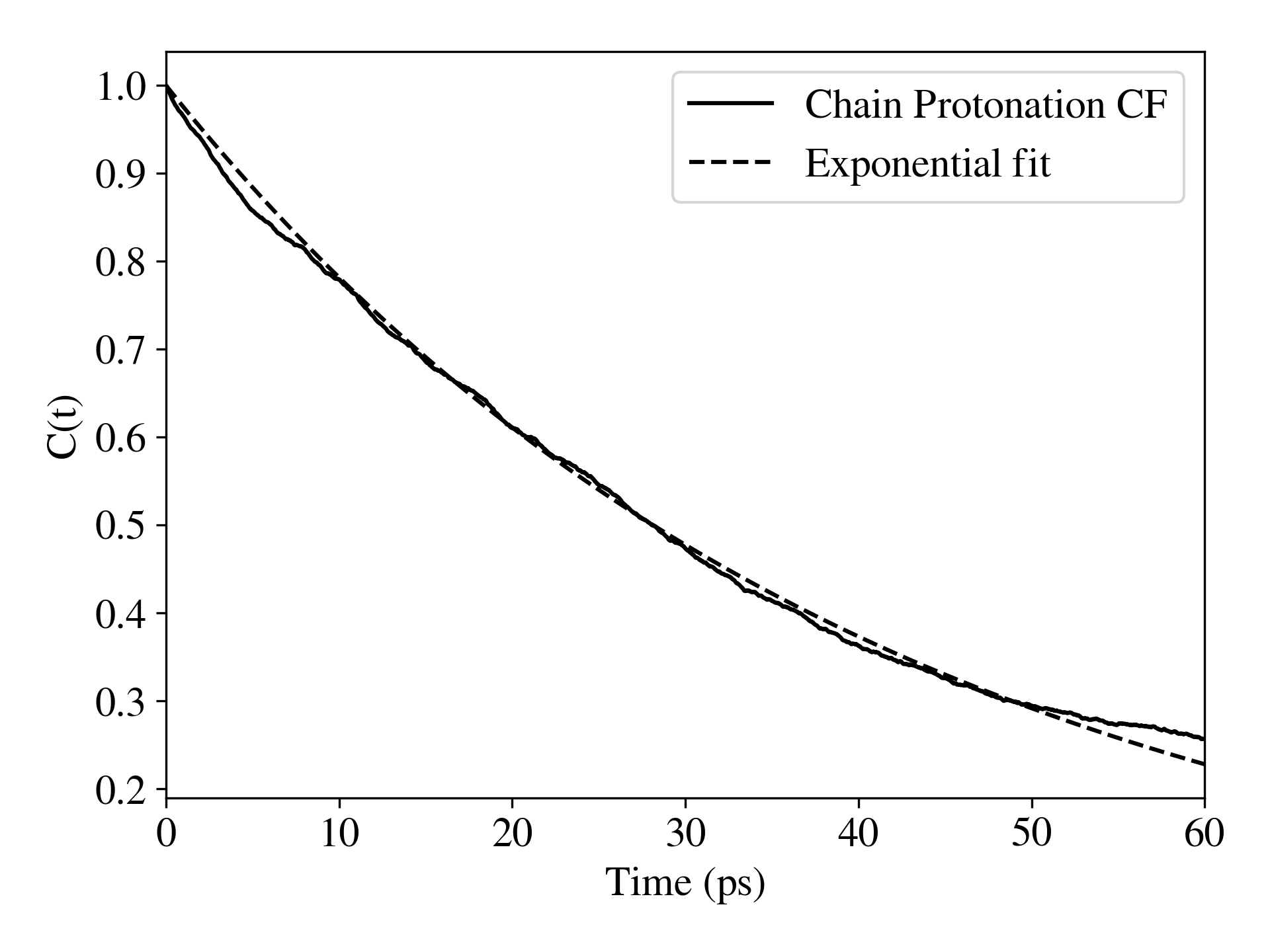}
        \caption{}
        \label{fig:cf_h_prescence}
    \end{subfigure}
        \begin{subfigure}{0.3\textwidth}
        \includegraphics[width=1\textwidth]{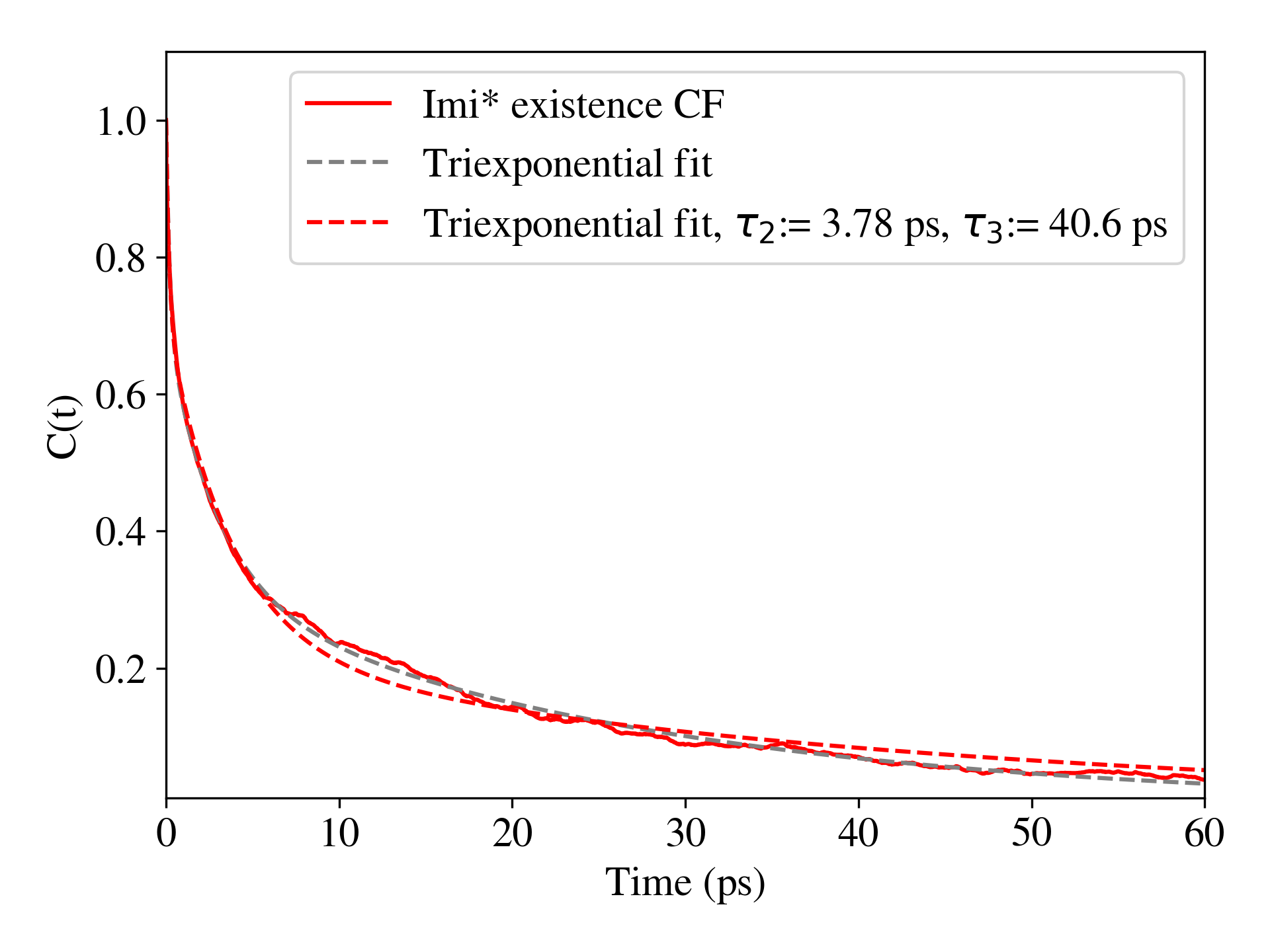}
        \caption{}
        \label{fig:cf_back_fit}
    \end{subfigure}
    \caption{a) The protonation population correlation function, including rattling (continuous red line) and excluding rattling (continuous blue line). Also shown are exponential (dotted blue line) and biexponential (dashed blue line) fits to the correlation function with rattling excluded. b) The chain protonation correlation function and its corresponding exponential fit. c) The protonation population correlation function (continuous red line) with its original triexponential fit (grey dotted line). The new triexponential fit (red dotted line) is obtained by fixing $\tau_2$ and $\tau_3$. $\tau_3$ (40.6 ps) is obtained from the H* chain presence correlation function, while $\tau_2$ is obtained from a biexpoential fit of the population correlation function excluding rattling, where the longer timescale has been fixed at 40.6 ps.}
\end{figure*}

We begin by analyzing the timescales associated with the different proton transport regimes. To achieve this, we employ the protonation population correlation function formalism of Chandra, {\it et al.}~\cite{Chandra_07,Tuckerman_10}. Specifically, we define an ``intermittent" correlation function $C(t)=\langle h(0)h(t)\rangle / \langle h\rangle $ where $h(0)$ and $h(t)$ are the population indicators at times $0$ and $t$, respectively. Here, $h(t)=1$ if an imidazole molecule is the proton defect (Imi$^*$) at time $t$, and $h(t)=0$ otherwise. Using this definition, the protonation population correlation function can be interpreted physically as giving the probability that an imidazole molecule that is protonated at $t = 0$ is still protonated at time $t$. 

Having defined the protonation population correlation function, we can now unravel the processes by which it decays. As shown in Fig.~\ref{fig:cf_biexp} the correlation function can be accurately fit to a tri-exponential form:
\begin{equation}
    C(t) = a_1 e^{-t/\tau_1} + a_2 e^{-t/\tau_2} + a_3 e^{-t/\tau_3}. 
    \label{eq:triex}
\end{equation}
Here $a_1 + a_2 + a_3 = 1$, and we order the processes such that $\tau_1$, $\tau_2$, and $\tau_3$ correspond to the short, intermediate, and long timescales, respectively. Such a fit is appropriate if the underlying mechanism of proton transport involves three first-order processes\cite{Chandra_07,Tuckerman_10}. Given the ability to fit the protonation correlation function using this form, we can investigate to which processes the various timescales correspond. The timescales extracted from the fit of the protonation correlation function are 0.21~ps, 2.9~ps, and 25.6~ps, each having a similar contribution to the $t=0$ correlation ($a_1 = 0.30, a_2 = 0.38,   a_3 = 0.32$). The shortest timescale is one that is commonly observed in hydrogen bonded systems that can undergo proton transfer and is usually associated with fast rattling events~\cite{Tuckerman_Nature_Hp,Chandra_07,Tuckerman_10}. In these events, the proton simply shuttles back and forth between Imi$^{*}$ and one of the two Imi molecules that it is hydrogen-bonded to. This sub-picosecond timescale is consistent with the low barrier along the proton transfer coordinate of $\sim2k_{\rm B}T$. To confirm such an assignment, we can remove rattling proton transfer events, i.e. events where the proton hop is reversed by the very next transfer event. If the sub-picosecond timescale indeed corresponds to back-and-forth hops between pairs of molecules, then excluding rattling events should eliminate this timescale and retain the longer ones. This is indeed what is observed in Fig.~\ref{fig:cf_biexp}, where the rattling excluded protonation correlation function is well captured by a bi-exponential fit with time constants 2.25~ps and 21.4~ps, which are close to the two longer timescales seen in the tri-exponential fit of the full protonation correlation function.

Having verified that the shortest time-scale process can be assigned to proton rattling events, we now consider the two longer timescales. As previously discussed, Imi forms hydrogen-bonded chains in the liquid. Fig.~\ref{fig:snapshots} shows a representative proton transfer event observed in our AIMD simulations. For clarity, only molecules involved in the proton transfer are shown. In this, one observes that the proton starts on the molecule labelled 0 at $t=0$, which is part of a hydrogen-bonded chain consisting of the Imi$^{*}$ and four Imi molecules (Fig.~\ref{fig:snapshots_1}). At time $t=2.3$~ps, (Fig.~\ref{fig:snapshots_2}) the proton has migrated to the adjacent molecule, and by 5.1~ps (Fig.~\ref{fig:snapshots_3}) the proton has shifted three molecules away from its initial site. At this time, the original Imi$^*$ (molecule 0) is no longer a member of the hydrogen-bonded chain that holds the proton, and molecules 5 and 6, which were not members of the original chain, have joined. When 35~ps have elapsed, the final snapshot (Fig.~\ref{fig:snapshots_4}) shows that the proton has migrated five molecules away from its original position and is now part of a largely reformed hydrogen-bonded chain consisting of many new members and few of the molecules that constituted the original chain. 

\begin{figure*}[ht]
\centering
    \begin{subfigure}{0.48\textwidth}
        \includegraphics[width=0.98\textwidth,trim=650 900 700 900, clip=true]{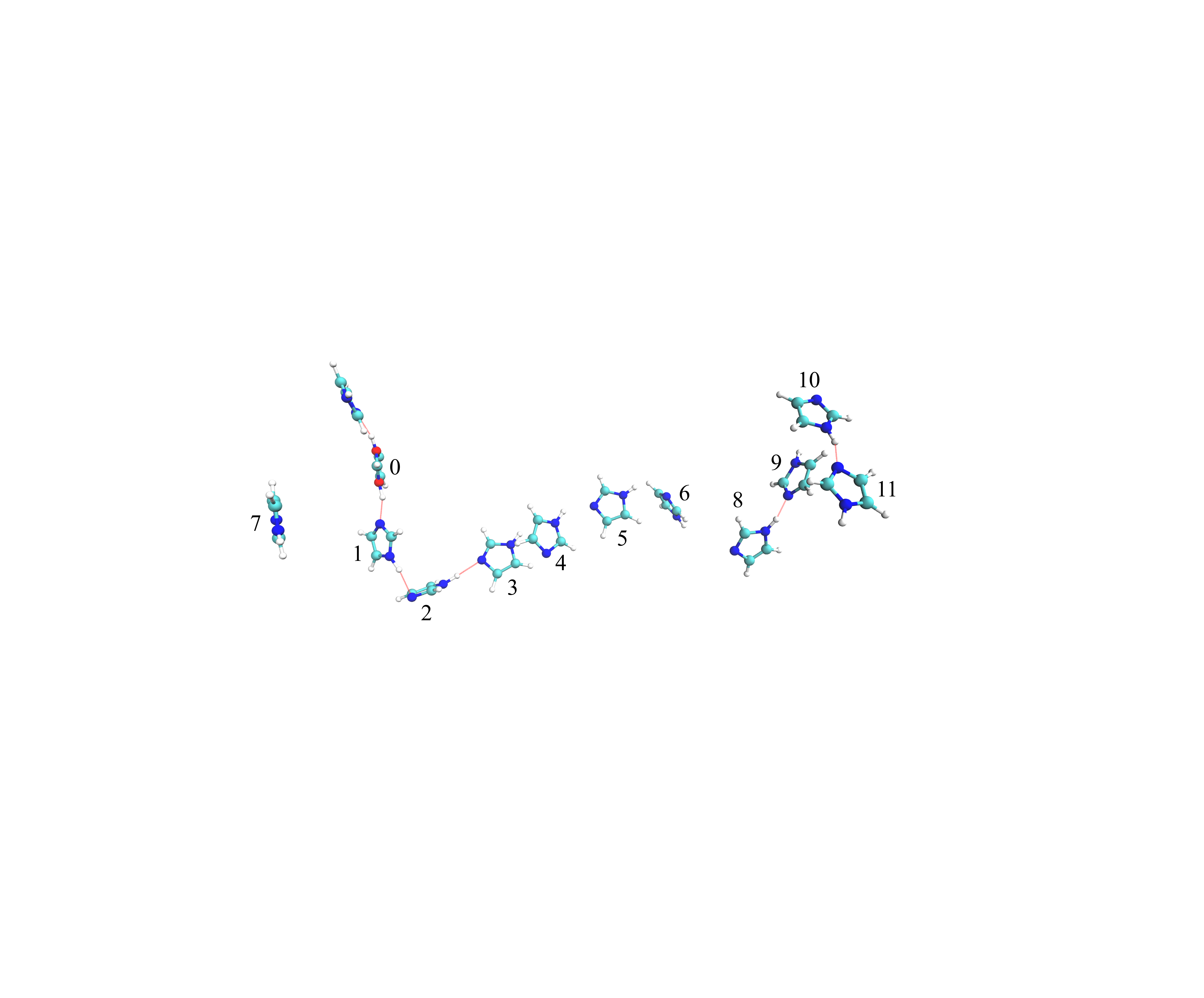}
        \caption{\(t=0.0~{\rm ps}\)}
        \label{fig:snapshots_1}
    \end{subfigure}
    \begin{subfigure}{0.48\textwidth}
        \includegraphics[width=0.98\textwidth,trim=650 900 700 900, clip=true]{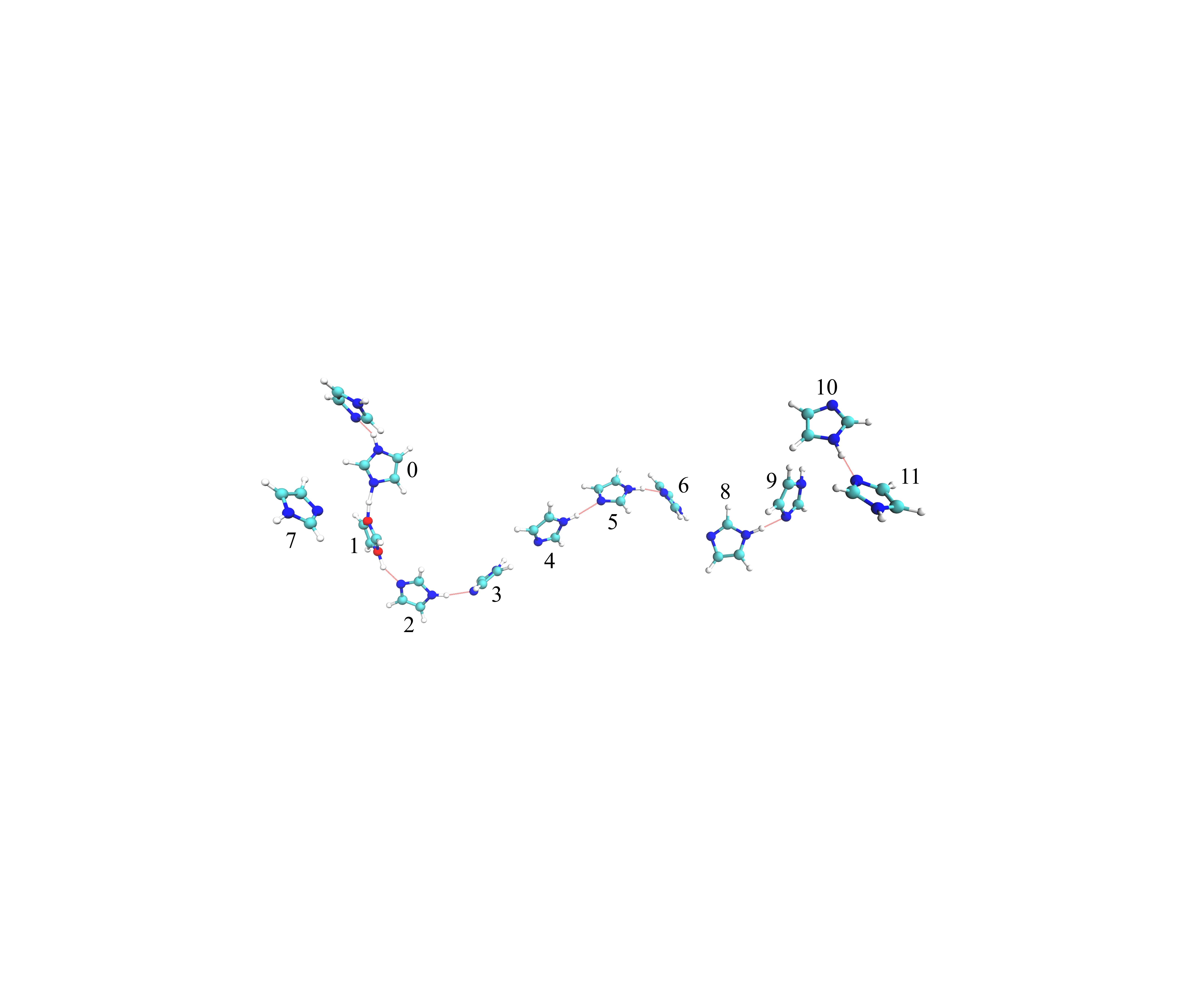}
        \caption{\(t=2.3~{\rm ps}\)}
        \label{fig:snapshots_2}
    \end{subfigure}
    \begin{subfigure}{0.48\textwidth}
        \includegraphics[width=0.98\textwidth,trim=650 900 700 900, clip=true]{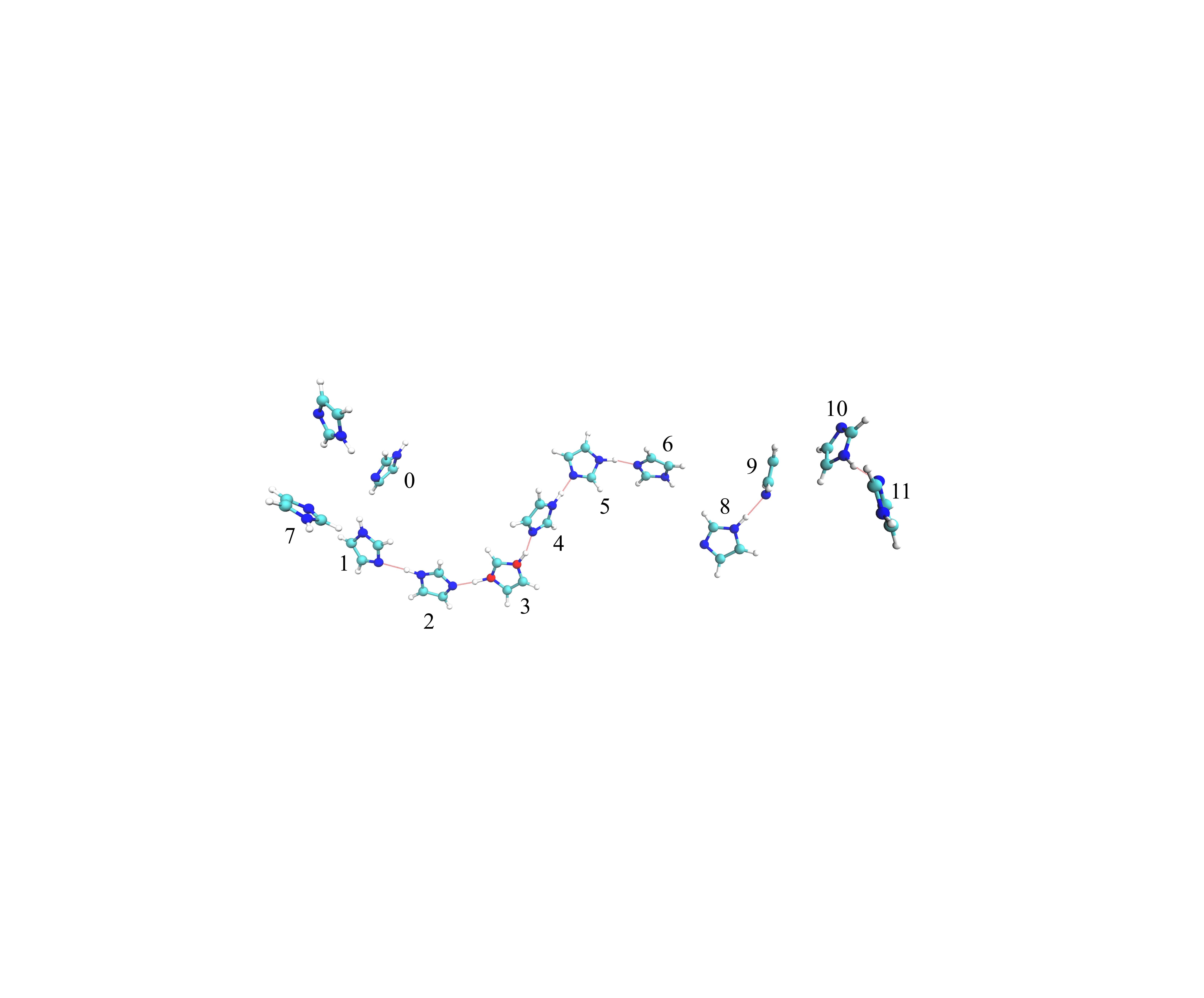}
        \caption{\(t=5.1~{\rm ps}\)}
        \label{fig:snapshots_3}
    \end{subfigure}
    \begin{subfigure}{0.48\textwidth}
        \includegraphics[width=0.98\textwidth,trim=650 900 700 900, clip=true]{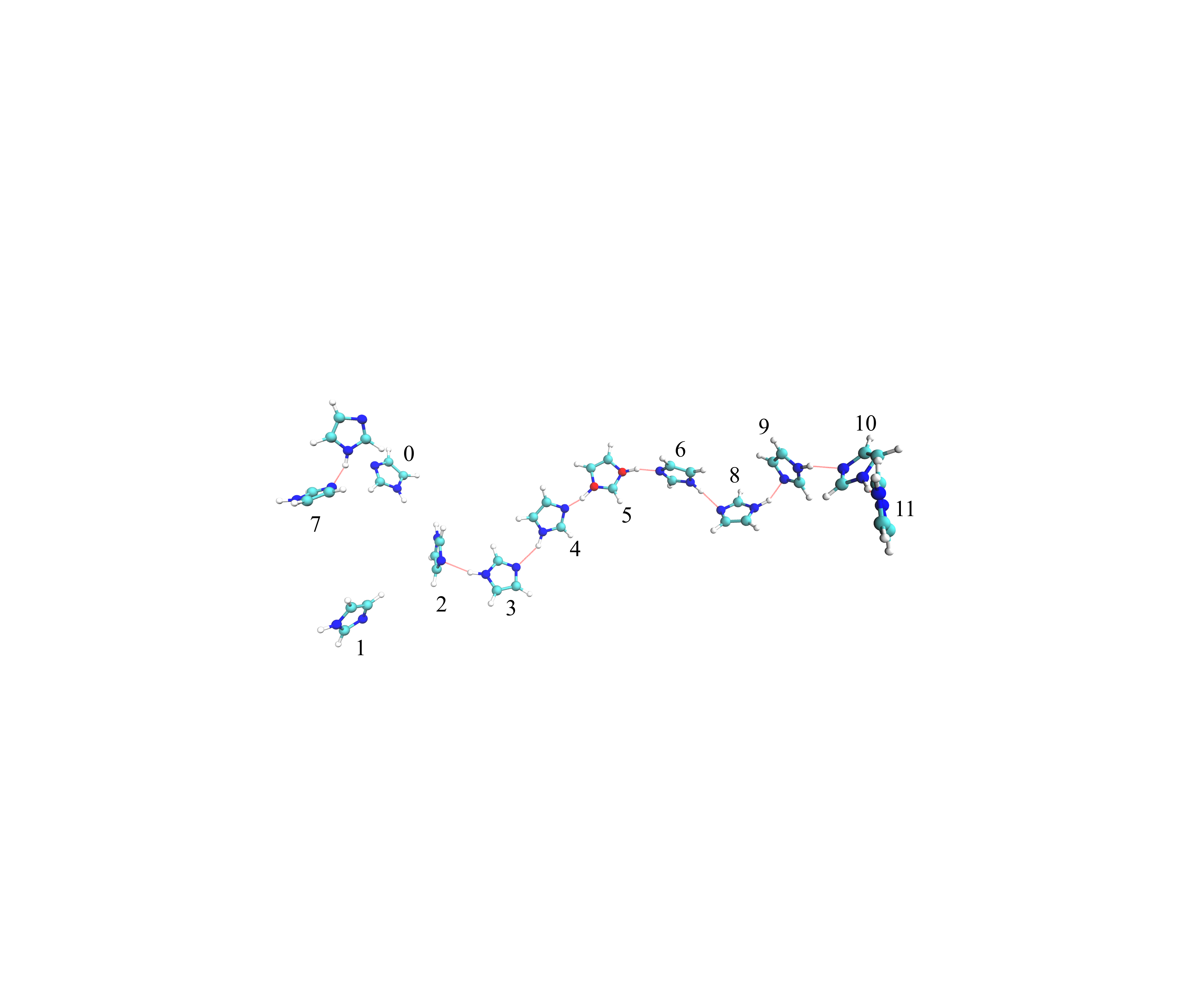}
        \caption{\(t=35.6~{\rm ps}\)}
        \label{fig:snapshots_4}
    \end{subfigure}
\caption{Snapshots showing local proton transfer events, proton transport within the chain and chain component change. The Imi$^*$ is marked with red sphere on the nitrogen atom. The corresponding time are listed below the snapshots.}
\label{fig:snapshots}
\end{figure*}

\begin{figure}
    \centering
    \begin{subfigure}{0.48\textwidth}
        \includegraphics[width=0.98\textwidth]{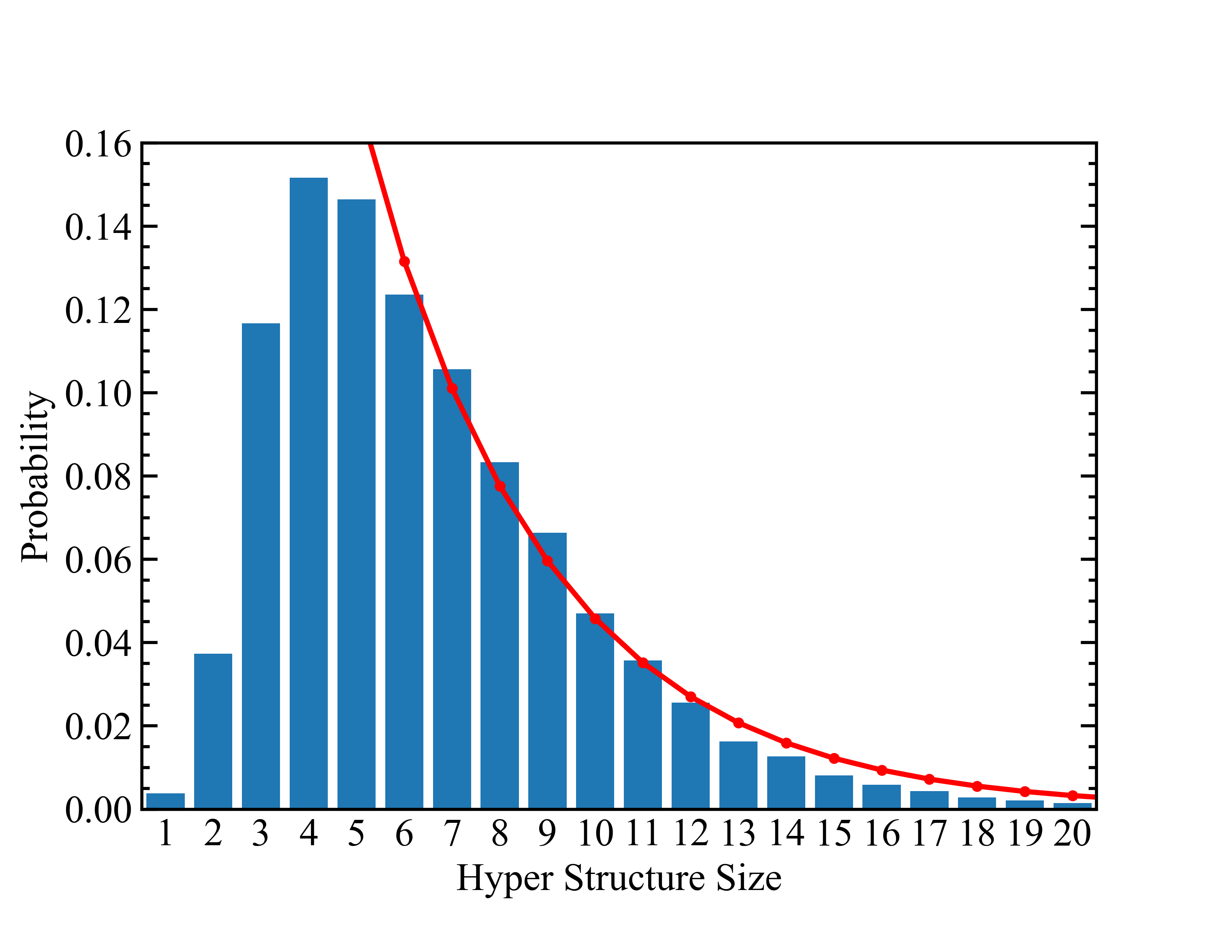}
        \caption{Size distribution of Imi$^*$ chain. The red line shows a fit for n $>$5 to $P(n)=aK^{n}$ with the parameters a=0.64 and K=0.77}
        \label{fig:h_imi_chain_size}
    \end{subfigure}
    \begin{subfigure}{0.48\textwidth}
        \includegraphics[width=0.98\textwidth]{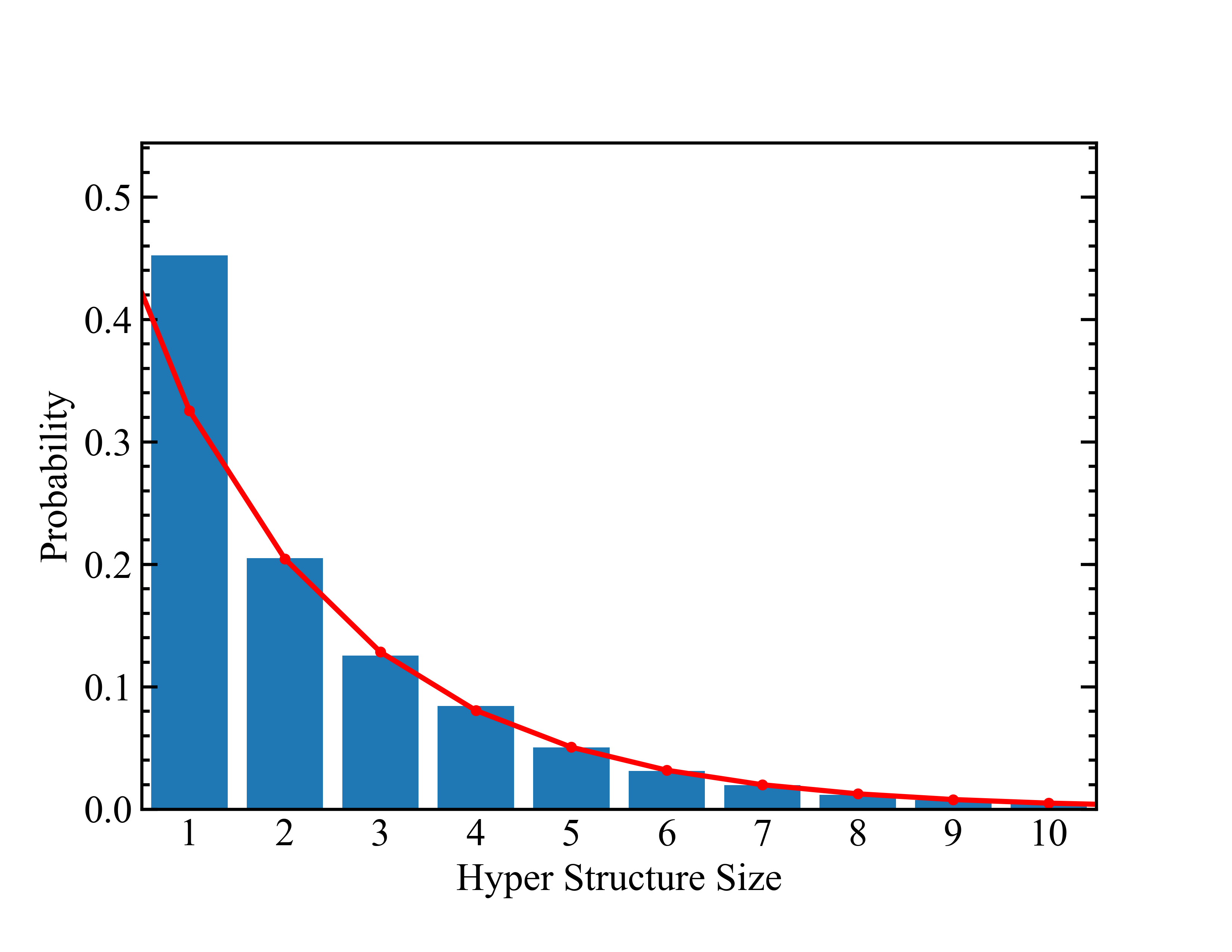}
        \caption{Size distirbution of Imi chain. he red line shows a fit for n $>$1 to $P(n)=aK^{n}$ with the parameters a=0.52 and K=0.63}
        \label{fig:normal_imi_chain_size}
    \end{subfigure}
    \caption{Size distribution of hydrogen-bonded chain sizes in liquid imidazole (blue bars) and theoretical fits of the distribution (solid red lines). Note that the distribution is computed by ``unwrapping" the periodic boundary conditions so as to reveal the putative diffusion pathways for the charge defect. In doing so, care was taken to ensure that in any of the chains counted, despite this unwrapping of the periodic boundaries, no molecule appears more than once.}
    \label{fig:hyper_struct_size}
\end{figure}

Figure~\ref{fig:snapshots} suggests that Imi$^*$ virtually always exists in long hydrogen bond chains, the distribution of which is shown in Fig~\ref{fig:h_imi_chain_size}.  Hydrogen bonds between two species (Imi-Imi or Imi$^*$-Imi) are defined by generating a two-dimensional free energy surface $F(r,\theta) = -k_{\rm B}T\ln P(r,\theta)$ as a function of the hydrogen-bond distance $r$ and angle $\theta$ and then using the contour at $k_{\rm B}T$ to determine the criteria for existence of a hydrogen bond (details are provided in the SI).  
While motion along these chains can occur on the picosecond timescale, for long range diffusion to occur, the hydrogen-bonded chain must reform to incorporate new members, and in so doing, lose former members, which happens on the tens-of-picoseconds timescale. This suggests an assignment of the two longer timescales extracted from the protonation correlation function, $\tau_2$ and $\tau_3$, as corresponding to the hopping of a proton along a hydrogen-bonded chain and the reformation of the hydrogen-bonded chain respectively. 

In order to confirm this assignment, we define a chain protonation correlation function. Rather than giving the probability that a molecule holding the proton at $t = 0$ still possesses it a a later time $t$, the chain protonation correlation function gives the probability that a hydrogen-bonded chain that held the proton at $t = 0$ still possesses it at a later time $t$.  Hence, in order for the chain protonation correlation function to decay, the proton must no longer be on {\it any} of the molecules that constituted the original chain. Since the chain protonation correlation function only decays once the excess proton leaves its original hydrogen-bonded chain, it should eliminate both the sub-picosecond and several-picosecond timescales, as these correspond to movements within the same hydrogen-bonded chain. Fig.~\ref{fig:cf_h_prescence} shows that this is indeed the case, and that the chain protonation correlation function decays with single exponential behavior and $\tau=$ 40.6~ps. This timescale is longer than the $\tau_3$ we extracted from a direct tri-exponential fit of the protonation correlation function of 25.6~ps. However, one can obtain an intermediate timescale by constraining $\tau_3$ to 40.6~ps and conducting a biexponential fit of the rattling excluded protonation population function, which yields a value of $\tau_2$ of 3.78~ps. Fixing $\tau_2$ to 3.78~ps and $\tau_3$ to 40.6~ps and conducting a triexponential fit of the remaining parameters to the protonation population function gives very good agreement, as shown in Fig.~\ref{fig:cf_back_fit}. This suggests that these timescales do arise from these processes, and by using the chain protonation correlation function and the rattling excluded protonation population function, one can unravel and assign them. It is also worth noting that the third timescale \(\tau_3\) is consistent with an estimate of 30~ps made in a previous study\cite{Seifert_01} for reorientation and proton hopping rates using the diffusion coefficient and proton-transfer distance. \(\tau_3\) is also consistent with the longest orientational correlation times extracted from different axes of the molecule as shown in SI Section~\ref{sec:ocf_si}.

Given the importance of the hydrogen-bonded chains of Imi molecules in the proton transport process, we now analyze the distribution of chain lengths since they determine the maximum lengthscale over which the proton can move in the intermediate ($\tau_2$) time regime. Figure~\ref{fig:h_imi_chain_size} shows the size distribution of chains containing Imi$^*$ obtained from our simulations. This distribution peaks at 4, and the average number of molecules in a chain containing Imi$^*$ is 6.6 for the hydrogen bond definition employed in this analysis (see SI). This is considerably longer than the reported value of 4 in a previous study~\cite{Shen_12} where a less restrictive hydrogen bond definition was used. Using the definition in that study~\cite{Shen_12}, the average chain size in our simulations would be $\sim$11. In contrast, Imi chains not containing Imi$^*$ are considerably shorter (see Fig.~\ref{fig:normal_imi_chain_size}), with chains consisting of just 1 molecule being the most common and the average chain length being 2.5 molecules. The distribution of neutral imidazole chains can be elucidated using a simple model assuming a single equilibrium constant. In particular, for a hydrogen-bonded Imi chain consisting of $n$ molecules, denoted as Imi$_n$, one can consider adding an imidazole to the chain (on either end) as a reversible reaction:
\begin{equation}
    {\rm Imi}_{n}+{\rm Imi}\rightleftharpoons{\rm Imi}_{n+1}
\end{equation}
where the equilibrium constant is,
7\begin{equation}
    K=\frac{[{\rm Imi}_{n+1}]}{[{\rm Imi}_{n}]}
\end{equation}
Assuming such a model with an equilibrium constant that does not depend on $n$ results in the prediction that $[{\rm Imi}_n]\propto K^n$. To test this model, the red line in Fig.~\ref{fig:normal_imi_chain_size} shows that the simulated distribution of neutral Imi chains can be fit well to the expression $P(n)=aK^n$, yielding values of $a=0.52$ and $K=0.63$, with the latter corresponding to a Helmholtz free energy of adding a molecule to the chain of 0.36 kcal/mol. Such a fit is clearly not appropriate in the case of Imi$^*$ (Fig.~\ref{fig:h_imi_chain_size}) since its distribution shows a turnover with respect to $n$ rather than a simple power law decay. This reflects the fact that Imi$^*$ is a charge defect and therefore perturbs the distribution at low $n$, where it forms significantly stronger hydrogen bonds. This leads to a substantial depletion in the probability of finding chains consisting of less than 5 molecules, i.e. an Imi$^*$ stabilized by 2 donors on each side, giving a chain of total length $n=5$. Beyond $n=5$, the distribution decays according to $P(n)=aK^n$ (red line in Fig.~\ref{fig:h_imi_chain_size}) with $a=0.64$ and $K=0.77$, which is close to that for pure Imi chains. We note that this prediction of chain size distribution is also robust under different hydrogen bond criteria (SI Section~\ref{sec:hb_chain_si}). 
The probability distributions in Fig.~\ref{fig:hyper_struct_size} show that chain formation around Imi$^*$ is favorable up to around $n = 5$ while formation around Imi is disfavored as the chain size increases.  If we take a hydrogen-bonded pair involving each species as the core, then adding a third member to the chain, i.e., increasing from $n = 2$ to $n = 3$ has an associated probability ratio of 0.12/0.04 for Imi$^*$ and 0.13/0.2 for Imi. If we use these ratios to compute a free energy difference $\Delta\Delta F_{2,3}$ favoring chain formation around Imi$^*$ using $\Delta F_{2,3} = -k_{\rm B}T\ln (P(3)/P(2))$ for each case, we find it is favored by $\Delta\Delta F_{2,3} \approx 1.2$ kcal/mol. This can be contrasted with the free energy difference characterizing the increased strength of the Imi$^*$-Imi hydrogen bond over that of the Imi-Imi pair, obtained from the radial distribution functions in Fig. 2 of the SI, which is roughly 0.4 kcal/mol.

Given the clear signature of an intermediate time scale for motion of the charge defect, Imi$^*$, along quasi one-dimensional chains of hydrogen-bonded imidazole molecules, it is instructive to analyze this motion using a simple one-dimensional random-walk diffusion model.  If individual proton transfer events along this chain are random and uncorrelated, such a model should be able to predict this.  The model we consider is a simple one-dimensional random walk on a finite grid of $n+1$ points with reflecting boundaries.  In this discrete-time Markov chain, the probability to hop from between neighboring points is $p = \exp(-t/\tau)$. This model can be worked out analytically by constructing the transition matrix
$p_{i+1,i} = p_{i-1,i} = p/2$, $p_{ii} = 1-p$, 
$p_{i,0} = p_{n,n-1} = p$. In the limit $t\rightarrow\infty$, 
this matrix gives the steady-state probability distribution 
as the eigenvector with a unit eigenvalue.  The elements $L_i$ of this vector are $L_0 = L_n = 1/(2n)$, 
$L_i = 1/n$.  If $d$ is the distance between grid points, then the long-time mean-square displacement becomes
\begin{eqnarray}
        {\rm MSD}(\infty)  & = & \sum_{j=0}^n\sum_{i=0}^n L_j L_i (i-j)^2d^2 \nonumber \\
                   \nonumber \\
                          & = &  
                          \frac{n^2+2}{6} d^2
\end{eqnarray}
A more detailed derivation and discussion is provided in the SI. Following Ref.~\onlinecite{Agmon_CPL}, if we take the proton hopping distance to be roughly the length of the N$^*$-N$_a$ hydrogen bond (see Fig. 2 in the SI), i.e., 2.71 \AA\ and use the average chain length of 6.6, the one-dimensional model gives 
${\rm MSD}(\infty) = 40.8$ \AA$^2$.  Comparing this to the value at which the MSD in Fig. 1(b) of the SI becomes linear (at $t = 4.9$ ps), we obtain a value of 44.2 \AA$^2$. The two values are in very close agreement, suggesting that migration along the Imi$^*$ chains follows the statistics of a simple random-walk process. This link between the intermediate timescale extracted from the protonation population functions (3.78~ps) and those observed in the mean square displacement of the proton (4.9~ps) leads to a consistent picture of the proton transport mechanism.

Investigating proton transport in liquid imidazole provides an opportunity to examine a system that contains relatively long Grotthuss-like chains, which have also been observed in phosphoric acid~\cite{Kreuer_12} and methanol~\cite{Morrone_02}. The result presented here from nanosecond-long AIMD trajectories, enabled by multiple time-stepping techniques, is that processes such as proton migration along these chains and local molecular reorientation only capture part of the complete picture of proton transport along the hydrogen bond network. The key to long-range transport is the ability of the system to form relatively long protonated hydrogen-bonded chains in the first place (see Fig.~\ref{fig:h_imi_chain_size}) and then to scramble the membership of these chains continually on a $\sim$40~ps timescale corresponding to the reorientation times such that a constantly evolving conduit of proton transport propels the defect to diffuse rather than become locally trapped on the original chain.

The notion that molecular reorientation is critical for driving proton transport in hydrogen-bonded media can be traced back to early work from the 1960s\cite{Riehl_65}. The idea that proton transport in an infinite one-dimensional chain of hydrogen-bonded imidazole molecules must be followed by a roughly concerted rate-determining molecular reorientation in order to return the system to its starting state\cite{Daycock_68,Kawada_70} has motivated subsequent simulation studies of imidazole crystals\cite{Seifert_01,Parrinello_04,Iannuzzi_06} and imidazole-terminating oligomers (imidazole 2-ethylene, Imi-2EO)\cite{Parrinello_04,Iannuzzi_06}. However, solid state NMR studies show that ordered crystalline domains do not participate in long-range proton transport, but rather, it is disordered or amorphous domains, grain boundaries, and crystal defects, etc., where no consistent hydrogen-bond ordering can be assigned, that contribute most to the transport process.\cite{Goward_02,Goward_04}  

This observation about imidazole-based solids informs what is seen here in the simulation of proton transport in the liquid state. The lability of the hydrogen-bond network, which allows for frequent reassignment of members of long protonated Grotthuss-like chains in the system, provides a dynamic network of pathways for the charge defect to move through the system on a time scale that is consistent with molecular reorientation. This step, in a sense, constitutes a missing piece in the idealized Grotthuss mechanism~\cite{Grot_1806,Marx_Rev_2006}.  
It is worth noting that molecular reorientation due to hydrogen bond lability remains an important factor for sustainable proton conductivity in imidazole-based systems, as has been observed in previous experiments of imidazole-tethered polymers\cite{Herz_03} and confined systems\cite{Kitagawa_09_MOF,Stock_16_MOF}.
Ultimately, we expect the mechanistic picture uncovered here, which provides a more complete view of the idealized Grotthuss structural diffusion process, to govern the proton transport process in a variety of related proton-conducting liquids that favor formation of hydrogen-bonded chain structures.

\section*{Acknowledgments}
This material is based upon work supported by the National Science Foundation Phase I CCI: NSF Center for First Principles Design of Quantum Processes (CHE-1740645). T.E.M also acknowledges support from the Camille Dreyfus Teacher-Scholar Awards Program.

\clearpage
\onecolumngrid

\setcounter{equation}{0}
\setcounter{figure}{0}
\makeatletter 
\renewcommand{\thefigure}{S\@arabic\c@figure}
\renewcommand{\thetable}{S\@arabic\c@table}
\makeatother
\renewcommand{\theequation}{S\arabic{equation}}

\section*{Supporting Information}
\section{\label{sec:comp_details}{\it Ab initio} multiple timescale molecular dynamics}

In an AIMD simulation, the motion of $N$ nuclei on the ground-state Born-Oppenheimer potential energy surface $E_0(\uR_1,...,\uR_N) \equiv E_0(\uR)$, where $\uR_1,...,\uR\equiv \uR$ are the nuclear positions, is generated by obtaining $E_0(\uR)$ and the forces $\uF_I = -\nabla_I E_0(\uR)$ ``on the fly" directly from electronic structure calculations. $E_0(\uR)$ cannot be computed exactly for system sizes used in condensed-phase simulations. The most common approximate theory for $E_0(\uR)$ is density functional theory (DFT), where $E_0(\uR) \approx E_0^{{\rm (DFT)}}(\uR)$. However, the computational overhead of DFT calculations is high. Therefore, in order to access time scales needed to characterize proton transport in a hydrogen-bonded liquid like imidazole, a novel multiple time-step (MTS) approach is employed in which a density-functional tight-binding (DFTB) parameterization is used as a cheap approximation of the ground-state energy surface, and the difference between DFTB and DFT forces is used as a correction~\cite{tuckerman1992reversible, luehr2014multiple}. Particularly, the nuclear Hamiltonian at the DFT level is written as:

\begin{eqnarray}
    H(\uR,\uP) & = &  \frac{\uP^{\rm T}{\rm M}^{-1}\uP}{2} + E_0^{\rm (DFT)}(\uR)
    \nonumber \\
    \nonumber \\
    & = &  \frac{\uP^{\rm T}{\rm M}^{-1}\uP}{2} + E_0^{\rm (DFTB)}(\uR)
    + \left(E_0^{\rm (DFT)}(\uR) - E_0^{\rm (DFTB)}(\uR)\right)
    \nonumber \\
    \nonumber \\
    & \equiv & H_{\rm ref}(\uR,\uP) + \Delta E_0(\uR)
    \label{eq:AIMD}
\end{eqnarray}

where $\uP \equiv (\uP_1,...,\uP_N)$ are the nuclear momenta and M is a diagonal matrix of nuclear masses. Within the reversible reference system propagator algorithm (r-RESPA) approach to MTS integration~\cite{tuck3}, the first two terms on the second line of Eq.\ (\ref{eq:AIMD}) constitute a ``reference system" Hamiltonian, which we denote as $H_{\rm ref}(\uP,\uR)$, and which is integrated with a small time step $\delta t$ via velocity Verlet. The difference between the DFT and DFTB surfaces, $\Delta E_0(\uR) = E_0^{\rm (DFT)}(\uR) - E_0^{\rm (DFTB)}(\uR)$, generates difference forces $\Delta \uF_I = \uF_I^{\rm (DFT)} - \uF_I^{\rm (DFTB)}$ that are applied before and after the reference system update using a large time step $\Delta t = n\delta t$ for some integer $n$. If the DFTB parameterization closely matches the target DFT, then the difference should be small, and $n$ can be chosen reasonably large. In addition, since the computational cost of DFTB is negligible compared to that of the full DFT, the saving in computational time is very close to $n$.

We performed AIMD simulations of a liquid imidazole system containing 64 molecules and one excess proton (no counter ion) in a periodic cubic supercell of length 19.337 \AA\ at 384 K. These parameters correspond to a density of 1.00 g/cm$^3$, which is close to the experimentally observed density at that temperature of 1.03 g/cm$^3$ \cite{crchandbook}. 7 trajectories were simulated independently. Their lengths were 320 ps, 65 ps, 337 ps, 147 ps, 24 ps, 27 ps and 80 ps. All trajectories were used to calculate thermodynamic properties. Dynamics were extracted individually from each trajectory, and averages were taken over the two longest trajectories 1 and 3 for further analysis. Results from an AIMD simulation of liquid water were also analyzed for comparison. The water system contained 64 molecules and one excess proton in a periodic cubic supercell of length 12.420 \AA\ at 300 K, corresponding to a density of 1.00 g/cm$^3$. 2 water trajectories of length 350 ps each were used. 

In both cases, the reference system Hamiltonian was constructed using the self-consistent charge DFTB3 level of theory~\cite{gaus2011dftb3}, while full DFT was implemented using the revPBE generalized gradient approximation functional~\cite{perdew1996generalized,zhang1998comment} with the Grimme D3 dispersion correction~\cite{grimme2010consistent}. The Kohn-Sham orbitals were expanded in a TZV2P atom-centered basis set, while the density was expanded in a plane-wave basis up to a cutoff of 400 Ry. Core electrons were replaced by atomic pseudopotentials of the Goedecker-Teter-Hutter type~\cite{goedecker1996separable}. The MTS algorithm was employed, where the DFTB reference system was integrated with a time step of 0.5 fs and the DFT corrections were applied every 2.0 fs. This yielded a factor of about 4 in computational efficiency. Simulations were performed in the NVT ensemble using a local Langevin thermostat with a time constant of 25~fs for the initial equilibration. Production runs, from which the properties reported in the manuscript were calculated, used a global stochastic velocity rescaling (SVR) thermostat~\cite{Bussi2007/10.1063/1.2408420} with a time constant of 1~ps. The global coupling of the SVR thermostat with such a large time constant results in negligible perturbation to the dynamics of the system~\cite{Ceriotti2010/10.1063/1.3489925}.

We also performed path integral AIMD simulations of imidazole using ring polymer contraction (RPC)~\cite{markland_mano_1,markland2008refined,ondrej_markland} with centroid contraction ($P'=1$). Non-centroid normal modes were thermostatted via a white noise Langevin thermostat~\cite{pimd_thermostat}. We accumulated a total of 15 ps of trajectory. Other properties such as the system size, density, and MTS propagation scheme were identical to those of the classical imidazole simulation described above.

\vspace{0.2in}
\section{\label{sec:msd_si}Mean Square Displacement Plots}

Diffusion coefficients are calculated following the usual approach of fitting the long timescale regime ($>$5ps) of the mean square displacement (MSD) curves to a straight line. The MSD is obtained by tracking the centers of mass of imidazole molecules with unwrapped periodic boundary conditions. For Imi*, identity changes of the defect are included in the calculation of the MSD.

\begin{equation}
    \label{eqn:msd_si}
    D=\frac{1}{6}\frac{d}{dt}\lim_{t\to\infty}\langle\left|\Delta\bm{r}(t)\right|^2 \rangle
\end{equation}    

\begin{figure}[ht]
    \centering
    \begin{subfigure}{0.4\textwidth}
        \includegraphics[width=\textwidth,trim=0 20 60 60,clip=true] {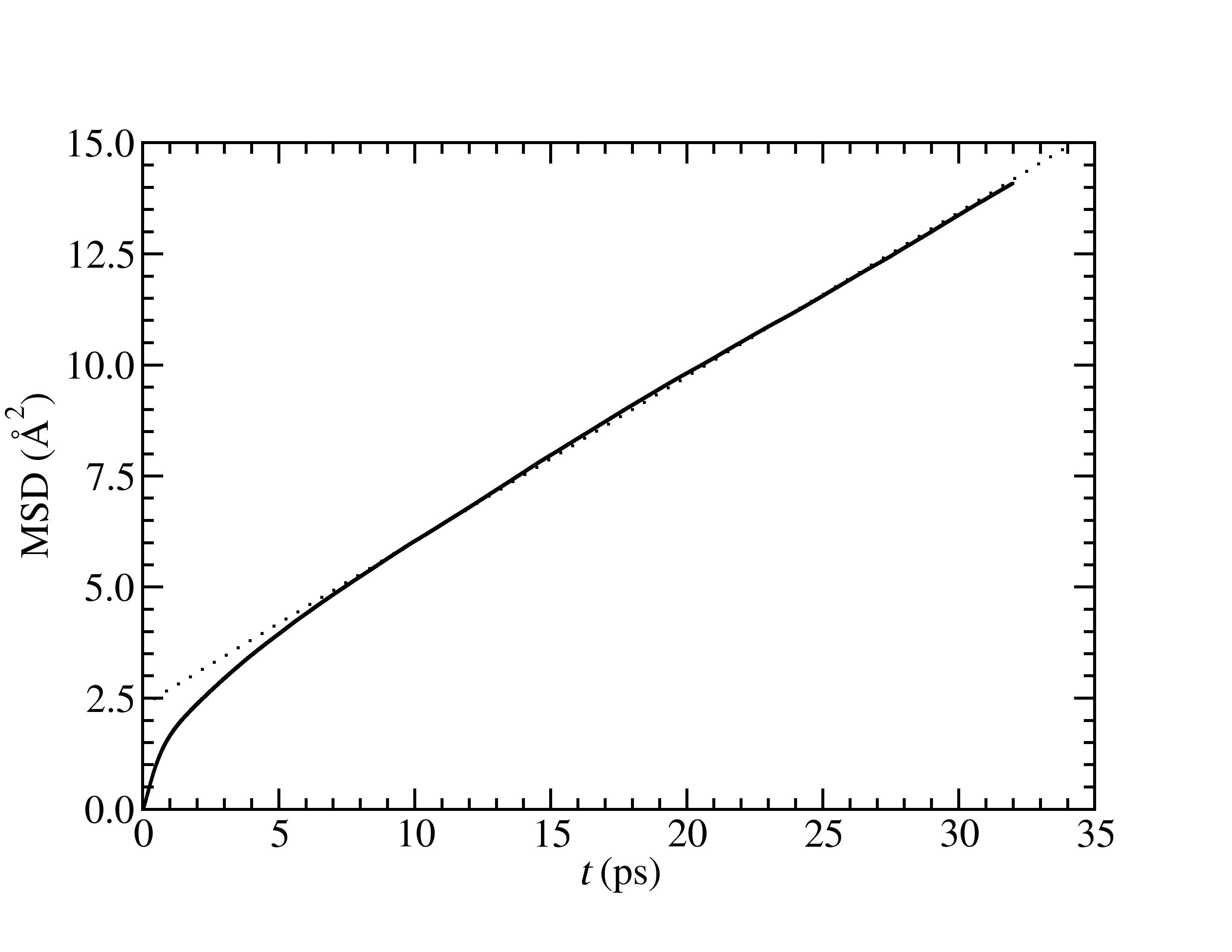}
        \caption{Neutral imidazole}
        \label{fig:normal_imi_msd_si}
    \end{subfigure}
    \begin{subfigure}{0.4\textwidth}
        \includegraphics[width=\textwidth,trim=0 20 60 60, clip=true] {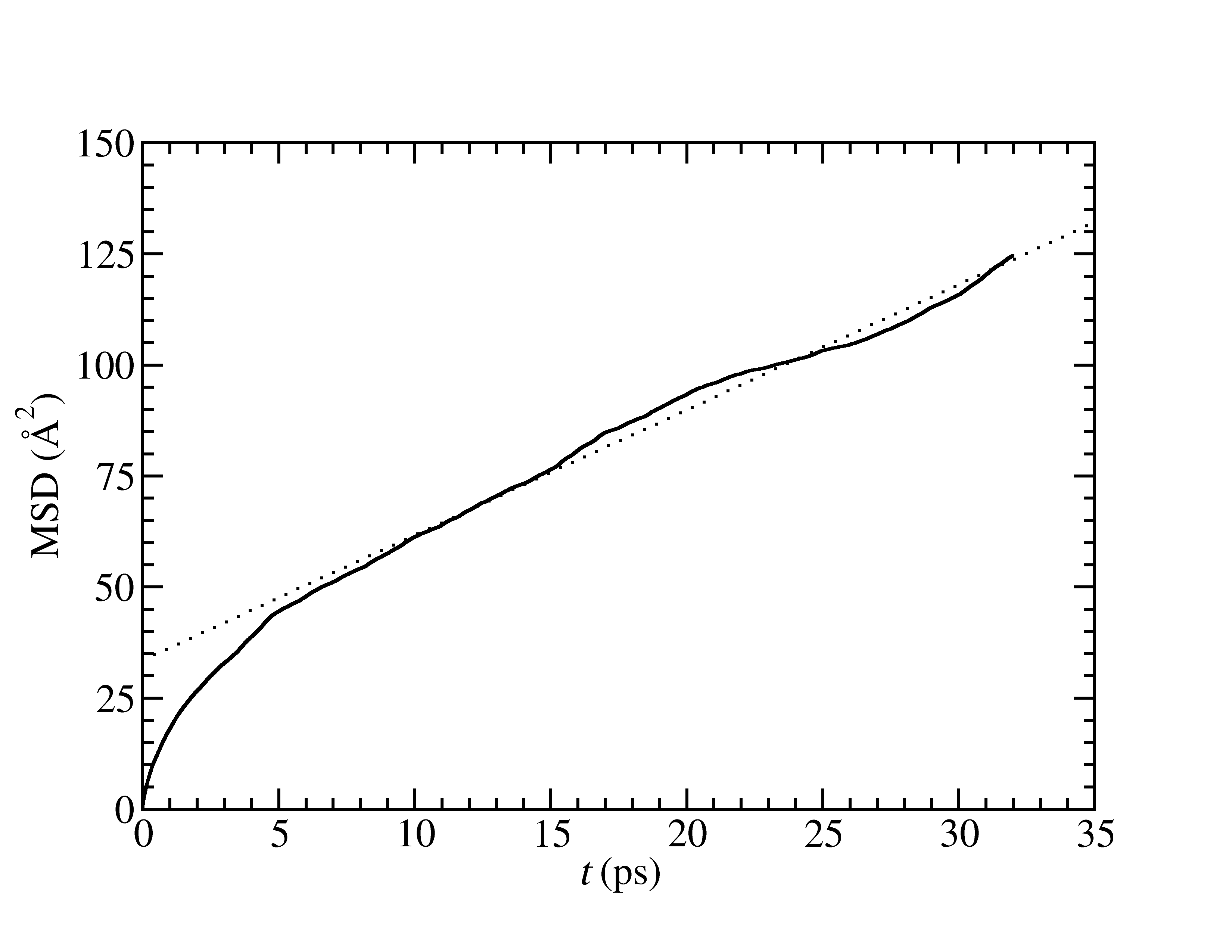}
        \caption{Imi*}
        \label{fig:h_imi_msd_si}
    \end{subfigure}
    \caption{Mean square displacement (MSD/6) plots and linear regression. The average of the MSD curve is taken from trajectories 1 and 3. Linear regression is carried out in the region \(t>5\)ps. The diffusion coefficient of neutral imidazole extracted from (a) is 0.062 \AA/ps$^2$ and that of Imi* extracted from (b) is 0.47 \AA/ps$^2$.}
    \label{fig:msd_plot_si}
\end{figure}

\vspace{0.2in}

\section{\label{sec:hb_criteria_si}Hydrogen Bond (HB) Geometry Cutoff}

The following geometric criteria are defined for the NHN HBs between imidazole molecules: N$_{a}$N$_{d}$ $\leq$ 3.09 \AA, N$_{a}$H$_{d}$ $\leq$ 2.11 \AA\, and N$_{a}$N$_{d}$H$_{d}$ $\leq$ 21.5$^\circ$. These are partly based on the radial distribution functions shown below.

\begin{figure}[ht]
    \centering
    \includegraphics[width=0.4\textwidth,trim=0 20 60 60, clip=true]{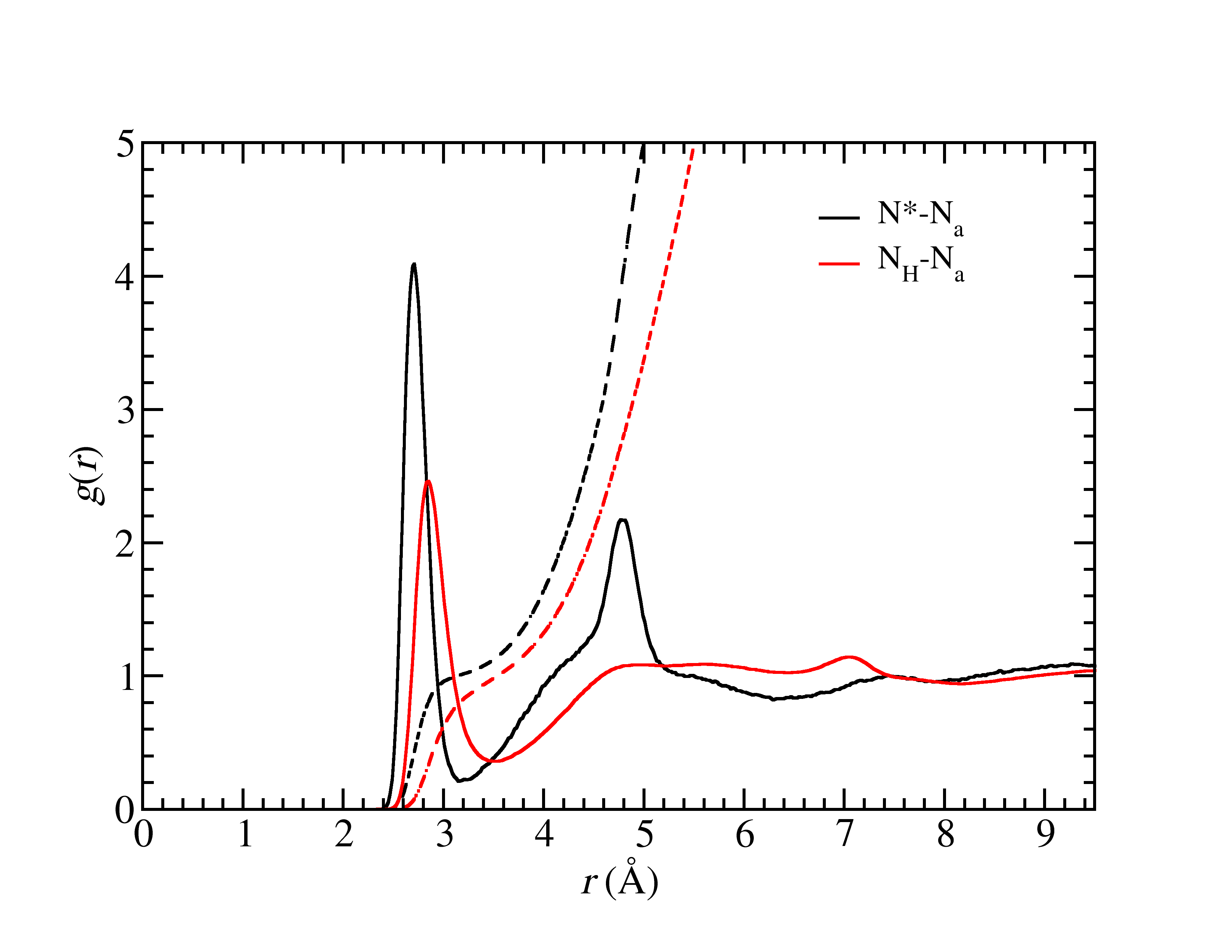}
    \caption{N$_d$-N$_a$ RDF. N$_d$, is the hydrogen bond donor nitrogen, and it includes either N* in Imi* or N$_{\rm H}$ in Imi. The first peak of the N*-N$_a$ curve at 2.71 $\AA$ is used the length of the hydrogen bond between Imi* and Imi, while the first peak of N$_{\rm H}$-N$_a$ curve at 2.85 $\AA$ is used as the length of the hydrogen bond between Imi and Imi. The dashed line is the running integration of the corresponding RDF defined in Eq.~\ref{eq:rdf_integrate}, and its value at the position of RDF first minimum is the coordination number of N$_a$ around N* or N$_{\rm H}$. }
    \label{fig:nd_na_identity}
\end{figure}

\begin{equation}
    \label{eq:rdf_integrate}
    n_{{\rm N}_d{\rm N}_a}(r) = 
        4\pi\rho_{N_a} \int_0^r g_{{\rm N}_d{\rm N}_a}(r) r^2{\rm d}r
\end{equation}

N$_d$ denotes the hydrogen bond donor nitrogen, usually either the N* of Imi* or N$_{\rm H}$ of Imi. The cutoff values were determined based on the maximum \(r\)  (N$_{a}$N$_{d}$ or N$_{a}$H$_{d}$) and maximum $\theta$ (N$_{a}$N$_{d}$H$_{d}$) of the free energy surface contour, $-k_BT\ln[P(r,\theta)]=k_BT$ ~\cite{Hayes_Tuckerman1,Hayes_Tuckerman2}, where $P(r,\theta)$ is the joint probability distribution of $r$ and $\theta$. This contour leads to a somewhat more restrictive definition than has been used in previous studies (N$_{a}$N$_{d}$H$_{d}$ $\leq$ 30$^\circ$ for Imi*-Imi HBs and N$_{a}$N$_{d}$H$_{d}$ $\leq$ 45$^\circ$ for Imi-Imi HBs; N$_{a}$N$_{d}$ $\leq$ 3.2~$\AA$ for both types)~\cite{Shen_12}. We applied the same criteria for Imi*-Imi HBs as Imi-Imi HBs since the former is stronger than the latter due to the excess charge and thus is well within the definition of the latter. The contour plot also shows a long N$_{a}$N$_{d}$H$_{d}$ angular tail, indicating a strong flexible and anharmonic character in the angular dependence of this NHN HB.

\begin{figure}[ht]
    \centering
    \includegraphics[width=0.5\textwidth]{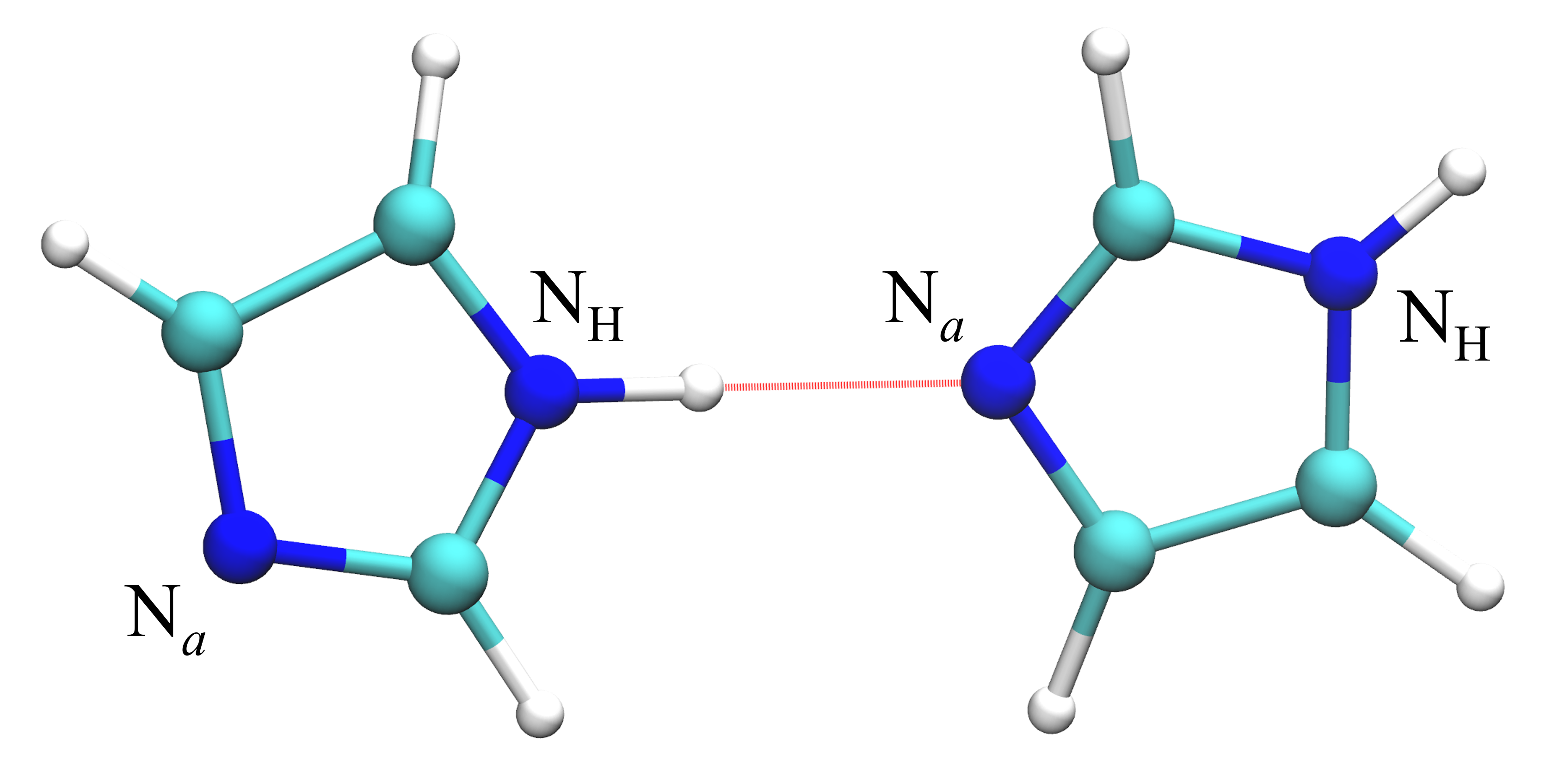}
    \caption{Snapshot from the simulated trajectory showing an Imi-Imi hydrogen bond.}
    \label{fig:imi_imi_hb_si}
\end{figure}

\begin{figure}[ht]
    \centering
    \begin{subfigure}{0.45\textwidth}
        \includegraphics[width=\textwidth,trim=0 20 60 60, clip=true]{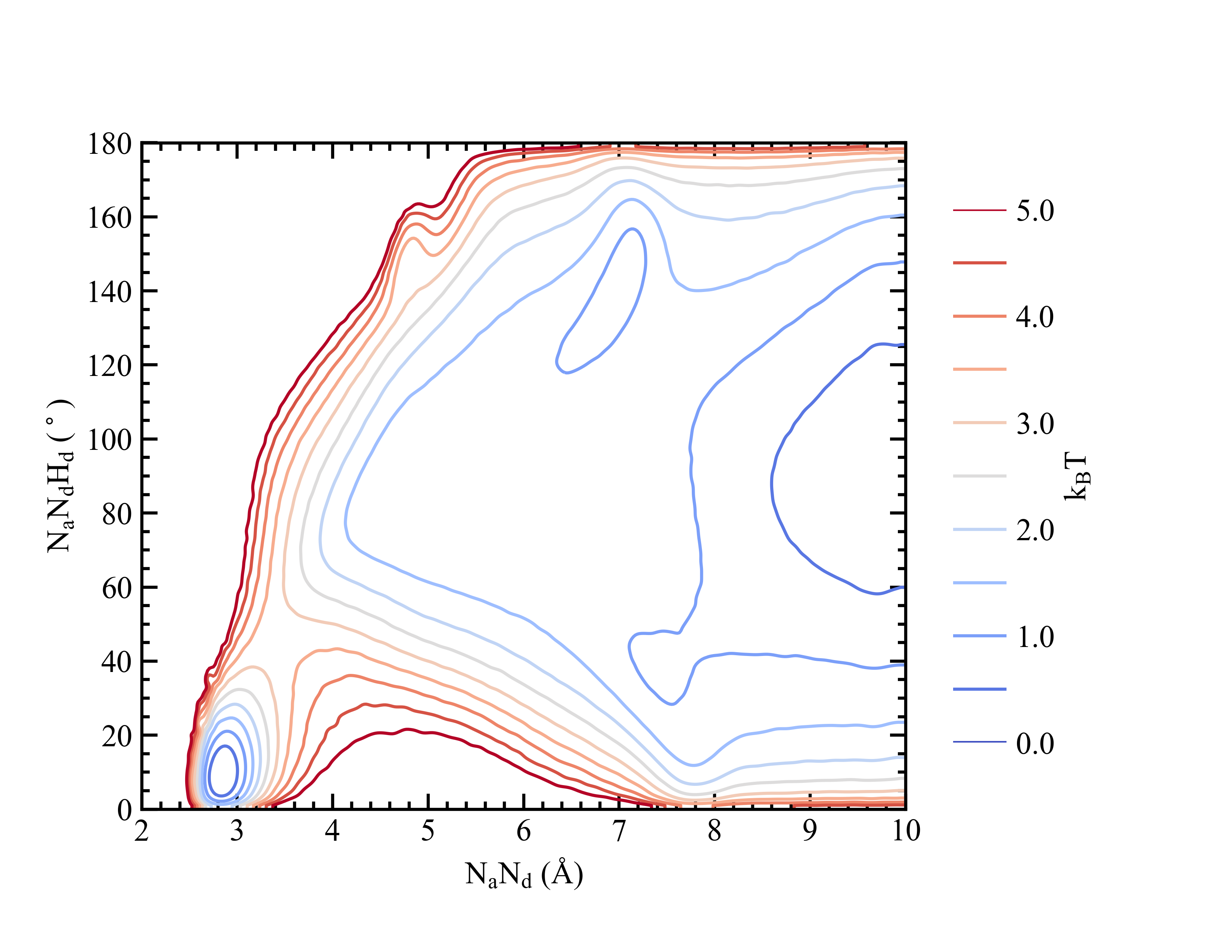}
        \caption{N$_a$N$_d$-N$_a$N$_d$H$_d$}
        \label{fig:nand_nandhd_si}
    \end{subfigure}
    \begin{subfigure}{0.45\textwidth}
        \includegraphics[width=\textwidth,trim=0 20 60 60, clip=true]{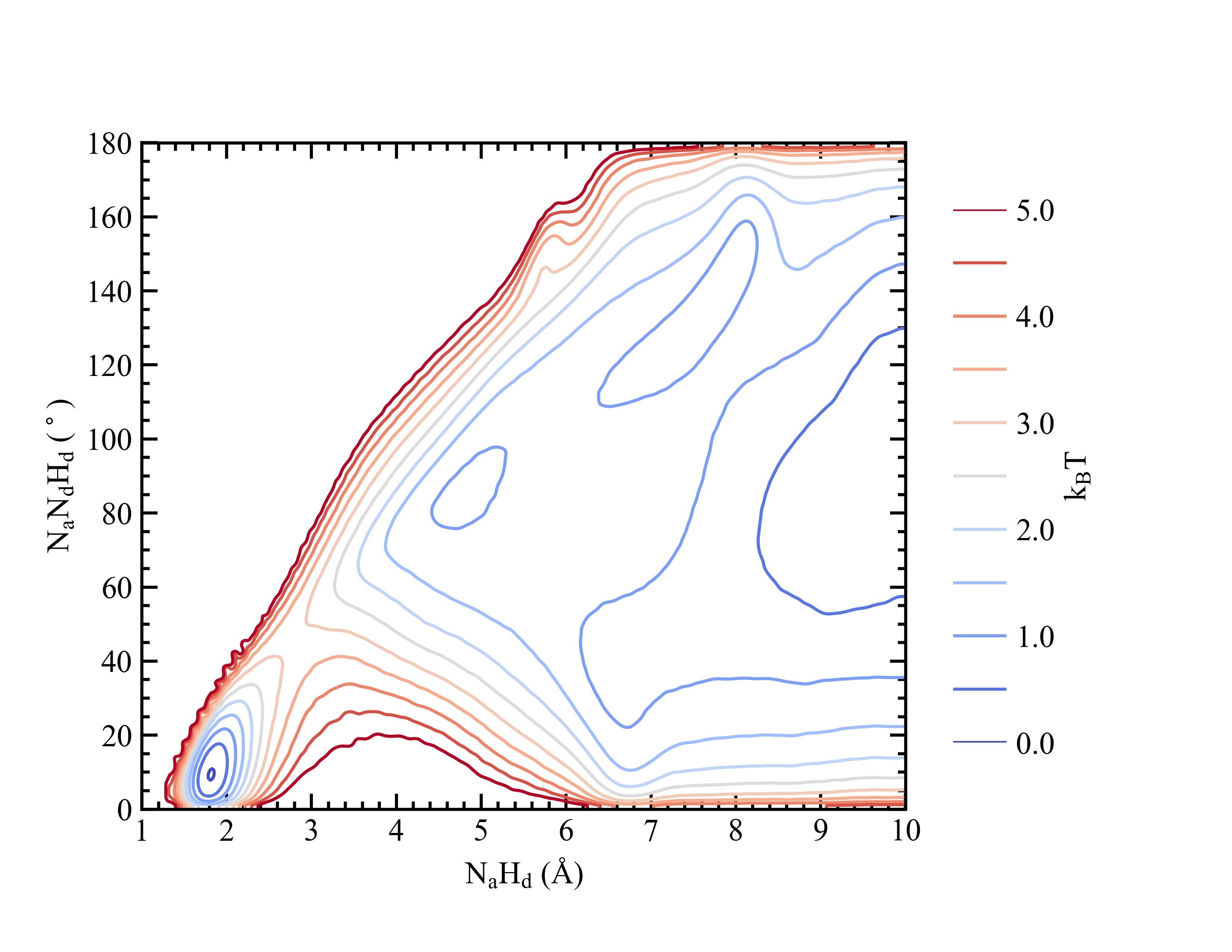}
        \caption{N$_a$H$_d$-N$_a$N$_d$H$_d$}
        \label{fig:nahd_nandhd_si}
    \end{subfigure}
    \caption{Free energy plot \(-k_BT\ln(P(r,\theta)\) (units $k_BT$) calculated from probability distribution \(P(r,\theta)\) for HB geometry cutoff.}
    \label{fig:hb_criteria_si}
\end{figure}

\begin{table}[ht]
\centering
\caption{HB Geometry Cutoff for Different Free Energy Criteria}
\label{tab:tab_hb_criteria_si}
\begin{ruledtabular}
\begin{tabular}{cccccc}
    & \(-k_BT\ln(P(r,\theta))=k_BT\) & \(-k_BT\ln(P(r,\theta))=1.5k_BT\) & \(-k_BT\ln(P(r,\theta))=2k_BT\) & 
    RDF 1$^{st}$ minimum \\
    \colrule
    N$_a$N$_d$ (\AA) & 3.09 & 3.17 & 3.25  & 3.49 \\
    N$_a$H$_d$ (\AA) & 2.11 & 2.20 & 2.30  & 2.91 \\
    N$_a$N$_d$H$_d$ ($^\circ$) & 21.5 & 25.5 & 29.5 & -- \\
\end{tabular}
\end{ruledtabular}
\end{table}
The different HB criteria will be referred as \(k_BT\), \(1.5k_BT\), and \(2k_BT\) in the following sections.

\begin{table}[ht]
\caption{Statistics Results for Hydrogen-Bonded Hyper Structure}
\label{tab:chain_statistics_si}
\centering
\begin{ruledtabular}
\begin{tabular}{ccccc}
     & \makecell{HB Criteria} & \makecell{Neutral Imidazole\\Chain} & \makecell{Neutral Imidazole\\Ring} & \makecell{Imi* Chain} \\
    \colrule
    \multirow{4}{*}{Average Number}
        & \(k_BT\) & 23.1 & 0.13 & 1 \\
        & \(1.5k_BT\) & 17.2 & 0.25 & 1 \\
        & \(2k_BT\) & 13.4 & 0.37 & 1 \\
    \colrule
    \multirow{4}{*}{Average Size}
        & \(k_BT\) & 2.5 & 4.7 & 6.6 \\
        & \(1.5k_BT\) & 3.2 & 5.0 & 8.6 \\
        & \(2k_BT\) & 3.8 & 5.4 & 10.6 \\
    \colrule
    \multirow{4}{*}{Maximum Size}
        & \(k_BT\) & 31 & 25 & 40 \\
        & \(1.5k_BT\) & 36 & 26 & 49 \\
        & \(2k_BT\) & 44 & 28 & 51 \\
    \colrule
    \multirow{4}{*}{Most Probable Size}
        & \(k_BT\) & 1 & 4 & 4 \\
        & \(1.5k_BT\) & 1 & 4 & 5 \\
        & \(2k_BT\) & 1 & 4 & 7 \\
\end{tabular}
\end{ruledtabular}
\end{table}

\vspace{0.2in}

\section{\label{sec:hb_chain_si}Hydrogen-Bonded Hyper Chain Statistics}

To compute the hydrogen bond chain lengths used to construct Fig.~5 in the main text, periodic boundary conditions were unwrapped in the same way as was done for the calculation of the diffusion constant. The purpose of doing so is to reveal actual diffusion pathways for the proton along hydrogen-bonded chains, which lead to the long tails in Fig.~5 of the main text. Unwrapping the periodic boundary conditions does not give rise to closed loops or rings in these chains: in even the longest chains seen in the distribution, no molecule appears more than once.

In Table \ref{tab:chain_statistics_si}, only the HBs of the most probable geometry in multiple donating/accepting HBs are kept to simplify the statistics. "Average Number" is the average number of hyper structures per configuration. Sizes refer to the number of imidazole moieties in the hydrogen-bonded hyper chain. Neutral imidazole chain and ring refer to the hyper chains/rings formed with neutral imidazole only. Imi* chain refers to chains including one Imi* and multiple neutral imidazole molecules (can be 0).

\begin{figure}[ht]
    \centering
    \begin{subfigure}{0.45\textwidth}
        \includegraphics[width=\textwidth]{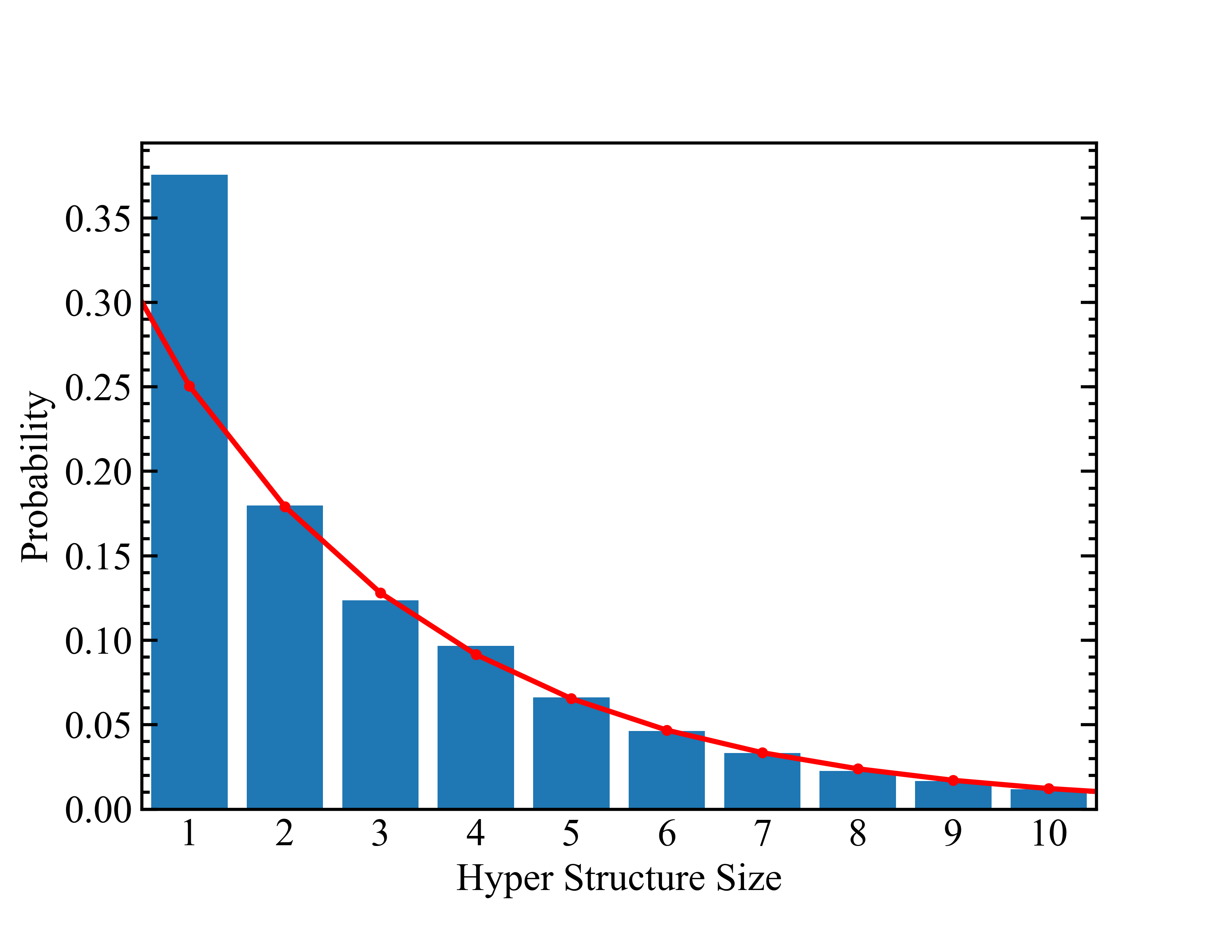}
        \caption{Imi chain, \(1.5k_BT\) criteria}
        \label{fig:imi_chain_size_15kt_si}
    \end{subfigure}
    \begin{subfigure}{0.45\textwidth}
        \includegraphics[width=\textwidth]{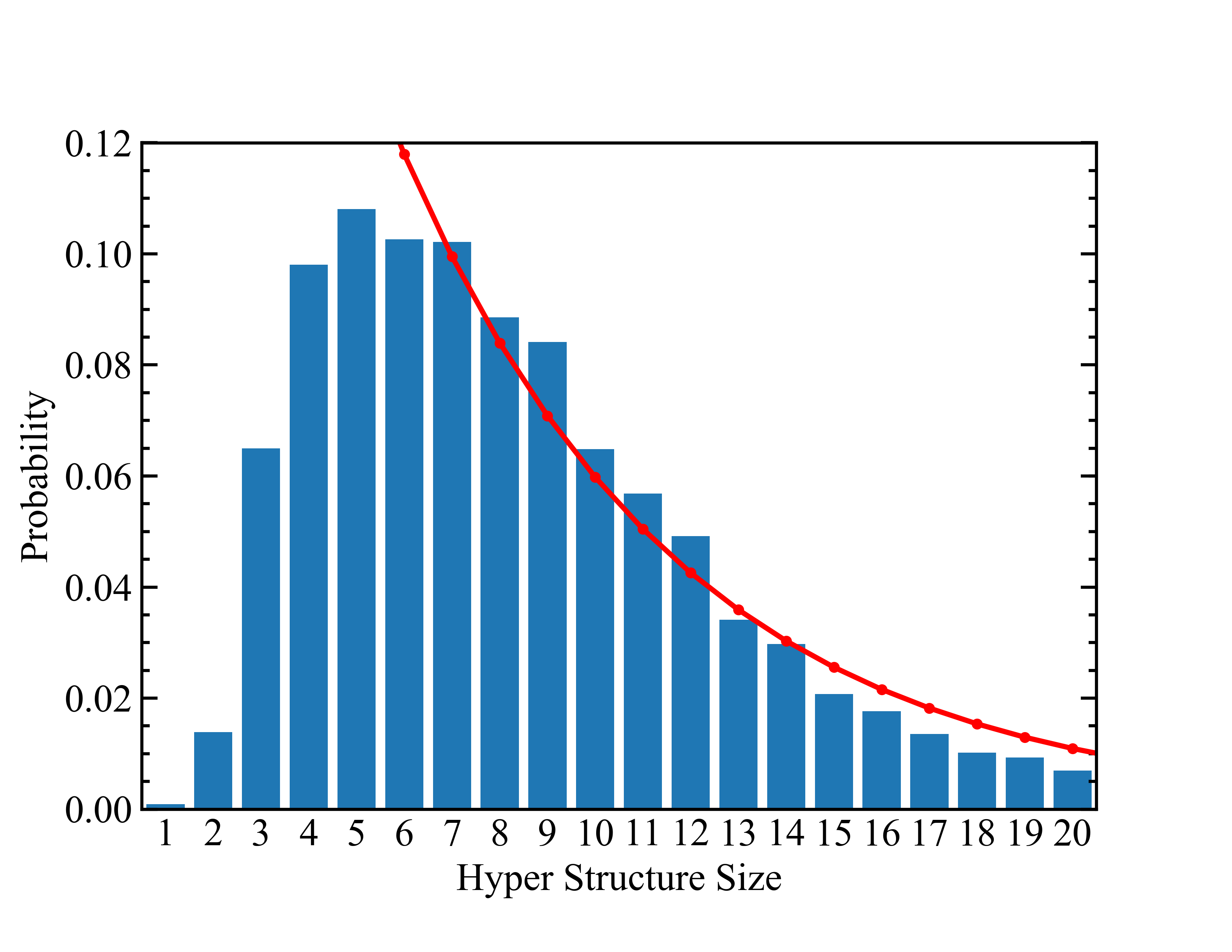}
        \caption{Imi* Chain, \(1.5k_BT\) criteria}
        \label{fig:h_imi_chain_size_15kt_si}
    \end{subfigure}
    
    \begin{subfigure}{0.45\textwidth}
        \includegraphics[width=\textwidth]{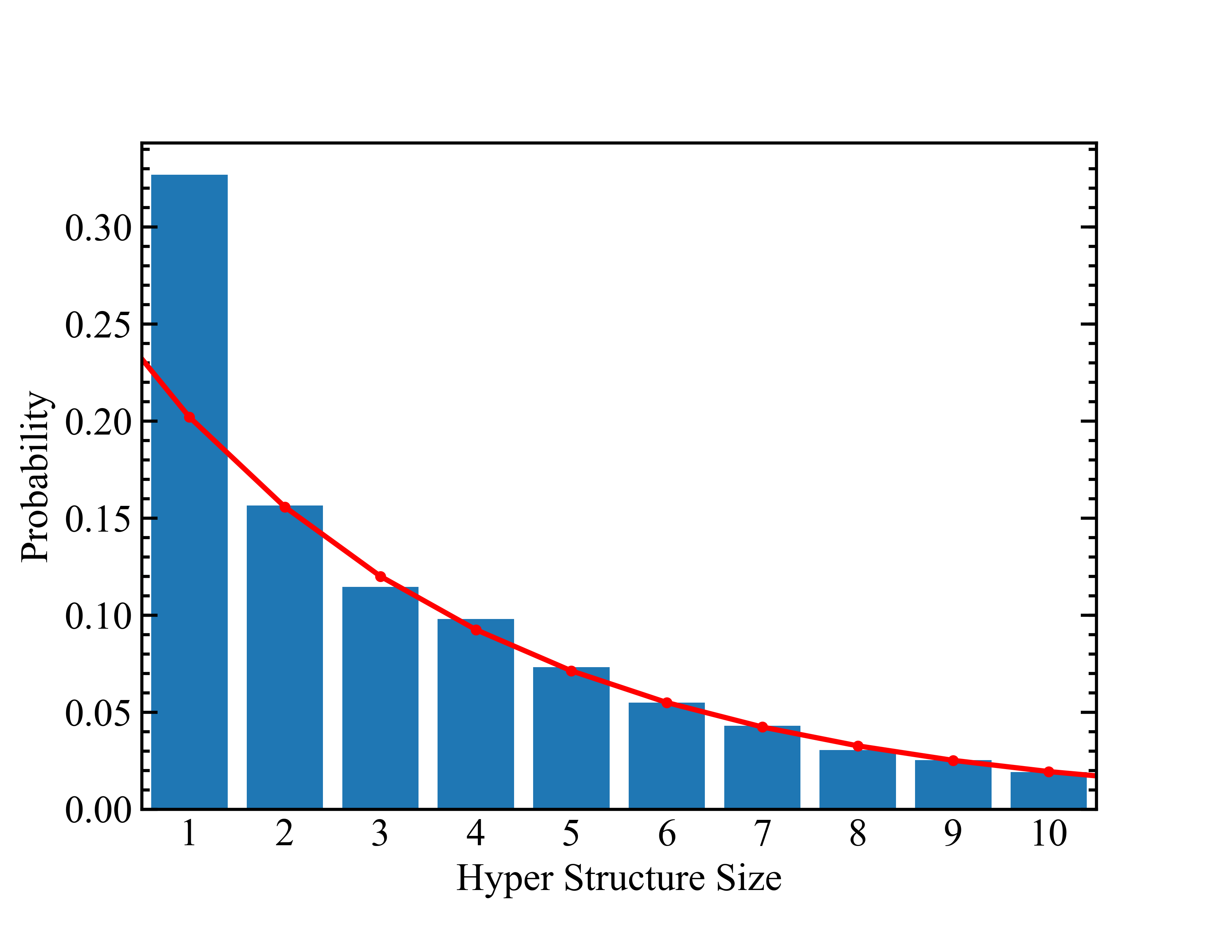}
        \caption{Imi chain, \(2k_BT\) criteria}
        \label{fig:imi_chain_size_2kt_si}
    \end{subfigure}
    \begin{subfigure}{0.45\textwidth}
        \includegraphics[width=\textwidth]{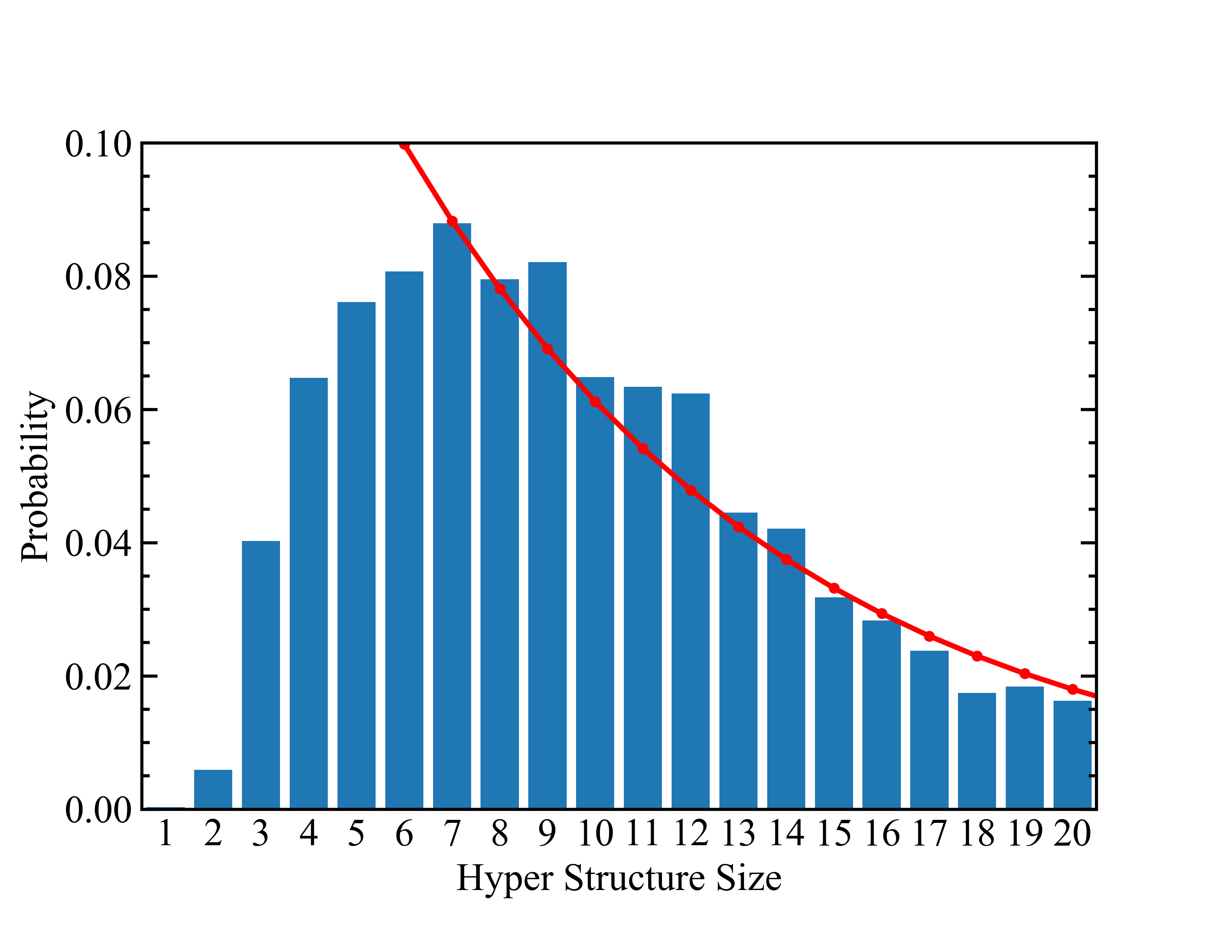}
        \caption{Imi* Chain, \(2k_BT\) criteria}
        \label{fig:h_imi_chain_size_2kt_si}
    \end{subfigure}
    
    \caption{Size distribution of hydrogen-bonded hyper structures. Blue bars are results obtained from trajectory. Red dotted lines are single power law fits, \(P(n)=aK^n\). The Imi chain size distribution is fitted for the range \(n>1\), while the Imi* chain size distribution is fitted for \(n>5\). The fitted parameters are shown in Table~\ref{tab:hyper_struc_fit}.}
    \label{fig:hyper_struct_size_si}
\end{figure}

\begin{table}[ht]
\centering
\caption{Fitted parameters for the hyper structure probability distribution}
\label{tab:hyper_struc_fit}
\begin{ruledtabular}
\begin{tabular}{ccccc}
    \multirow{2}{*}{HB Criteria} & \multicolumn{2}{c}{Imi Chain} &
        \multicolumn{2}{c}{Imi* Chain}\\
                     \cline{2-5} & a & K & a & K \\
    \colrule
    \(k_BT\) & 0.518 & 0.628 & 0.641 & 0.768 \\
    \colrule
    \(1.5k_BT\) & 0.350 & 0.715 & 0.327 & 0.844\\
    \colrule
    \(2k_BT\) & 0.262 & 0.771 & 0.208 & 0.885\\
\end{tabular}
\end{ruledtabular}
\end{table}

\vspace{0.2in}
\section{\label{sec:msd_finite}Mean Square Displacement for Random Walk within 1D Finite Grid}
Here we discuss about diffusion within a 1D finite grid containing \(n+1\) grid points. The grid points are marked 0, $\cdots$, \(n\). An object can leave its current position with a probability \(p\), which will change with the size of time step \(\delta t\) as \(p=e^{-\frac{\delta t}{\tau}}\). This random walk model (with reflection boundaries) can be treated with a discrete-time Markov chain. The transition probability matrix \(P\) has the elements \(p_{ij}\) as

\begin{align}
    & p_{ii}=1-p,  i=0, \cdots, n \\
    & p_{10}=p_{n-1,n}=p \\
    & p_{i+1,i}=p_{i-1,i}=p/2,  i=1, \cdots, n-1    
\end{align}

As it diffuses within the grid, taking the time average, the position of the object in the grid will follow a particular probability distribution. With \(t\rightarrow\infty\), a steady state can be reached. The steady state probability distribution \(L\), which is also the eigenvector of the transition probability matrix with an eigenvalue \(\lambda=1\), will have elements \(L_i\) as 
\begin{align}
    & L_0=L_n=\frac{1}{2n} \\
    & L_i=\frac{1}{n}, i=1, \cdots, n-1
\end{align}

From this probability distribution \(L\), we can calculate the mean square displacement MSD($\infty$) from each starting point \(j\), with \(d_m\) representing the distance of \(m\) separations.

\begin{equation}
    {\rm MSD}_j(\infty)=\sum_{i=0}^n L_i d^2_{\lvert i-j \rvert}
\end{equation}

We then take the average over all starting points, which also follows the steady state distribution.

\begin{equation}
    {\rm MSD}(\infty) = \sum_{j=0}^n L_j{\rm MSD}_j(\infty)
                      = \sum_{j=0}^n\sum_{i=0}^n L_j L_i  d^2_{\lvert i-j \rvert}
\end{equation}

The \(d^2_{\lvert i-j \rvert}\) term indicates that there will be multiple terms with the same \(d_m\) value. Combining terms with same \(d_m\) value, we get:

\begin{equation}
    \begin{split}
        {\rm MSD}(\infty) 
        & = 2\left(\frac{1}{2n}*\frac{1}{n}*2+(n-2)*\frac{1}{n}*\frac{1}{n}\right)d_1^2
            +2\left(\frac{1}{2n}*\frac{1}{n}*2+(n-3)*\frac{1}{n}*\frac{1}{n}\right)d_2^2 \\
        & +\cdots +2\left(\frac{1}{2n}*\frac{1}{n}*2\right)d_{n-1}^2 
          +2\left(\frac{1}{2n}\right)^2d_n^2 \\
        & = \frac{2}{n^2}\sum_{i=1}^{n-1}(n-i)d_i^2+\frac{1}{2n^2}d_n^2
    \end{split}
\end{equation}

If we have uniform spacing between grid points (which is the usual case), \(d_i=i*d\):

\begin{equation}
    {\rm MSD}(\infty) = \frac{2d^2}{n^2}\sum_{i=1}^{n-1}ni^2-i^3 + \frac{1}{2n^2}n^2d^2
                      = \frac{n^2+2}{6}d^2
\end{equation}

\section{\label{sec:ocf_si} Orientational Correlation Times}

\begin{figure}[ht]
    \centering
    \begin{subfigure}{0.2\textwidth}
        \includegraphics[width=0.9\textwidth,trim=50 200 50 200]{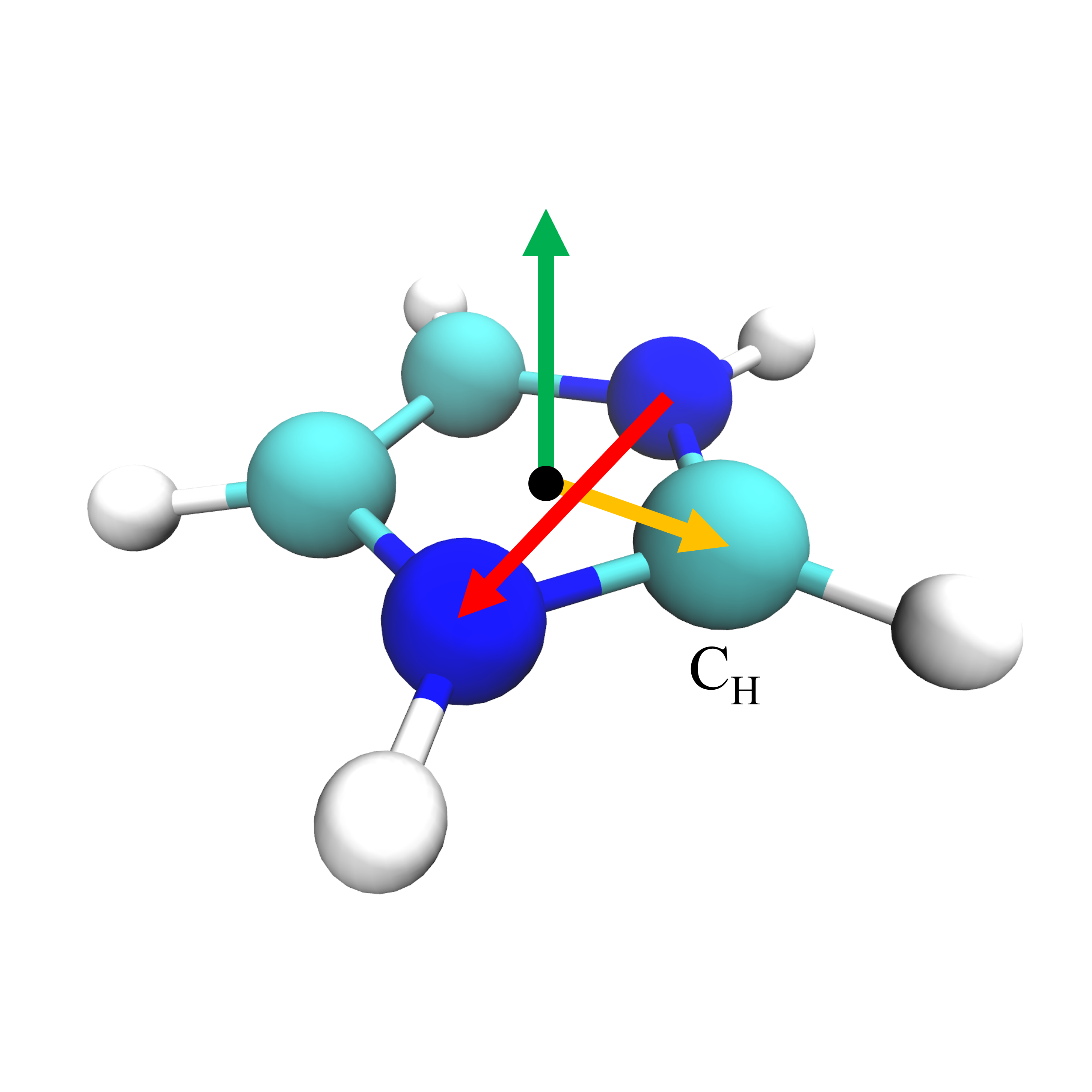}
        \caption{Imi*}
        \label{fig:rot_vector_h_imi}
    \end{subfigure}
    \begin{subfigure}{0.2\textwidth}
        \includegraphics[width=0.9\textwidth,trim=50 200 50 200]{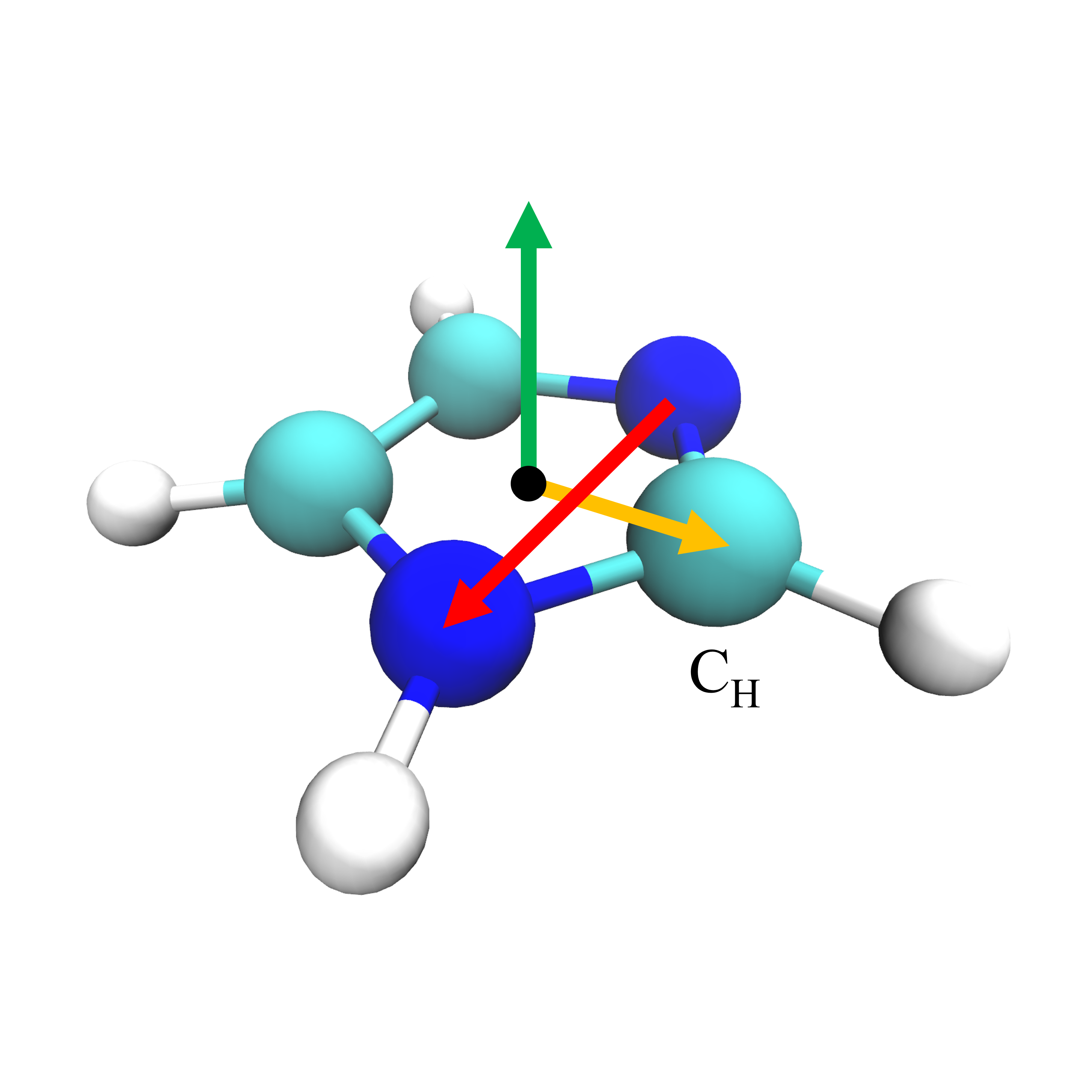}
        \caption{Imi}
        \label{fig:rot_vector_normal_imi}
    \end{subfigure}
    
    \caption{Schematic representation of vectors picked for orientational correlation function calculation. The black dot represents the center of mass or corresponding molecule. Red arrow represents the N-N vector, orange Center-C$_{\rm H}$ and green Molecular Plane Normal. C$_{\rm H}$ is the carbon atom bonded to two nitrogen atoms. Molecular Plane Normal calculated from C$_{\rm H}$ and two nitrogen atoms. This selection of vectors is a rough correspondence to the principle axes system of Imi*}
    \label{fig:rot_vector}
\end{figure}

\begin{figure}[ht]
    \centering
    \includegraphics[width=0.4\textwidth,trim=0 20 60 60, clip=true]{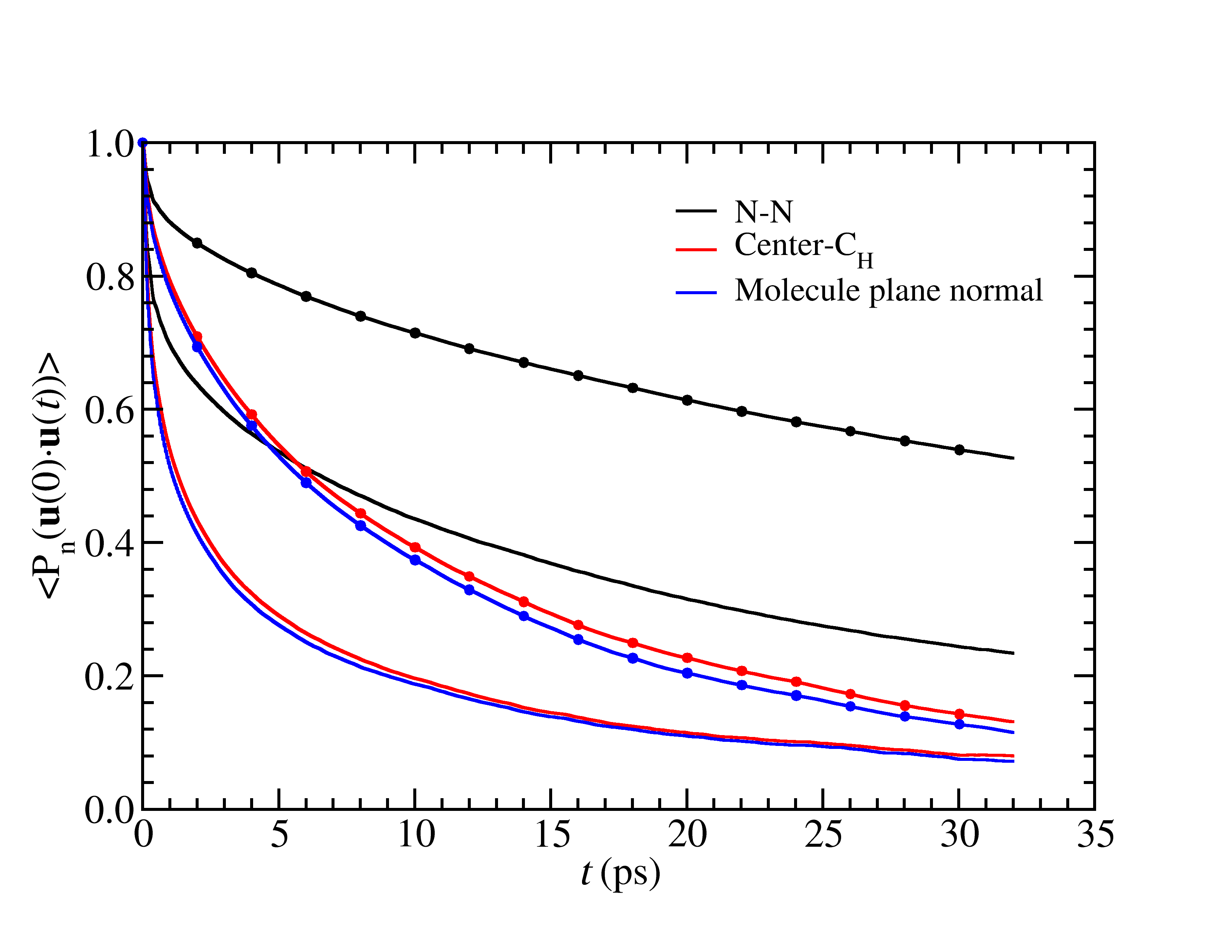}
    \caption{Rotational correlation function of different vectors. The choice of the vector is a rough correspondence to the principle axes system. Curves with circle marks are n=1 Legendre polynomials, and curves without marks are n=2. Molecule plane normal vectors are calculated from two N atoms and C$_\text{H}$ atom}
    \label{fig:rot_cf_all_si}
\end{figure}

Figure~\ref{fig:rot_cf_all_si} shows the P$_{1}$ and P$_{2}$ (corresponding to first and second order Legendre polynomials respectively) orientational correlation functions~\cite{doi:10.1080/00268978100102181,doi:10.1080/00268978200101361} for the three axes of imidazole shown in Fig.~\ref{fig:rot_vector} (N-N, Center-C$_{\rm H}$ and the Molecular Plane Normal). The time constants extracted from a triexponential fit to these correlation functions are shown in Table~\ref{tab:rot_cf_param_si}. 

\begin{table}[ht]
\centering
\caption{Orientational Correlation Times}
\label{tab:rot_cf_param_si}
\begin{ruledtabular}
\begin{tabular}{ccdddddddd}
    Vector & & 
    \multicolumn{1}{c}{\(a_1\)} & 
    \multicolumn{1}{c}{\(\tau_1\)(ps)} &
    \multicolumn{1}{c}{\(a_2\)} & 
    \multicolumn{1}{c}{\(\tau_2\)(ps)} &
    \multicolumn{1}{c}{\(a_3\)} & 
    \multicolumn{1}{c}{\(\tau_3\)(ps)} &
    \multicolumn{1}{c}{\(\tau_\text{fit}\)(ps)} &
    \multicolumn{1}{c}{\(R^2\)} \\
    \colrule
    \multirow{2}{*}{N-N}
        & \(P_1\) & 0.094 & 0.27 & 0.11 & 5.27 & 0.79 & 77.44 & 62.02 & 0.999\\
        & \(P_2\) & 0.26 & 0.24 & 0.20 & 4.34 & 0.54 & 37.03 & 21.02 & 0.999\\
    \colrule
    \multirow{2}{*}{Center-C$_\text{H}$}
        & \(P_1\) & 0.13 & 0.37 & 0.30 & 4.48 & 0.57 & 21.61 & 13.67 & 0.999\\
        & \(P_2\) & 0.39 & 0.32 & 0.36 & 3.37 & 0.26 & 25.84 & 7.94 & 0.999\\
    \colrule
    \multirow{2}{*}{Molecule Plane Normal}
        & \(P_1\) & 0.16 & 0.38 & 0.33 & 5.27 & 0.51 & 21.20 & 12.69 & 0.999\\
        & \(P_2\) & 0.39 & 0.27 & 0.35 & 2.92 & 0.26 & 23.61 & 7.34 & 0.999\\
\end{tabular}
\end{ruledtabular}
\end{table}


\begin{thebibliography}{60}%
\makeatletter
\providecommand \@ifxundefined [1]{%
 \@ifx{#1\undefined}
}%
\providecommand \@ifnum [1]{%
 \ifnum #1\expandafter \@firstoftwo
 \else \expandafter \@secondoftwo
 \fi
}%
\providecommand \@ifx [1]{%
 \ifx #1\expandafter \@firstoftwo
 \else \expandafter \@secondoftwo
 \fi
}%
\providecommand \natexlab [1]{#1}%
\providecommand \enquote  [1]{``#1''}%
\providecommand \bibnamefont  [1]{#1}%
\providecommand \bibfnamefont [1]{#1}%
\providecommand \citenamefont [1]{#1}%
\providecommand \href@noop [0]{\@secondoftwo}%
\providecommand \href [0]{\begingroup \@sanitize@url \@href}%
\providecommand \@href[1]{\@@startlink{#1}\@@href}%
\providecommand \@@href[1]{\endgroup#1\@@endlink}%
\providecommand \@sanitize@url [0]{\catcode `\\12\catcode `\$12\catcode
  `\&12\catcode `\#12\catcode `\^12\catcode `\_12\catcode `\%12\relax}%
\providecommand \@@startlink[1]{}%
\providecommand \@@endlink[0]{}%
\providecommand \url  [0]{\begingroup\@sanitize@url \@url }%
\providecommand \@url [1]{\endgroup\@href {#1}{\urlprefix }}%
\providecommand \urlprefix  [0]{URL }%
\providecommand \Eprint [0]{\href }%
\providecommand \doibase [0]{http://dx.doi.org/}%
\providecommand \selectlanguage [0]{\@gobble}%
\providecommand \bibinfo  [0]{\@secondoftwo}%
\providecommand \bibfield  [0]{\@secondoftwo}%
\providecommand \translation [1]{[#1]}%
\providecommand \BibitemOpen [0]{}%
\providecommand \bibitemStop [0]{}%
\providecommand \bibitemNoStop [0]{.\EOS\space}%
\providecommand \EOS [0]{\spacefactor3000\relax}%
\providecommand \BibitemShut  [1]{\csname bibitem#1\endcsname}%
\let\auto@bib@innerbib\@empty
\bibitem [{\citenamefont {de~Grotthuss}(1806)}]{Grot_1806}%
  \BibitemOpen
  \bibfield  {author} {\bibinfo {author} {\bibfnamefont {C.~J.~T.}\
  \bibnamefont {de~Grotthuss}},\ }\href@noop {} {\bibfield  {journal} {\bibinfo
   {journal} {Ann. Chim.}\ }\textbf {\bibinfo {volume} {1806}} (\bibinfo {year}
  {1806})}\BibitemShut {NoStop}%
\bibitem [{\citenamefont {Marx}(2006)}]{Marx_Rev_2006}%
  \BibitemOpen
  \bibfield  {author} {\bibinfo {author} {\bibfnamefont {D.}~\bibnamefont
  {Marx}},\ }\href@noop {} {\bibfield  {journal} {\bibinfo  {journal}
  {ChemPhysChem}\ }\textbf {\bibinfo {volume} {7}} (\bibinfo {year}
  {2006})}\BibitemShut {NoStop}%
\bibitem [{\citenamefont {Tuckerman}\ \emph
  {et~al.}(1995{\natexlab{a}})\citenamefont {Tuckerman}, \citenamefont
  {Laasonen}, \citenamefont {Sprik},\ and\ \citenamefont
  {Parrinello}}]{Tuckerman_JPC_1995}%
  \BibitemOpen
  \bibfield  {author} {\bibinfo {author} {\bibfnamefont {M.}~\bibnamefont
  {Tuckerman}}, \bibinfo {author} {\bibfnamefont {K.}~\bibnamefont {Laasonen}},
  \bibinfo {author} {\bibfnamefont {M.}~\bibnamefont {Sprik}}, \ and\ \bibinfo
  {author} {\bibfnamefont {M.}~\bibnamefont {Parrinello}},\ }\href@noop {}
  {\bibfield  {journal} {\bibinfo  {journal} {J. Phys. Chem.}\ }\textbf
  {\bibinfo {volume} {99}} (\bibinfo {year} {1995}{\natexlab{a}})}\BibitemShut
  {NoStop}%
\bibitem [{\citenamefont {Agmon}(1995)}]{Agmon_CPL}%
  \BibitemOpen
  \bibfield  {author} {\bibinfo {author} {\bibfnamefont {N.}~\bibnamefont
  {Agmon}},\ }\href@noop {} {\bibfield  {journal} {\bibinfo  {journal} {Chem.
  Phys. Lett.}\ }\textbf {\bibinfo {volume} {244}} (\bibinfo {year}
  {1995})}\BibitemShut {NoStop}%
\bibitem [{\citenamefont {Tuckerman}\ \emph
  {et~al.}(1995{\natexlab{b}})\citenamefont {Tuckerman}, \citenamefont
  {Laasonen}, \citenamefont {Sprik},\ and\ \citenamefont
  {Parrinello}}]{Tuckerman_JCP_1995}%
  \BibitemOpen
  \bibfield  {author} {\bibinfo {author} {\bibfnamefont {M.}~\bibnamefont
  {Tuckerman}}, \bibinfo {author} {\bibfnamefont {K.}~\bibnamefont {Laasonen}},
  \bibinfo {author} {\bibfnamefont {M.}~\bibnamefont {Sprik}}, \ and\ \bibinfo
  {author} {\bibfnamefont {M.}~\bibnamefont {Parrinello}},\ }\href@noop {}
  {\bibfield  {journal} {\bibinfo  {journal} {J. Chem. Phys.}\ }\textbf
  {\bibinfo {volume} {103}} (\bibinfo {year} {1995}{\natexlab{b}})}\BibitemShut
  {NoStop}%
\bibitem [{\citenamefont {Marx}\ \emph {et~al.}(1999)\citenamefont {Marx},
  \citenamefont {Tuckerman}, \citenamefont {Hutter},\ and\ \citenamefont
  {Parrinello}}]{Tuckerman_Nature_Hp}%
  \BibitemOpen
  \bibfield  {author} {\bibinfo {author} {\bibfnamefont {D.}~\bibnamefont
  {Marx}}, \bibinfo {author} {\bibfnamefont {M.~E.}\ \bibnamefont {Tuckerman}},
  \bibinfo {author} {\bibfnamefont {J.}~\bibnamefont {Hutter}}, \ and\ \bibinfo
  {author} {\bibfnamefont {M.}~\bibnamefont {Parrinello}},\ }\href@noop {}
  {\bibfield  {journal} {\bibinfo  {journal} {Nature}\ }\textbf {\bibinfo
  {volume} {397}} (\bibinfo {year} {1999})}\BibitemShut {NoStop}%
\bibitem [{\citenamefont {Marx}, \citenamefont {Chandra},\ and\ \citenamefont
  {Tuckerman}(2010)}]{Tuckerman_rev_2010}%
  \BibitemOpen
  \bibfield  {author} {\bibinfo {author} {\bibfnamefont {D.}~\bibnamefont
  {Marx}}, \bibinfo {author} {\bibfnamefont {A.}~\bibnamefont {Chandra}}, \
  and\ \bibinfo {author} {\bibfnamefont {M.~E.}\ \bibnamefont {Tuckerman}},\
  }\href@noop {} {\bibfield  {journal} {\bibinfo  {journal} {Chem. Rev.}\
  }\textbf {\bibinfo {volume} {110}} (\bibinfo {year} {2010})}\BibitemShut
  {NoStop}%
\bibitem [{\citenamefont {Agmon}\ \emph {et~al.}(2016)\citenamefont {Agmon},
  \citenamefont {Bakker}, \citenamefont {Campen}, \citenamefont {Henchman},
  \citenamefont {Pohl}, \citenamefont {Roke}, \citenamefont {Thamer},\ and\
  \citenamefont {Hassanali}}]{Agmon_rev_2016}%
  \BibitemOpen
  \bibfield  {author} {\bibinfo {author} {\bibfnamefont {N.}~\bibnamefont
  {Agmon}}, \bibinfo {author} {\bibfnamefont {H.~J.}\ \bibnamefont {Bakker}},
  \bibinfo {author} {\bibfnamefont {R.~K.}\ \bibnamefont {Campen}}, \bibinfo
  {author} {\bibfnamefont {R.~H.}\ \bibnamefont {Henchman}}, \bibinfo {author}
  {\bibfnamefont {P.}~\bibnamefont {Pohl}}, \bibinfo {author} {\bibfnamefont
  {S.}~\bibnamefont {Roke}}, \bibinfo {author} {\bibfnamefont {M.}~\bibnamefont
  {Thamer}}, \ and\ \bibinfo {author} {\bibfnamefont {A.}~\bibnamefont
  {Hassanali}},\ }\href@noop {} {\bibfield  {journal} {\bibinfo  {journal}
  {Chem. Rev.}\ }\textbf {\bibinfo {volume} {116}} (\bibinfo {year}
  {2016})}\BibitemShut {NoStop}%
\bibitem [{\citenamefont {Morrone}\ and\ \citenamefont
  {Tuckerman}(2002)}]{Morrone_02}%
  \BibitemOpen
  \bibfield  {author} {\bibinfo {author} {\bibfnamefont {J.~A.}\ \bibnamefont
  {Morrone}}\ and\ \bibinfo {author} {\bibfnamefont {M.~E.}\ \bibnamefont
  {Tuckerman}},\ }\href@noop {} {\bibfield  {journal} {\bibinfo  {journal} {J.
  Chem. Phys.}\ }\textbf {\bibinfo {volume} {117}} (\bibinfo {year}
  {2002})}\BibitemShut {NoStop}%
\bibitem [{\citenamefont {Stoyanov}, \citenamefont {Stoyanova},\ and\
  \citenamefont {Reed}(2008)}]{Reed_08}%
  \BibitemOpen
  \bibfield  {author} {\bibinfo {author} {\bibfnamefont {E.~S.}\ \bibnamefont
  {Stoyanov}}, \bibinfo {author} {\bibfnamefont {F.~V.}\ \bibnamefont
  {Stoyanova}}, \ and\ \bibinfo {author} {\bibfnamefont {C.~A.}\ \bibnamefont
  {Reed}},\ }\href@noop {} {\bibfield  {journal} {\bibinfo  {journal} {Chem.
  Euro. J.}\ }\textbf {\bibinfo {volume} {14}} (\bibinfo {year}
  {2008})}\BibitemShut {NoStop}%
\bibitem [{\citenamefont {Lee}, \citenamefont {Son},\ and\ \citenamefont
  {Park}(2015)}]{Park_15}%
  \BibitemOpen
  \bibfield  {author} {\bibinfo {author} {\bibfnamefont {C.}~\bibnamefont
  {Lee}}, \bibinfo {author} {\bibfnamefont {H.}~\bibnamefont {Son}}, \ and\
  \bibinfo {author} {\bibfnamefont {S.}~\bibnamefont {Park}},\ }\href@noop {}
  {\bibfield  {journal} {\bibinfo  {journal} {Phys. Chem. Chem. Phys.}\
  }\textbf {\bibinfo {volume} {17}} (\bibinfo {year} {2015})}\BibitemShut
  {NoStop}%
\bibitem [{\citenamefont {Lee}\ \emph {et~al.}(2010)\citenamefont {Lee},
  \citenamefont {Ogawa}, \citenamefont {Kanno}, \citenamefont {Nakamoto},
  \citenamefont {Yasuda},\ and\ \citenamefont {Watanabe}}]{Watanabe_10}%
  \BibitemOpen
  \bibfield  {author} {\bibinfo {author} {\bibfnamefont {S.~Y.}\ \bibnamefont
  {Lee}}, \bibinfo {author} {\bibfnamefont {A.}~\bibnamefont {Ogawa}}, \bibinfo
  {author} {\bibfnamefont {M.}~\bibnamefont {Kanno}}, \bibinfo {author}
  {\bibfnamefont {H.}~\bibnamefont {Nakamoto}}, \bibinfo {author}
  {\bibfnamefont {T.}~\bibnamefont {Yasuda}}, \ and\ \bibinfo {author}
  {\bibfnamefont {M.}~\bibnamefont {Watanabe}},\ }\href@noop {} {\bibfield
  {journal} {\bibinfo  {journal} {J. Am. Chem. Soc.}\ }\textbf {\bibinfo
  {volume} {132}} (\bibinfo {year} {2010})}\BibitemShut {NoStop}%
\bibitem [{\citenamefont {Vilciauskas}\ \emph {et~al.}(2012)\citenamefont
  {Vilciauskas}, \citenamefont {Tuckerman}, \citenamefont {Bester},
  \citenamefont {Paddison},\ and\ \citenamefont {Kreuer}}]{Kreuer_12}%
  \BibitemOpen
  \bibfield  {author} {\bibinfo {author} {\bibfnamefont {L.}~\bibnamefont
  {Vilciauskas}}, \bibinfo {author} {\bibfnamefont {M.~E.}\ \bibnamefont
  {Tuckerman}}, \bibinfo {author} {\bibfnamefont {G.}~\bibnamefont {Bester}},
  \bibinfo {author} {\bibfnamefont {S.~J.}\ \bibnamefont {Paddison}}, \ and\
  \bibinfo {author} {\bibfnamefont {K.~D.}\ \bibnamefont {Kreuer}},\
  }\href@noop {} {\bibfield  {journal} {\bibinfo  {journal} {Nature Chem.}\
  }\textbf {\bibinfo {volume} {4}} (\bibinfo {year} {2012})}\BibitemShut
  {NoStop}%
\bibitem [{\citenamefont {Vilciauskas}\ \emph {et~al.}(2013)\citenamefont
  {Vilciauskas}, \citenamefont {Tuckerman}, \citenamefont {Melchior},
  \citenamefont {Bester},\ and\ \citenamefont {Kreuer}}]{Kreuer_13}%
  \BibitemOpen
  \bibfield  {author} {\bibinfo {author} {\bibfnamefont {L.}~\bibnamefont
  {Vilciauskas}}, \bibinfo {author} {\bibfnamefont {M.~E.}\ \bibnamefont
  {Tuckerman}}, \bibinfo {author} {\bibfnamefont {J.~P.}\ \bibnamefont
  {Melchior}}, \bibinfo {author} {\bibfnamefont {G.}~\bibnamefont {Bester}}, \
  and\ \bibinfo {author} {\bibfnamefont {K.~D.}\ \bibnamefont {Kreuer}},\
  }\href@noop {} {\bibfield  {journal} {\bibinfo  {journal} {Solid State
  Ionics}\ }\textbf {\bibinfo {volume} {252}} (\bibinfo {year}
  {2013})}\BibitemShut {NoStop}%
\bibitem [{\citenamefont {Chandra}\ \emph {et~al.}(2014)\citenamefont
  {Chandra}, \citenamefont {Kundu}, \citenamefont {Kandambeth}, \citenamefont
  {BabaRao}, \citenamefont {Marathe}, \citenamefont {Kunjir},\ and\
  \citenamefont {Banerjee}}]{Banerjee_14}%
  \BibitemOpen
  \bibfield  {author} {\bibinfo {author} {\bibfnamefont {S.}~\bibnamefont
  {Chandra}}, \bibinfo {author} {\bibfnamefont {T.}~\bibnamefont {Kundu}},
  \bibinfo {author} {\bibfnamefont {S.}~\bibnamefont {Kandambeth}}, \bibinfo
  {author} {\bibfnamefont {R.}~\bibnamefont {BabaRao}}, \bibinfo {author}
  {\bibfnamefont {Y.}~\bibnamefont {Marathe}}, \bibinfo {author} {\bibfnamefont
  {S.~M.}\ \bibnamefont {Kunjir}}, \ and\ \bibinfo {author} {\bibfnamefont
  {R.}~\bibnamefont {Banerjee}},\ }\href@noop {} {\bibfield  {journal}
  {\bibinfo  {journal} {J. Am. Chem. Soc.}\ }\textbf {\bibinfo {volume} {136}}
  (\bibinfo {year} {2014})}\BibitemShut {NoStop}%
\bibitem [{\citenamefont {Chen}, \citenamefont {Yan},\ and\ \citenamefont
  {Voth}(2009)}]{Voth_09}%
  \BibitemOpen
  \bibfield  {author} {\bibinfo {author} {\bibfnamefont {H.~N.}\ \bibnamefont
  {Chen}}, \bibinfo {author} {\bibfnamefont {T.~Y.}\ \bibnamefont {Yan}}, \
  and\ \bibinfo {author} {\bibfnamefont {G.~A.}\ \bibnamefont {Voth}},\
  }\href@noop {} {\bibfield  {journal} {\bibinfo  {journal} {J. Phys. Chem. A}\
  }\textbf {\bibinfo {volume} {113}} (\bibinfo {year} {2009})}\BibitemShut
  {NoStop}%
\bibitem [{\citenamefont {Li}\ \emph {et~al.}(2012)\citenamefont {Li},
  \citenamefont {Li}, \citenamefont {Yan},\ and\ \citenamefont
  {Shen}}]{Shen_12}%
  \BibitemOpen
  \bibfield  {author} {\bibinfo {author} {\bibfnamefont {A.~L.}\ \bibnamefont
  {Li}}, \bibinfo {author} {\bibfnamefont {Y.}~\bibnamefont {Li}}, \bibinfo
  {author} {\bibfnamefont {T.~Y.}\ \bibnamefont {Yan}}, \ and\ \bibinfo
  {author} {\bibfnamefont {P.~W.}\ \bibnamefont {Shen}},\ }\href@noop {}
  {\bibfield  {journal} {\bibinfo  {journal} {J. Phys. Chem. B}\ }\textbf
  {\bibinfo {volume} {116}} (\bibinfo {year} {2012})}\BibitemShut {NoStop}%
\bibitem [{\citenamefont {Yaghini}\ \emph {et~al.}(2016)\citenamefont
  {Yaghini}, \citenamefont {Gomez-Gonzalez}, \citenamefont {Varela},\ and\
  \citenamefont {Martinelli}}]{Martinelli_16}%
  \BibitemOpen
  \bibfield  {author} {\bibinfo {author} {\bibfnamefont {N.}~\bibnamefont
  {Yaghini}}, \bibinfo {author} {\bibfnamefont {V.}~\bibnamefont
  {Gomez-Gonzalez}}, \bibinfo {author} {\bibfnamefont {L.~M.}\ \bibnamefont
  {Varela}}, \ and\ \bibinfo {author} {\bibfnamefont {A.}~\bibnamefont
  {Martinelli}},\ }\href@noop {} {\bibfield  {journal} {\bibinfo  {journal}
  {Phys. Chem. Chem. Phys.}\ }\textbf {\bibinfo {volume} {18}} (\bibinfo {year}
  {2016})}\BibitemShut {NoStop}%
\bibitem [{\citenamefont {Ponomareva}, \citenamefont {Shutova},\ and\
  \citenamefont {Matvienko}(2004)}]{Matvienko_04}%
  \BibitemOpen
  \bibfield  {author} {\bibinfo {author} {\bibfnamefont {V.~G.}\ \bibnamefont
  {Ponomareva}}, \bibinfo {author} {\bibfnamefont {E.~S.}\ \bibnamefont
  {Shutova}}, \ and\ \bibinfo {author} {\bibfnamefont {A.~A.}\ \bibnamefont
  {Matvienko}},\ }\href@noop {} {\bibfield  {journal} {\bibinfo  {journal}
  {Inorg. Mater.}\ }\textbf {\bibinfo {volume} {40}} (\bibinfo {year}
  {2004})}\BibitemShut {NoStop}%
\bibitem [{\citenamefont {Wood}\ and\ \citenamefont
  {Marzari}(2007)}]{Marzari_07}%
  \BibitemOpen
  \bibfield  {author} {\bibinfo {author} {\bibfnamefont {B.~C.}\ \bibnamefont
  {Wood}}\ and\ \bibinfo {author} {\bibfnamefont {N.}~\bibnamefont {Marzari}},\
  }\href@noop {} {\bibfield  {journal} {\bibinfo  {journal} {Phys. Rev. B}\
  }\textbf {\bibinfo {volume} {76}} (\bibinfo {year} {2007})}\BibitemShut
  {NoStop}%
\bibitem [{\citenamefont {Haile}\ \emph {et~al.}(2001)\citenamefont {Haile},
  \citenamefont {Boysen}, \citenamefont {Chisholm},\ and\ \citenamefont
  {Merle}}]{Haile_01}%
  \BibitemOpen
  \bibfield  {author} {\bibinfo {author} {\bibfnamefont {S.~M.}\ \bibnamefont
  {Haile}}, \bibinfo {author} {\bibfnamefont {D.~A.}\ \bibnamefont {Boysen}},
  \bibinfo {author} {\bibfnamefont {C.~R.~I.}\ \bibnamefont {Chisholm}}, \ and\
  \bibinfo {author} {\bibfnamefont {R.~B.}\ \bibnamefont {Merle}},\ }\href@noop
  {} {\bibfield  {journal} {\bibinfo  {journal} {Nature}\ }\textbf {\bibinfo
  {volume} {1410}} (\bibinfo {year} {2001})}\BibitemShut {NoStop}%
\bibitem [{\citenamefont {Lee}\ and\ \citenamefont {Tuckerman}(2008)}]{MET_08}%
  \BibitemOpen
  \bibfield  {author} {\bibinfo {author} {\bibfnamefont {H.-S.}\ \bibnamefont
  {Lee}}\ and\ \bibinfo {author} {\bibfnamefont {M.~E.}\ \bibnamefont
  {Tuckerman}},\ }\href@noop {} {\bibfield  {journal} {\bibinfo  {journal} {J.
  Phys. Chem. C}\ }\textbf {\bibinfo {volume} {112}} (\bibinfo {year}
  {2008})}\BibitemShut {NoStop}%
\bibitem [{\citenamefont {Kim}\ \emph {et~al.}(2013)\citenamefont {Kim},
  \citenamefont {Blanc}, \citenamefont {Hu},\ and\ \citenamefont
  {Grey}}]{Grey_13}%
  \BibitemOpen
  \bibfield  {author} {\bibinfo {author} {\bibfnamefont {G.}~\bibnamefont
  {Kim}}, \bibinfo {author} {\bibfnamefont {F.}~\bibnamefont {Blanc}}, \bibinfo
  {author} {\bibfnamefont {Y.~Y.}\ \bibnamefont {Hu}}, \ and\ \bibinfo {author}
  {\bibfnamefont {C.~P.}\ \bibnamefont {Grey}},\ }\href@noop {} {\bibfield
  {journal} {\bibinfo  {journal} {J. Phys. Chem. C}\ }\textbf {\bibinfo
  {volume} {117}} (\bibinfo {year} {2013})}\BibitemShut {NoStop}%
\bibitem [{\citenamefont {Andrio}\ \emph {et~al.}(2019)\citenamefont {Andrio},
  \citenamefont {Hernandez}, \citenamefont {Garcia-Alcantara}, \citenamefont
  {del Castillo}, \citenamefont {Compan},\ and\ \citenamefont
  {Santamaria-Holek}}]{Santamaria_19}%
  \BibitemOpen
  \bibfield  {author} {\bibinfo {author} {\bibfnamefont {A.}~\bibnamefont
  {Andrio}}, \bibinfo {author} {\bibfnamefont {S.~I.}\ \bibnamefont
  {Hernandez}}, \bibinfo {author} {\bibfnamefont {C.}~\bibnamefont
  {Garcia-Alcantara}}, \bibinfo {author} {\bibfnamefont {L.~F.}\ \bibnamefont
  {del Castillo}}, \bibinfo {author} {\bibfnamefont {V.}~\bibnamefont
  {Compan}}, \ and\ \bibinfo {author} {\bibfnamefont {I.}~\bibnamefont
  {Santamaria-Holek}},\ }\href@noop {} {\bibfield  {journal} {\bibinfo
  {journal} {Phys. Chem. Chem. Phys.}\ }\textbf {\bibinfo {volume} {21}}
  (\bibinfo {year} {2019})}\BibitemShut {NoStop}%
\bibitem [{\citenamefont {Munch}\ \emph {et~al.}(2001)\citenamefont {Munch},
  \citenamefont {Kreuer}, \citenamefont {Silvestri}, \citenamefont {Maier},\
  and\ \citenamefont {Seifert}}]{Seifert_01}%
  \BibitemOpen
  \bibfield  {author} {\bibinfo {author} {\bibfnamefont {W.}~\bibnamefont
  {Munch}}, \bibinfo {author} {\bibfnamefont {K.~D.}\ \bibnamefont {Kreuer}},
  \bibinfo {author} {\bibfnamefont {W.}~\bibnamefont {Silvestri}}, \bibinfo
  {author} {\bibfnamefont {J.}~\bibnamefont {Maier}}, \ and\ \bibinfo {author}
  {\bibfnamefont {G.}~\bibnamefont {Seifert}},\ }\href@noop {} {\bibfield
  {journal} {\bibinfo  {journal} {Solid State Ionics}\ }\textbf {\bibinfo
  {volume} {145}} (\bibinfo {year} {2001})}\BibitemShut {NoStop}%
\bibitem [{\citenamefont {Iannuzzi}\ and\ \citenamefont
  {Parrinello}(2004)}]{Parrinello_04}%
  \BibitemOpen
  \bibfield  {author} {\bibinfo {author} {\bibfnamefont {M.}~\bibnamefont
  {Iannuzzi}}\ and\ \bibinfo {author} {\bibfnamefont {M.}~\bibnamefont
  {Parrinello}},\ }\href@noop {} {\bibfield  {journal} {\bibinfo  {journal}
  {Phys. Rev. Lett.}\ }\textbf {\bibinfo {volume} {93}} (\bibinfo {year}
  {2004})}\BibitemShut {NoStop}%
\bibitem [{\citenamefont {Iannuzzi}(2006)}]{Iannuzzi_06}%
  \BibitemOpen
  \bibfield  {author} {\bibinfo {author} {\bibfnamefont {M.}~\bibnamefont
  {Iannuzzi}},\ }\href@noop {} {\bibfield  {journal} {\bibinfo  {journal} {J.
  Chem. Phys.}\ }\textbf {\bibinfo {volume} {124}} (\bibinfo {year}
  {2006})}\BibitemShut {NoStop}%
\bibitem [{\citenamefont {Kreuer}(1997)}]{Kreuer_96}%
  \BibitemOpen
  \bibfield  {author} {\bibinfo {author} {\bibfnamefont {K.~D.}\ \bibnamefont
  {Kreuer}},\ }\href@noop {} {\bibfield  {journal} {\bibinfo  {journal} {Solid
  State Ionics}\ }\textbf {\bibinfo {volume} {97}} (\bibinfo {year}
  {1997})}\BibitemShut {NoStop}%
\bibitem [{\citenamefont {Kreuer}(1999)}]{Kreuer_99}%
  \BibitemOpen
  \bibfield  {author} {\bibinfo {author} {\bibfnamefont {K.~D.}\ \bibnamefont
  {Kreuer}},\ }\href@noop {} {\bibfield  {journal} {\bibinfo  {journal} {Solid
  State Ionics}\ }\textbf {\bibinfo {volume} {125}} (\bibinfo {year}
  {1999})}\BibitemShut {NoStop}%
\bibitem [{\citenamefont {Kreuer}(2003)}]{Kreuer_03}%
  \BibitemOpen
  \bibfield  {author} {\bibinfo {author} {\bibfnamefont {K.~D.}\ \bibnamefont
  {Kreuer}},\ }\href@noop {} {\bibfield  {journal} {\bibinfo  {journal} {Ann.
  Rev. Mat. Sci.}\ }\textbf {\bibinfo {volume} {33}} (\bibinfo {year}
  {2003})}\BibitemShut {NoStop}%
\bibitem [{\citenamefont {Malavasi}, \citenamefont {Fisher},\ and\
  \citenamefont {Islam}(2010)}]{Islam_10}%
  \BibitemOpen
  \bibfield  {author} {\bibinfo {author} {\bibfnamefont {L.}~\bibnamefont
  {Malavasi}}, \bibinfo {author} {\bibfnamefont {C.~A.~J.}\ \bibnamefont
  {Fisher}}, \ and\ \bibinfo {author} {\bibfnamefont {M.~S.}\ \bibnamefont
  {Islam}},\ }\href@noop {} {\bibfield  {journal} {\bibinfo  {journal} {Chem.
  Soc. Rev.}\ }\textbf {\bibinfo {volume} {39}} (\bibinfo {year}
  {2010})}\BibitemShut {NoStop}%
\bibitem [{\citenamefont {Kreuer}(2001)}]{Kreuer_01}%
  \BibitemOpen
  \bibfield  {author} {\bibinfo {author} {\bibfnamefont {K.~D.}\ \bibnamefont
  {Kreuer}},\ }\href@noop {} {\bibfield  {journal} {\bibinfo  {journal} {J.
  Membrane Sci.}\ }\textbf {\bibinfo {volume} {185}} (\bibinfo {year}
  {2001})}\BibitemShut {NoStop}%
\bibitem [{\citenamefont {Schuster}\ \emph {et~al.}(2001)\citenamefont
  {Schuster}, \citenamefont {Meyer}, \citenamefont {Wegner}, \citenamefont
  {Herz}, \citenamefont {Ise}, \citenamefont {Schuster}, \citenamefont
  {Kreuer},\ and\ \citenamefont {Maier}}]{Maier_01}%
  \BibitemOpen
  \bibfield  {author} {\bibinfo {author} {\bibfnamefont {M.}~\bibnamefont
  {Schuster}}, \bibinfo {author} {\bibfnamefont {W.~H.}\ \bibnamefont {Meyer}},
  \bibinfo {author} {\bibfnamefont {G.}~\bibnamefont {Wegner}}, \bibinfo
  {author} {\bibfnamefont {H.~G.}\ \bibnamefont {Herz}}, \bibinfo {author}
  {\bibfnamefont {M.}~\bibnamefont {Ise}}, \bibinfo {author} {\bibfnamefont
  {M.}~\bibnamefont {Schuster}}, \bibinfo {author} {\bibfnamefont {K.~D.}\
  \bibnamefont {Kreuer}}, \ and\ \bibinfo {author} {\bibfnamefont
  {J.}~\bibnamefont {Maier}},\ }\href@noop {} {\bibfield  {journal} {\bibinfo
  {journal} {Solid State Ionics}\ }\textbf {\bibinfo {volume} {145}} (\bibinfo
  {year} {2001})}\BibitemShut {NoStop}%
\bibitem [{\citenamefont {Schuster}\ \emph {et~al.}(2004)\citenamefont
  {Schuster}, \citenamefont {Meyer}, \citenamefont {Schuster},\ and\
  \citenamefont {Kreuer}}]{Kreuer_04}%
  \BibitemOpen
  \bibfield  {author} {\bibinfo {author} {\bibfnamefont {M.~F.~H.}\
  \bibnamefont {Schuster}}, \bibinfo {author} {\bibfnamefont {W.~H.}\
  \bibnamefont {Meyer}}, \bibinfo {author} {\bibfnamefont {M.}~\bibnamefont
  {Schuster}}, \ and\ \bibinfo {author} {\bibfnamefont {K.~D.}\ \bibnamefont
  {Kreuer}},\ }\href@noop {} {\bibfield  {journal} {\bibinfo  {journal} {Chem.
  Mater.}\ }\textbf {\bibinfo {volume} {16}} (\bibinfo {year}
  {2004})}\BibitemShut {NoStop}%
\bibitem [{\citenamefont {Paddison}, \citenamefont {Kreuer},\ and\
  \citenamefont {Maier}(2006)}]{Maier_06}%
  \BibitemOpen
  \bibfield  {author} {\bibinfo {author} {\bibfnamefont {S.~J.}\ \bibnamefont
  {Paddison}}, \bibinfo {author} {\bibfnamefont {K.~D.}\ \bibnamefont
  {Kreuer}}, \ and\ \bibinfo {author} {\bibfnamefont {J.}~\bibnamefont
  {Maier}},\ }\href@noop {} {\bibfield  {journal} {\bibinfo  {journal} {Phys.
  Chem. Chem. Phys.}\ }\textbf {\bibinfo {volume} {8}} (\bibinfo {year}
  {2006})}\BibitemShut {NoStop}%
\bibitem [{\citenamefont {Yoon}\ \emph {et~al.}(2013)\citenamefont {Yoon},
  \citenamefont {Suh}, \citenamefont {Natarajan},\ and\ \citenamefont
  {Kim}}]{Kim_13}%
  \BibitemOpen
  \bibfield  {author} {\bibinfo {author} {\bibfnamefont {M.}~\bibnamefont
  {Yoon}}, \bibinfo {author} {\bibfnamefont {K.}~\bibnamefont {Suh}}, \bibinfo
  {author} {\bibfnamefont {S.}~\bibnamefont {Natarajan}}, \ and\ \bibinfo
  {author} {\bibfnamefont {K.}~\bibnamefont {Kim}},\ }\href@noop {} {\bibfield
  {journal} {\bibinfo  {journal} {Angew Minirev.}\ }\textbf {\bibinfo {volume}
  {52}} (\bibinfo {year} {2013})}\BibitemShut {NoStop}%
\bibitem [{\citenamefont {Horike}, \citenamefont {Umeyama},\ and\ \citenamefont
  {Kitagawa}(2013)}]{Kitagawa_13}%
  \BibitemOpen
  \bibfield  {author} {\bibinfo {author} {\bibfnamefont {S.}~\bibnamefont
  {Horike}}, \bibinfo {author} {\bibfnamefont {D.}~\bibnamefont {Umeyama}}, \
  and\ \bibinfo {author} {\bibfnamefont {S.}~\bibnamefont {Kitagawa}},\
  }\href@noop {} {\bibfield  {journal} {\bibinfo  {journal} {Acc. Chem. Rev.}\
  }\textbf {\bibinfo {volume} {46}} (\bibinfo {year} {2013})}\BibitemShut
  {NoStop}%
\bibitem [{\citenamefont {Wu}\ \emph {et~al.}(2014)\citenamefont {Wu},
  \citenamefont {Ge}, \citenamefont {Lin}, \citenamefont {Wu}, \citenamefont
  {Luo},\ and\ \citenamefont {Xu}}]{Xu_14}%
  \BibitemOpen
  \bibfield  {author} {\bibinfo {author} {\bibfnamefont {B.}~\bibnamefont
  {Wu}}, \bibinfo {author} {\bibfnamefont {L.}~\bibnamefont {Ge}}, \bibinfo
  {author} {\bibfnamefont {X.}~\bibnamefont {Lin}}, \bibinfo {author}
  {\bibfnamefont {L.}~\bibnamefont {Wu}}, \bibinfo {author} {\bibfnamefont
  {J.}~\bibnamefont {Luo}}, \ and\ \bibinfo {author} {\bibfnamefont
  {T.}~\bibnamefont {Xu}},\ }\href@noop {} {\bibfield  {journal} {\bibinfo
  {journal} {J. Membrane Sci.}\ }\textbf {\bibinfo {volume} {458}} (\bibinfo
  {year} {2014})}\BibitemShut {NoStop}%
\bibitem [{\citenamefont {Luo}\ \emph {et~al.}(2019)\citenamefont {Luo},
  \citenamefont {Ren}, \citenamefont {Wang}, \citenamefont {Zhang},
  \citenamefont {Wang},\ and\ \citenamefont {Ren}}]{Ren_19}%
  \BibitemOpen
  \bibfield  {author} {\bibinfo {author} {\bibfnamefont {H.~B.}\ \bibnamefont
  {Luo}}, \bibinfo {author} {\bibfnamefont {Q.}~\bibnamefont {Ren}}, \bibinfo
  {author} {\bibfnamefont {P.}~\bibnamefont {Wang}}, \bibinfo {author}
  {\bibfnamefont {J.}~\bibnamefont {Zhang}}, \bibinfo {author} {\bibfnamefont
  {L.~F.}\ \bibnamefont {Wang}}, \ and\ \bibinfo {author} {\bibfnamefont
  {X.~M.}\ \bibnamefont {Ren}},\ }\href@noop {} {\bibfield  {journal} {\bibinfo
   {journal} {ACS Appl. Mater. Interfaces}\ }\textbf {\bibinfo {volume} {11}}
  (\bibinfo {year} {2019})}\BibitemShut {NoStop}%
\bibitem [{\citenamefont {Daycock}\ \emph {et~al.}(1968)\citenamefont
  {Daycock}, \citenamefont {Jones}, \citenamefont {Evans},\ and\ \citenamefont
  {Thomas}}]{Daycock_68}%
  \BibitemOpen
  \bibfield  {author} {\bibinfo {author} {\bibfnamefont {J.~T.}\ \bibnamefont
  {Daycock}}, \bibinfo {author} {\bibfnamefont {G.~P.}\ \bibnamefont {Jones}},
  \bibinfo {author} {\bibfnamefont {J.~R.~N.}\ \bibnamefont {Evans}}, \ and\
  \bibinfo {author} {\bibfnamefont {J.~M.}\ \bibnamefont {Thomas}},\
  }\href@noop {} {\bibfield  {journal} {\bibinfo  {journal} {Nature}\ }\textbf
  {\bibinfo {volume} {218}} (\bibinfo {year} {1968})}\BibitemShut {NoStop}%
\bibitem [{\citenamefont {Kawada}, \citenamefont {McGhie},\ and\ \citenamefont
  {Labes}(1970)}]{Kawada_70}%
  \BibitemOpen
  \bibfield  {author} {\bibinfo {author} {\bibfnamefont {A.}~\bibnamefont
  {Kawada}}, \bibinfo {author} {\bibfnamefont {A.~R.}\ \bibnamefont {McGhie}},
  \ and\ \bibinfo {author} {\bibfnamefont {M.~M.}\ \bibnamefont {Labes}},\
  }\href@noop {} {\bibfield  {journal} {\bibinfo  {journal} {J. Chem. Phys.}\ }
  (\bibinfo {year} {1970})}\BibitemShut {NoStop}%
\bibitem [{\citenamefont {Car}\ and\ \citenamefont {Parrinello}(1985)}]{CP_85}%
  \BibitemOpen
  \bibfield  {author} {\bibinfo {author} {\bibfnamefont {R.}~\bibnamefont
  {Car}}\ and\ \bibinfo {author} {\bibfnamefont {M.}~\bibnamefont
  {Parrinello}},\ }\href@noop {} {\bibfield  {journal} {\bibinfo  {journal}
  {Phys. Rev. Lett.}\ }\textbf {\bibinfo {volume} {55}},\ \bibinfo {pages}
  {2471} (\bibinfo {year} {1985})}\BibitemShut {NoStop}%
\bibitem [{\citenamefont {Tuckerman}(2002)}]{Tuckerman_AIMD_rev}%
  \BibitemOpen
  \bibfield  {author} {\bibinfo {author} {\bibfnamefont {M.}~\bibnamefont
  {Tuckerman}},\ }\href@noop {} {\bibfield  {journal} {\bibinfo  {journal}
  {Journal of Physics: Condensed Matter}\ }\textbf {\bibinfo {volume} {14}},\
  \bibinfo {pages} {R1297} (\bibinfo {year} {2002})}\BibitemShut {NoStop}%
\bibitem [{\citenamefont {Marx}\ and\ \citenamefont
  {Hutter}(2009)}]{Marx_book}%
  \BibitemOpen
  \bibfield  {author} {\bibinfo {author} {\bibfnamefont {D.}~\bibnamefont
  {Marx}}\ and\ \bibinfo {author} {\bibfnamefont {J.}~\bibnamefont {Hutter}},\
  }\href@noop {} {\emph {\bibinfo {title} {Ab Initio Molecular Dynamics: Basic
  Theory and Advanced Methods}}}\ (\bibinfo  {publisher} {Cambridge University
  Press},\ \bibinfo {address} {Cambridge},\ \bibinfo {year} {2009})\BibitemShut
  {NoStop}%
\bibitem [{\citenamefont {Tuckerman}, \citenamefont {Martyna},\ and\
  \citenamefont {Berne}(1992)}]{tuck3}%
  \BibitemOpen
  \bibfield  {author} {\bibinfo {author} {\bibfnamefont {M.}~\bibnamefont
  {Tuckerman}}, \bibinfo {author} {\bibfnamefont {G.}~\bibnamefont {Martyna}},
  \ and\ \bibinfo {author} {\bibfnamefont {B.}~\bibnamefont {Berne}},\
  }\href@noop {} {\bibfield  {journal} {\bibinfo  {journal} {J. Chem. Phys.}\
  }\textbf {\bibinfo {volume} {97}},\ \bibinfo {pages} {1990} (\bibinfo {year}
  {1992})}\BibitemShut {NoStop}%
\bibitem [{\citenamefont {Luehr}, \citenamefont {Markland},\ and\ \citenamefont
  {Martinez}(2014)}]{tom3}%
  \BibitemOpen
  \bibfield  {author} {\bibinfo {author} {\bibfnamefont {N.}~\bibnamefont
  {Luehr}}, \bibinfo {author} {\bibfnamefont {T.~E.}\ \bibnamefont {Markland}},
  \ and\ \bibinfo {author} {\bibfnamefont {T.~J.}\ \bibnamefont {Martinez}},\
  }\href@noop {} {\bibfield  {journal} {\bibinfo  {journal} {J. Chem. Phys.}\
  }\textbf {\bibinfo {volume} {140}},\ \bibinfo {pages} {2014} (\bibinfo {year}
  {2014})}\BibitemShut {NoStop}%
\bibitem [{\citenamefont {Marsalek}\ and\ \citenamefont
  {Markland}(2016)}]{Marsalek2016}%
  \BibitemOpen
  \bibfield  {author} {\bibinfo {author} {\bibfnamefont {O.}~\bibnamefont
  {Marsalek}}\ and\ \bibinfo {author} {\bibfnamefont {T.~E.}\ \bibnamefont
  {Markland}},\ }\href@noop {} {\bibfield  {journal} {\bibinfo  {journal} {The
  Journal of Chemical Physics}\ }\textbf {\bibinfo {volume} {144}},\ \bibinfo
  {pages} {054112} (\bibinfo {year} {2016})}\BibitemShut {NoStop}%
\bibitem [{\citenamefont {Kreuer}\ \emph {et~al.}(1998)\citenamefont {Kreuer},
  \citenamefont {Fuchs}, \citenamefont {Ise}, \citenamefont {Spaeth},\ and\
  \citenamefont {Maier}}]{kreuer1998imidazole}%
  \BibitemOpen
  \bibfield  {author} {\bibinfo {author} {\bibfnamefont {K.}~\bibnamefont
  {Kreuer}}, \bibinfo {author} {\bibfnamefont {A.}~\bibnamefont {Fuchs}},
  \bibinfo {author} {\bibfnamefont {M.}~\bibnamefont {Ise}}, \bibinfo {author}
  {\bibfnamefont {M.}~\bibnamefont {Spaeth}}, \ and\ \bibinfo {author}
  {\bibfnamefont {J.}~\bibnamefont {Maier}},\ }\href@noop {} {\bibfield
  {journal} {\bibinfo  {journal} {Electrochimica Acta}\ }\textbf {\bibinfo
  {volume} {43}},\ \bibinfo {pages} {1281} (\bibinfo {year}
  {1998})}\BibitemShut {NoStop}%
\bibitem [{ics(2008)}]{icsc_data}%
  \BibitemOpen
  \href
  {http://www.ilo.org/dyn/icsc/showcard.display?p_version=2&p_card_id=1721}
  {\enquote {\bibinfo {title} {International chemical safety cards (icscs)},}\
  } (\bibinfo {year} {2008})\BibitemShut {NoStop}%
\bibitem [{\citenamefont {Napoli}, \citenamefont {Marsalek},\ and\
  \citenamefont {Markland}(2018)}]{Napoli2018}%
  \BibitemOpen
  \bibfield  {author} {\bibinfo {author} {\bibfnamefont {J.~A.}\ \bibnamefont
  {Napoli}}, \bibinfo {author} {\bibfnamefont {O.}~\bibnamefont {Marsalek}}, \
  and\ \bibinfo {author} {\bibfnamefont {T.~E.}\ \bibnamefont {Markland}},\
  }\href@noop {} {\bibfield  {journal} {\bibinfo  {journal} {The Journal of
  Chemical Physics}\ }\textbf {\bibinfo {volume} {148}},\ \bibinfo {pages}
  {222833} (\bibinfo {year} {2018})}\BibitemShut {NoStop}%
\bibitem [{\citenamefont {Markland}\ and\ \citenamefont
  {Manolopoulos}(2008{\natexlab{a}})}]{Markland2008}%
  \BibitemOpen
  \bibfield  {author} {\bibinfo {author} {\bibfnamefont {T.~E.}\ \bibnamefont
  {Markland}}\ and\ \bibinfo {author} {\bibfnamefont {D.~E.}\ \bibnamefont
  {Manolopoulos}},\ }\href@noop {} {\bibfield  {journal} {\bibinfo  {journal}
  {The Journal of Chemical Physics}\ }\textbf {\bibinfo {volume} {129}},\
  \bibinfo {pages} {024105} (\bibinfo {year} {2008}{\natexlab{a}})}\BibitemShut
  {NoStop}%
\bibitem [{\citenamefont {Markland}\ and\ \citenamefont
  {Manolopoulos}(2008{\natexlab{b}})}]{Markland2008a}%
  \BibitemOpen
  \bibfield  {author} {\bibinfo {author} {\bibfnamefont {T.~E.}\ \bibnamefont
  {Markland}}\ and\ \bibinfo {author} {\bibfnamefont {D.~E.}\ \bibnamefont
  {Manolopoulos}},\ }\href@noop {} {\bibfield  {journal} {\bibinfo  {journal}
  {Chemical Physics Letters}\ }\textbf {\bibinfo {volume} {464}},\ \bibinfo
  {pages} {256 } (\bibinfo {year} {2008}{\natexlab{b}})}\BibitemShut {NoStop}%
\bibitem [{\citenamefont {Chandra}, \citenamefont {Tuckerman},\ and\
  \citenamefont {Marx}(2007)}]{Chandra_07}%
  \BibitemOpen
  \bibfield  {author} {\bibinfo {author} {\bibfnamefont {A.}~\bibnamefont
  {Chandra}}, \bibinfo {author} {\bibfnamefont {M.~E.}\ \bibnamefont
  {Tuckerman}}, \ and\ \bibinfo {author} {\bibfnamefont {D.}~\bibnamefont
  {Marx}},\ }\href@noop {} {\bibfield  {journal} {\bibinfo  {journal} {Phys.
  Rev. Lett.}\ }\textbf {\bibinfo {volume} {99}} (\bibinfo {year}
  {2007})}\BibitemShut {NoStop}%
\bibitem [{\citenamefont {Tuckerman}, \citenamefont {Chandra},\ and\
  \citenamefont {Marx}(2010)}]{Tuckerman_10}%
  \BibitemOpen
  \bibfield  {author} {\bibinfo {author} {\bibfnamefont {M.~E.}\ \bibnamefont
  {Tuckerman}}, \bibinfo {author} {\bibfnamefont {A.}~\bibnamefont {Chandra}},
  \ and\ \bibinfo {author} {\bibfnamefont {D.}~\bibnamefont {Marx}},\
  }\href@noop {} {\bibfield  {journal} {\bibinfo  {journal} {J. Chem. Phys.}\
  }\textbf {\bibinfo {volume} {133}} (\bibinfo {year} {2010})}\BibitemShut
  {NoStop}%
\bibitem [{\citenamefont {Riehl}(1965)}]{Riehl_65}%
  \BibitemOpen
  \bibfield  {author} {\bibinfo {author} {\bibfnamefont {N.}~\bibnamefont
  {Riehl}},\ }\href@noop {} {\bibfield  {journal} {\bibinfo  {journal} {Trans.
  N. Y. Acad. Sci.}\ }\textbf {\bibinfo {volume} {27}} (\bibinfo {year}
  {1965})}\BibitemShut {NoStop}%
\bibitem [{\citenamefont {Goward}\ \emph {et~al.}(2002)\citenamefont {Goward},
  \citenamefont {Schuster}, \citenamefont {Sebastiani}, \citenamefont
  {Schnell},\ and\ \citenamefont {Spiess}}]{Goward_02}%
  \BibitemOpen
  \bibfield  {author} {\bibinfo {author} {\bibfnamefont {G.~R.}\ \bibnamefont
  {Goward}}, \bibinfo {author} {\bibfnamefont {M.~F.~H.}\ \bibnamefont
  {Schuster}}, \bibinfo {author} {\bibfnamefont {D.}~\bibnamefont
  {Sebastiani}}, \bibinfo {author} {\bibfnamefont {I.}~\bibnamefont {Schnell}},
  \ and\ \bibinfo {author} {\bibfnamefont {H.~W.}\ \bibnamefont {Spiess}},\
  }\href@noop {} {\bibfield  {journal} {\bibinfo  {journal} {J. Phys. Chem. B}\
  }\textbf {\bibinfo {volume} {106}} (\bibinfo {year} {2002})}\BibitemShut
  {NoStop}%
\bibitem [{\citenamefont {Fischbach}\ \emph {et~al.}(2004)\citenamefont
  {Fischbach}, \citenamefont {Spiess}, \citenamefont {Saalw\"{a}chter},\ and\
  \citenamefont {Goward}}]{Goward_04}%
  \BibitemOpen
  \bibfield  {author} {\bibinfo {author} {\bibfnamefont {I.}~\bibnamefont
  {Fischbach}}, \bibinfo {author} {\bibfnamefont {H.~W.}\ \bibnamefont
  {Spiess}}, \bibinfo {author} {\bibfnamefont {K.}~\bibnamefont
  {Saalw\"{a}chter}}, \ and\ \bibinfo {author} {\bibfnamefont {G.~R.}\
  \bibnamefont {Goward}},\ }\href@noop {} {\bibfield  {journal} {\bibinfo
  {journal} {J. Phys. Chem. B}\ }\textbf {\bibinfo {volume} {108}} (\bibinfo
  {year} {2004})}\BibitemShut {NoStop}%
\bibitem [{\citenamefont {Herz}\ \emph {et~al.}(2003)\citenamefont {Herz},
  \citenamefont {Kreuer}, \citenamefont {Maier}, \citenamefont
  {Scharfenberger}, \citenamefont {Schuster},\ and\ \citenamefont
  {Meyer}}]{Herz_03}%
  \BibitemOpen
  \bibfield  {author} {\bibinfo {author} {\bibfnamefont {H.~G.}\ \bibnamefont
  {Herz}}, \bibinfo {author} {\bibfnamefont {K.~D.}\ \bibnamefont {Kreuer}},
  \bibinfo {author} {\bibfnamefont {J.}~\bibnamefont {Maier}}, \bibinfo
  {author} {\bibfnamefont {G.}~\bibnamefont {Scharfenberger}}, \bibinfo
  {author} {\bibfnamefont {M.~F.~H.}\ \bibnamefont {Schuster}}, \ and\ \bibinfo
  {author} {\bibfnamefont {W.~H.}\ \bibnamefont {Meyer}},\ }\href@noop {}
  {\bibfield  {journal} {\bibinfo  {journal} {Electrochim. Acta}\ }\textbf
  {\bibinfo {volume} {48}} (\bibinfo {year} {2003})}\BibitemShut {NoStop}%
\bibitem [{\citenamefont {Bureekaew}\ \emph {et~al.}(2009)\citenamefont
  {Bureekaew}, \citenamefont {Horike}, \citenamefont {Higuchi}, \citenamefont
  {Mizuno}, \citenamefont {Kawamura}, \citenamefont {Tanaka}, \citenamefont
  {Yanai},\ and\ \citenamefont {Kitagawa}}]{Kitagawa_09_MOF}%
  \BibitemOpen
  \bibfield  {author} {\bibinfo {author} {\bibfnamefont {S.}~\bibnamefont
  {Bureekaew}}, \bibinfo {author} {\bibfnamefont {S.}~\bibnamefont {Horike}},
  \bibinfo {author} {\bibfnamefont {M.}~\bibnamefont {Higuchi}}, \bibinfo
  {author} {\bibfnamefont {M.}~\bibnamefont {Mizuno}}, \bibinfo {author}
  {\bibfnamefont {T.}~\bibnamefont {Kawamura}}, \bibinfo {author}
  {\bibfnamefont {D.}~\bibnamefont {Tanaka}}, \bibinfo {author} {\bibfnamefont
  {N.}~\bibnamefont {Yanai}}, \ and\ \bibinfo {author} {\bibfnamefont
  {S.}~\bibnamefont {Kitagawa}},\ }\href@noop {} {\bibfield  {journal}
  {\bibinfo  {journal} {Nat. Mater.}\ }\textbf {\bibinfo {volume} {8}}
  (\bibinfo {year} {2009})}\BibitemShut {NoStop}%
\bibitem [{\citenamefont {Homburg}\ \emph {et~al.}(2016)\citenamefont
  {Homburg}, \citenamefont {Hartwig}, \citenamefont {Reinsch}, \citenamefont
  {Wark},\ and\ \citenamefont {Stock}}]{Stock_16_MOF}%
  \BibitemOpen
  \bibfield  {author} {\bibinfo {author} {\bibfnamefont {T.}~\bibnamefont
  {Homburg}}, \bibinfo {author} {\bibfnamefont {C.}~\bibnamefont {Hartwig}},
  \bibinfo {author} {\bibfnamefont {H.}~\bibnamefont {Reinsch}}, \bibinfo
  {author} {\bibfnamefont {M.}~\bibnamefont {Wark}}, \ and\ \bibinfo {author}
  {\bibfnamefont {N.}~\bibnamefont {Stock}},\ }\href@noop {} {\bibfield
  {journal} {\bibinfo  {journal} {Dalton Trans.}\ }\textbf {\bibinfo {volume}
  {45}} (\bibinfo {year} {2016})}\BibitemShut {NoStop}%
\end{thebibliography}

\begin{thebibliography}{20}%
\makeatletter
\providecommand \@ifxundefined [1]{%
 \@ifx{#1\undefined}
}%
\providecommand \@ifnum [1]{%
 \ifnum #1\expandafter \@firstoftwo
 \else \expandafter \@secondoftwo
 \fi
}%
\providecommand \@ifx [1]{%
 \ifx #1\expandafter \@firstoftwo
 \else \expandafter \@secondoftwo
 \fi
}%
\providecommand \natexlab [1]{#1}%
\providecommand \enquote  [1]{``#1''}%
\providecommand \bibnamefont  [1]{#1}%
\providecommand \bibfnamefont [1]{#1}%
\providecommand \citenamefont [1]{#1}%
\providecommand \href@noop [0]{\@secondoftwo}%
\providecommand \href [0]{\begingroup \@sanitize@url \@href}%
\providecommand \@href[1]{\@@startlink{#1}\@@href}%
\providecommand \@@href[1]{\endgroup#1\@@endlink}%
\providecommand \@sanitize@url [0]{\catcode `\\12\catcode `\$12\catcode
  `\&12\catcode `\#12\catcode `\^12\catcode `\_12\catcode `\%12\relax}%
\providecommand \@@startlink[1]{}%
\providecommand \@@endlink[0]{}%
\providecommand \url  [0]{\begingroup\@sanitize@url \@url }%
\providecommand \@url [1]{\endgroup\@href {#1}{\urlprefix }}%
\providecommand \urlprefix  [0]{URL }%
\providecommand \Eprint [0]{\href }%
\providecommand \doibase [0]{http://dx.doi.org/}%
\providecommand \selectlanguage [0]{\@gobble}%
\providecommand \bibinfo  [0]{\@secondoftwo}%
\providecommand \bibfield  [0]{\@secondoftwo}%
\providecommand \translation [1]{[#1]}%
\providecommand \BibitemOpen [0]{}%
\providecommand \bibitemStop [0]{}%
\providecommand \bibitemNoStop [0]{.\EOS\space}%
\providecommand \EOS [0]{\spacefactor3000\relax}%
\providecommand \BibitemShut  [1]{\csname bibitem#1\endcsname}%
\let\auto@bib@innerbib\@empty
\bibitem [{\citenamefont {Tuckerman}\ \emph
  {et~al.}(1992{\natexlab{a}})\citenamefont {Tuckerman}, \citenamefont
  {Berne},\ and\ \citenamefont {Martyna}}]{tuckerman1992reversible}%
  \BibitemOpen
  \bibfield  {author} {\bibinfo {author} {\bibfnamefont {M.}~\bibnamefont
  {Tuckerman}}, \bibinfo {author} {\bibfnamefont {B.~J.}\ \bibnamefont
  {Berne}}, \ and\ \bibinfo {author} {\bibfnamefont {G.~J.}\ \bibnamefont
  {Martyna}},\ }\href@noop {} {\bibfield  {journal} {\bibinfo  {journal} {J.
  Chem. Phys.}\ }\textbf {\bibinfo {volume} {97}},\ \bibinfo {pages} {1990}
  (\bibinfo {year} {1992}{\natexlab{a}})}\BibitemShut {NoStop}%
\bibitem [{\citenamefont {Luehr}\ \emph {et~al.}(2014)\citenamefont {Luehr},
  \citenamefont {Markland},\ and\ \citenamefont
  {Mart{\'\i}nez}}]{luehr2014multiple}%
  \BibitemOpen
  \bibfield  {author} {\bibinfo {author} {\bibfnamefont {N.}~\bibnamefont
  {Luehr}}, \bibinfo {author} {\bibfnamefont {T.~E.}\ \bibnamefont {Markland}},
  \ and\ \bibinfo {author} {\bibfnamefont {T.~J.}\ \bibnamefont
  {Mart{\'\i}nez}},\ }\href@noop {} {\bibfield  {journal} {\bibinfo  {journal}
  {J. Chem. Phys.}\ }\textbf {\bibinfo {volume} {140}},\ \bibinfo {pages}
  {084116} (\bibinfo {year} {2014})}\BibitemShut {NoStop}%
\bibitem [{\citenamefont {Tuckerman}\ \emph
  {et~al.}(1992{\natexlab{b}})\citenamefont {Tuckerman}, \citenamefont
  {Martyna},\ and\ \citenamefont {Berne}}]{tuck3}%
  \BibitemOpen
  \bibfield  {author} {\bibinfo {author} {\bibfnamefont {M.}~\bibnamefont
  {Tuckerman}}, \bibinfo {author} {\bibfnamefont {G.}~\bibnamefont {Martyna}},
  \ and\ \bibinfo {author} {\bibfnamefont {B.}~\bibnamefont {Berne}},\
  }\href@noop {} {\bibfield  {journal} {\bibinfo  {journal} {J. Chem. Phys.}\
  }\textbf {\bibinfo {volume} {97}},\ \bibinfo {pages} {1990} (\bibinfo {year}
  {1992}{\natexlab{b}})}\BibitemShut {NoStop}%
\bibitem [{\citenamefont {Lide}(2007)}]{crchandbook}%
  \BibitemOpen
  \bibfield  {author} {\bibinfo {author} {\bibfnamefont {D.~R.}\ \bibnamefont
  {Lide}},\ }in\ \href@noop {} {\emph {\bibinfo {booktitle} {Handbook of
  Chemistry and Physics}}}\ (\bibinfo  {publisher} {CRC Press},\ \bibinfo
  {address} {Boca Raton, FL},\ \bibinfo {year} {2007})\ \bibinfo {edition}
  {88th}\ ed.\BibitemShut {Stop}%
\bibitem [{\citenamefont {Gaus}\ \emph {et~al.}(2011)\citenamefont {Gaus},
  \citenamefont {Cui},\ and\ \citenamefont {Elstner}}]{gaus2011dftb3}%
  \BibitemOpen
  \bibfield  {author} {\bibinfo {author} {\bibfnamefont {M.}~\bibnamefont
  {Gaus}}, \bibinfo {author} {\bibfnamefont {Q.}~\bibnamefont {Cui}}, \ and\
  \bibinfo {author} {\bibfnamefont {M.}~\bibnamefont {Elstner}},\ }\href@noop
  {} {\bibfield  {journal} {\bibinfo  {journal} {Journal of chemical theory and
  computation}\ }\textbf {\bibinfo {volume} {7}},\ \bibinfo {pages} {931}
  (\bibinfo {year} {2011})}\BibitemShut {NoStop}%
\bibitem [{\citenamefont {Perdew}\ \emph {et~al.}(1996)\citenamefont {Perdew},
  \citenamefont {Burke},\ and\ \citenamefont
  {Ernzerhof}}]{perdew1996generalized}%
  \BibitemOpen
  \bibfield  {author} {\bibinfo {author} {\bibfnamefont {J.~P.}\ \bibnamefont
  {Perdew}}, \bibinfo {author} {\bibfnamefont {K.}~\bibnamefont {Burke}}, \
  and\ \bibinfo {author} {\bibfnamefont {M.}~\bibnamefont {Ernzerhof}},\
  }\href@noop {} {\bibfield  {journal} {\bibinfo  {journal} {Phys. Rev. Lett.}\
  }\textbf {\bibinfo {volume} {77}},\ \bibinfo {pages} {3865} (\bibinfo {year}
  {1996})}\BibitemShut {NoStop}%
\bibitem [{\citenamefont {Zhang}\ and\ \citenamefont
  {Yang}(1998)}]{zhang1998comment}%
  \BibitemOpen
  \bibfield  {author} {\bibinfo {author} {\bibfnamefont {Y.}~\bibnamefont
  {Zhang}}\ and\ \bibinfo {author} {\bibfnamefont {W.}~\bibnamefont {Yang}},\
  }\href@noop {} {\bibfield  {journal} {\bibinfo  {journal} {Phys. Rev. Lett.}\
  }\textbf {\bibinfo {volume} {80}},\ \bibinfo {pages} {890} (\bibinfo {year}
  {1998})}\BibitemShut {NoStop}%
\bibitem [{\citenamefont {Grimme}\ \emph {et~al.}(2010)\citenamefont {Grimme},
  \citenamefont {Antony}, \citenamefont {Ehrlich},\ and\ \citenamefont
  {Krieg}}]{grimme2010consistent}%
  \BibitemOpen
  \bibfield  {author} {\bibinfo {author} {\bibfnamefont {S.}~\bibnamefont
  {Grimme}}, \bibinfo {author} {\bibfnamefont {J.}~\bibnamefont {Antony}},
  \bibinfo {author} {\bibfnamefont {S.}~\bibnamefont {Ehrlich}}, \ and\
  \bibinfo {author} {\bibfnamefont {H.}~\bibnamefont {Krieg}},\ }\href@noop {}
  {\bibfield  {journal} {\bibinfo  {journal} {J. Chem. Phys.}\ }\textbf
  {\bibinfo {volume} {132}},\ \bibinfo {pages} {154104} (\bibinfo {year}
  {2010})}\BibitemShut {NoStop}%
\bibitem [{\citenamefont {Goedecker}\ \emph {et~al.}(1996)\citenamefont
  {Goedecker}, \citenamefont {Teter},\ and\ \citenamefont
  {Hutter}}]{goedecker1996separable}%
  \BibitemOpen
  \bibfield  {author} {\bibinfo {author} {\bibfnamefont {S.}~\bibnamefont
  {Goedecker}}, \bibinfo {author} {\bibfnamefont {M.}~\bibnamefont {Teter}}, \
  and\ \bibinfo {author} {\bibfnamefont {J.}~\bibnamefont {Hutter}},\
  }\href@noop {} {\bibfield  {journal} {\bibinfo  {journal} {Physical Review
  B}\ }\textbf {\bibinfo {volume} {54}},\ \bibinfo {pages} {1703} (\bibinfo
  {year} {1996})}\BibitemShut {NoStop}%
\bibitem [{\citenamefont {Bussi}\ \emph {et~al.}(2007)\citenamefont {Bussi},
  \citenamefont {Donadio},\ and\ \citenamefont
  {Parrinello}}]{Bussi2007/10.1063/1.2408420}%
  \BibitemOpen
  \bibfield  {author} {\bibinfo {author} {\bibfnamefont {G.}~\bibnamefont
  {Bussi}}, \bibinfo {author} {\bibfnamefont {D.}~\bibnamefont {Donadio}}, \
  and\ \bibinfo {author} {\bibfnamefont {M.}~\bibnamefont {Parrinello}},\
  }\href {\doibase 10.1063/1.2408420} {\bibfield  {journal} {\bibinfo
  {journal} {J. Chem. Phys.}\ }\textbf {\bibinfo {volume} {126}},\ \bibinfo
  {pages} {014101} (\bibinfo {year} {2007})}\BibitemShut {NoStop}%
\bibitem [{\citenamefont {Ceriotti}\ \emph
  {et~al.}(2010{\natexlab{a}})\citenamefont {Ceriotti}, \citenamefont
  {Parrinello}, \citenamefont {Markland},\ and\ \citenamefont
  {Manolopoulos}}]{Ceriotti2010/10.1063/1.3489925}%
  \BibitemOpen
  \bibfield  {author} {\bibinfo {author} {\bibfnamefont {M.}~\bibnamefont
  {Ceriotti}}, \bibinfo {author} {\bibfnamefont {M.}~\bibnamefont
  {Parrinello}}, \bibinfo {author} {\bibfnamefont {T.~E.}\ \bibnamefont
  {Markland}}, \ and\ \bibinfo {author} {\bibfnamefont {D.~E.}\ \bibnamefont
  {Manolopoulos}},\ }\href {\doibase 10.1063/1.3489925} {\bibfield  {journal}
  {\bibinfo  {journal} {J. Chem. Phys.}\ }\textbf {\bibinfo {volume} {133}},\
  \bibinfo {pages} {124104} (\bibinfo {year} {2010}{\natexlab{a}})}\BibitemShut
  {NoStop}%
\bibitem [{\citenamefont {Markland}\ and\ \citenamefont
  {Manolopoulos}(2008{\natexlab{a}})}]{markland_mano_1}%
  \BibitemOpen
  \bibfield  {author} {\bibinfo {author} {\bibfnamefont {T.~E.}\ \bibnamefont
  {Markland}}\ and\ \bibinfo {author} {\bibfnamefont {D.~E.}\ \bibnamefont
  {Manolopoulos}},\ }\href {\doibase 10.1063/1.2953308} {\bibfield  {journal}
  {\bibinfo  {journal} {The Journal of Chemical Physics}\ }\textbf {\bibinfo
  {volume} {129}},\ \bibinfo {pages} {024105} (\bibinfo {year}
  {2008}{\natexlab{a}})},\ \Eprint
  {http://arxiv.org/abs/https://doi.org/10.1063/1.2953308}
  {https://doi.org/10.1063/1.2953308} \BibitemShut {NoStop}%
\bibitem [{\citenamefont {Markland}\ and\ \citenamefont
  {Manolopoulos}(2008{\natexlab{b}})}]{markland2008refined}%
  \BibitemOpen
  \bibfield  {author} {\bibinfo {author} {\bibfnamefont {T.~E.}\ \bibnamefont
  {Markland}}\ and\ \bibinfo {author} {\bibfnamefont {D.~E.}\ \bibnamefont
  {Manolopoulos}},\ }\href@noop {} {\bibfield  {journal} {\bibinfo  {journal}
  {Chemical Physics Letters}\ }\textbf {\bibinfo {volume} {464}},\ \bibinfo
  {pages} {256} (\bibinfo {year} {2008}{\natexlab{b}})}\BibitemShut {NoStop}%
\bibitem [{\citenamefont {Marsalek}\ and\ \citenamefont
  {Markland}(2016)}]{ondrej_markland}%
  \BibitemOpen
  \bibfield  {author} {\bibinfo {author} {\bibfnamefont {O.}~\bibnamefont
  {Marsalek}}\ and\ \bibinfo {author} {\bibfnamefont {T.~E.}\ \bibnamefont
  {Markland}},\ }\href {\doibase 10.1063/1.4941093} {\bibfield  {journal}
  {\bibinfo  {journal} {The Journal of Chemical Physics}\ }\textbf {\bibinfo
  {volume} {144}},\ \bibinfo {pages} {054112} (\bibinfo {year} {2016})},\
  \Eprint {http://arxiv.org/abs/https://doi.org/10.1063/1.4941093}
  {https://doi.org/10.1063/1.4941093} \BibitemShut {NoStop}%
\bibitem [{\citenamefont {Ceriotti}\ \emph
  {et~al.}(2010{\natexlab{b}})\citenamefont {Ceriotti}, \citenamefont
  {Parrinello}, \citenamefont {Markland},\ and\ \citenamefont
  {Manolopoulos}}]{pimd_thermostat}%
  \BibitemOpen
  \bibfield  {author} {\bibinfo {author} {\bibfnamefont {M.}~\bibnamefont
  {Ceriotti}}, \bibinfo {author} {\bibfnamefont {M.}~\bibnamefont
  {Parrinello}}, \bibinfo {author} {\bibfnamefont {T.~E.}\ \bibnamefont
  {Markland}}, \ and\ \bibinfo {author} {\bibfnamefont {D.~E.}\ \bibnamefont
  {Manolopoulos}},\ }\href {\doibase 10.1063/1.3489925} {\bibfield  {journal}
  {\bibinfo  {journal} {The Journal of Chemical Physics}\ }\textbf {\bibinfo
  {volume} {133}},\ \bibinfo {pages} {124104} (\bibinfo {year}
  {2010}{\natexlab{b}})},\ \Eprint
  {http://arxiv.org/abs/https://doi.org/10.1063/1.3489925}
  {https://doi.org/10.1063/1.3489925} \BibitemShut {NoStop}%
\bibitem [{\citenamefont {Hayes}\ \emph {et~al.}(2009)\citenamefont {Hayes},
  \citenamefont {Paddison},\ and\ \citenamefont
  {Tuckerman}}]{Hayes_Tuckerman1}%
  \BibitemOpen
  \bibfield  {author} {\bibinfo {author} {\bibfnamefont {R.~L.}\ \bibnamefont
  {Hayes}}, \bibinfo {author} {\bibfnamefont {S.~J.}\ \bibnamefont {Paddison}},
  \ and\ \bibinfo {author} {\bibfnamefont {M.~E.}\ \bibnamefont {Tuckerman}},\
  }\href@noop {} {\bibfield  {journal} {\bibinfo  {journal} {J. Phys. Chem. B}\
  }\textbf {\bibinfo {volume} {113}} (\bibinfo {year} {2009})}\BibitemShut
  {NoStop}%
\bibitem [{\citenamefont {Hayes}\ \emph {et~al.}(2011)\citenamefont {Hayes},
  \citenamefont {Paddison},\ and\ \citenamefont
  {Tuckerman}}]{Hayes_Tuckerman2}%
  \BibitemOpen
  \bibfield  {author} {\bibinfo {author} {\bibfnamefont {R.~L.}\ \bibnamefont
  {Hayes}}, \bibinfo {author} {\bibfnamefont {S.~J.}\ \bibnamefont {Paddison}},
  \ and\ \bibinfo {author} {\bibfnamefont {M.~E.}\ \bibnamefont {Tuckerman}},\
  }\href@noop {} {\bibfield  {journal} {\bibinfo  {journal} {J. Phys. Chem. A}\
  }\textbf {\bibinfo {volume} {115}} (\bibinfo {year} {2011})}\BibitemShut
  {NoStop}%
\bibitem [{\citenamefont {Li}\ \emph {et~al.}(2012)\citenamefont {Li},
  \citenamefont {Li}, \citenamefont {Yan},\ and\ \citenamefont
  {Shen}}]{Shen_12}%
  \BibitemOpen
  \bibfield  {author} {\bibinfo {author} {\bibfnamefont {A.~L.}\ \bibnamefont
  {Li}}, \bibinfo {author} {\bibfnamefont {Y.}~\bibnamefont {Li}}, \bibinfo
  {author} {\bibfnamefont {T.~Y.}\ \bibnamefont {Yan}}, \ and\ \bibinfo
  {author} {\bibfnamefont {P.~W.}\ \bibnamefont {Shen}},\ }\href@noop {}
  {\bibfield  {journal} {\bibinfo  {journal} {J. Phys. Chem. B}\ }\textbf
  {\bibinfo {volume} {116}} (\bibinfo {year} {2012})}\BibitemShut {NoStop}%
\bibitem [{\citenamefont {Lynden-Bell}\ and\ \citenamefont
  {McDonald}(1981)}]{doi:10.1080/00268978100102181}%
  \BibitemOpen
  \bibfield  {author} {\bibinfo {author} {\bibfnamefont {R.}~\bibnamefont
  {Lynden-Bell}}\ and\ \bibinfo {author} {\bibfnamefont {I.}~\bibnamefont
  {McDonald}},\ }\href {\doibase 10.1080/00268978100102181} {\bibfield
  {journal} {\bibinfo  {journal} {Molecular Physics}\ }\textbf {\bibinfo
  {volume} {43}},\ \bibinfo {pages} {1429} (\bibinfo {year} {1981})},\ \Eprint
  {http://arxiv.org/abs/https://doi.org/10.1080/00268978100102181}
  {https://doi.org/10.1080/00268978100102181} \BibitemShut {NoStop}%
\bibitem [{\citenamefont {Impey}\ \emph {et~al.}(1982)\citenamefont {Impey},
  \citenamefont {Madden},\ and\ \citenamefont
  {McDonald}}]{doi:10.1080/00268978200101361}%
  \BibitemOpen
  \bibfield  {author} {\bibinfo {author} {\bibfnamefont {R.}~\bibnamefont
  {Impey}}, \bibinfo {author} {\bibfnamefont {P.}~\bibnamefont {Madden}}, \
  and\ \bibinfo {author} {\bibfnamefont {I.}~\bibnamefont {McDonald}},\ }\href
  {\doibase 10.1080/00268978200101361} {\bibfield  {journal} {\bibinfo
  {journal} {Molecular Physics}\ }\textbf {\bibinfo {volume} {46}},\ \bibinfo
  {pages} {513} (\bibinfo {year} {1982})},\ \Eprint
  {http://arxiv.org/abs/https://doi.org/10.1080/00268978200101361}
  {https://doi.org/10.1080/00268978200101361} \BibitemShut {NoStop}%
\end{thebibliography}
\end{document}